\documentclass[aps,pra,superscriptaddress,floatfix,twocolumn,longbibliography]{revtex4-1}

\usepackage[final]{graphicx}
\graphicspath{{figures/}}       

\usepackage{dcolumn} 
\usepackage{bm} 
\usepackage{amsmath}
\usepackage{amsfonts} 
\usepackage{amssymb}
\usepackage[utf8]{inputenc}
\usepackage{textcomp}
\usepackage[english]{babel}
\usepackage[usenames,dvipsnames]{xcolor}
\usepackage{courier}
\usepackage{etoolbox}
\usepackage[normalem]{ulem}
\usepackage{enumerate}
\usepackage{hyperref}
\usepackage{soul}

\usepackage{listings}
\usepackage{color}

\begin{document}
\title{Nonlinear dynamics and quantum chaos of a family of kicked $p$-spin models}

\author{Manuel H. Muñoz-Arias}
\email{mhmunoz@unm.edu}
\affiliation{Center for Quantum Information and Control, Department of Physics 
and Astronomy, University of New Mexico, Albuquerque, New Mexico 87131, USA}
\author{Pablo M. Poggi}
\affiliation{Center for Quantum Information and Control, Department of Physics 
and Astronomy, University of New Mexico, Albuquerque, New Mexico 87131, USA}
\author{Ivan H. Deutsch}
\affiliation{Center for Quantum Information and Control, Department of Physics 
and Astronomy, University of New Mexico, Albuquerque, New Mexico 87131, USA}

\begin{abstract}
We introduce kicked $p$-spin models describing a family of transverse Ising-like models for an ensemble of spin-$1/2$ particles with all-to-all $p$-body interaction terms occurring periodically in time as delta-kicks. This is the natural generalization of the well-studied quantum kicked top ($p$=2)~\cite{Haake1987}. We fully characterize the classical nonlinear dynamics of these models, including the transition to global Hamiltonian chaos. The classical analysis allows us to build a classification for this family of models, distinguishing between $p=2$ and $p>2$, and  between models with odd and even $p$'s. Quantum chaos in these models is characterized in both kinematic and dynamic signatures. For the latter we show numerically that the growth rate of the out-of-time-order correlator is dictated by the classical Lyapunov exponent. Finally, we argue that the classification of these models constructed in the classical system applies to the quantum system as well.
\end{abstract}

\date{\today}

\maketitle

\section{Introduction}
Ising-like models play a central role in quantum information science at the interface of statistical physics and computation~\cite{Nakahara2013}.  Fundamental areas of  research include Hamiltonian complexity~\cite{Lucas2014}, optimization~\cite{Albash2018}, machine learning~\cite{Biamonte2017}, spin glasses~\cite{Kirkpatrick1987}, and critical phenomena in many-body systems such as quantum ground-state phase transitions~\cite{Sachdev2011,Peng2005,Filippone2011} and dynamical phase transitions~\cite{Zhang2017,Jurcevic2017,Zunkovic2018,Heyl2013}.  Understanding dynamics in such systems is essential for studies of nonequilibrium physics, such as many-body quantum chaos \cite{Gubin2012,Kos2018}, and thermalization~\cite{Gogolin2016,Rigol2016}.

Today, quantum simulation offers the prospect of studying Ising-like models by encoding spins in qubits and engineering the desired interactions in a controlled way~\cite{Simon2011, Zeiher2017, Blatt2012, Monroe2019, Scholl2020, Ebadi2020}. One approach to quantum simulation is to employ a gate-based model in order to implement a desired unitary evolution of the many-body system.  The seminal work of Lloyd~\cite{Lloyd1996} showed that through a Trotter-Suzuki expansion one can approximate any desired unitary map on $N$ qubits with $k$-local interactions through an appropriate sequence of gates acting on no more than $k$ qubits at a time. While, such a gate-based protocol is often called ``digital quantum simulation," when implemented in a non-fault-tolerant manner, the operation is fundamentally ``analog," with gates chosen for a continuum of possible duration.  As such, the resulting map can exhibit dynamical instabilities and quantum chaos, which can lead to a proliferation of errors~\cite{Heyl2019, Sieberer2019}.

Of particular importance in this context is the fact that Trotterization introduces a hidden time-dependent driving force.  Explicitly, given a generic time independent Hamiltonian $H=H_y+H_z$ where $[H_y,H_z] \neq 0$, the unitary map up to time $t$ can be simulated as $U(t)=e^{-iHt}\approx U_{\rm Trot}^{n_\tau}(\tau)$, where $n_{\tau} = t/\tau$ is the number of Trotter steps and the single-step Trotter approximated map is $U_{\rm Trot}(\tau)=e^{-i H_z \tau}e^{-i H_y \tau}=\mathcal{T}\left(\exp\{-i \int_0^\tau H_{\rm kicked} (t) dt \}\right)$. The effective single-step simulated Hamiltonian $H_{\rm kicked} (t) = H_y + \tau H_z\sum_n \delta(t-n\tau)$ describes a periodically ``delta-kicked" system, and $U_{\rm Trot}(\tau)$ is its respective Floquet map.

For example, given a transverse Ising model, $H = -h \sum_i \sigma^{(i)}_y -\sum_{i,j}\Lambda_{ij} \sigma^{(i)}_z \sigma^{(i)}_z$, Heyl {\em et al.} studied the Trotterized approximation arising from a gate-based simulation, and showed that above a critical Trotter step size, $\tau$, the resulting Floquet operator is characterized by a many-body quantum chaotic regime, where Trotter errors proliferate and become uncontrollable~\cite{Heyl2019}. This is true even for integrable systems described by a single degree of freedom encoded in the collective spin of $N_s$ spin-1/2 particles, $\mathbf{J} = \sum_{i=1}^{N_s} \Vec{\sigma}^{(i)}/2$.   For the Lipkin-Meshkov-Glick (LMG) model~\cite{Lipkin1965}, $H_{LMG} = -B J_y - \frac{\Omega}{2 J} J_z^2$, the Trotterized map is the famous quantum kicked-top model, $U_{QKT}=\exp\{i\frac{k}{2 J} J_z^2\} \exp\{i\alpha J_y\}$, with $\alpha = B\tau$ and $k=\Omega \tau$~\cite{Haake1987}. Haake {\em et al.} introduced this model as a paradigm for quantum chaos, and in their seminal work~\cite{Haake1987}, systematically studied the classical chaos (in the thermodynamic limit, $N_s\rightarrow \infty$) and quantum signatures of chaos for finite $N_s$. After this pioneering work a plethora of theoretical and experimental developments in quantum chaos~\cite{Schack1994,Ghose2008,Kumari2018,Kumari2019,Chaudhury2009,Neill2016,Trail2008,Herman2020,Lombardi2011,munoz2019} have been facilitated by direct or indirect usage of the kicked top. Recently Sieberer and coworkers showed that the quantum chaos in the kicked top can lead to proliferation of errors in Trotterized simulation of the LMG model~\cite{Heyl2019,Sieberer2019}.



In the present work we study the quantum and classical chaos of a family of delta-kicked transverse Ising models with all-to-all connectivity for $N_s$ spin 1/2 particles, generalizing Haake's pioneering work~\cite{Haake1987} to models with arbitrary $p$-body interactions. Following from our discussion above, these delta-kicked systems correspond to the effective time-dependent Hamiltonian description of the Trotterization of a family of completely connected transverse Ising models, usually called ``$p$-spin'' models,
\begin{eqnarray}
\label{eqn:intro_p_spin_hamil}
\hat{H}_p &=& -B \sum_ {i=1}^{N_s}\frac{\hat{\sigma}_y^{(i)}}{2} - \frac{\Omega}{p}\sum_{i_1,i_2,\ldots,i_p= 1}^{N_s}      \frac{\hat{\sigma}_z^{(i_1)}\hat{\sigma}_z^{(i_2)}\ldots\hat{\sigma}_z^{(i_p)}}{2N_s^{p-1}} \nonumber \\
&=& -J\left[ B\left( \frac{\hat{J}_y}{J} \right)+ \frac{\Omega}{p} \left(\frac{\hat{J}_z}{J}\right)^p \right].
\end{eqnarray}
The dependencies on $N_s$ and $p$ are chosen to ensure that the model is extensive and of a universal form in the (mean-field) thermodynamic limit. 

The two-body case ($p=2$) is the LMG model mentioned above, featuring a continuous quantum phase transition between paramagnetic and ferromagnetic phases. The generalization for $p>2$ gained prominence in the context of quantum information in the work of J\"org {\em et al.}, who showed that for $p>2$ this system undergoes a first-order (discontinuous) quantum phase transition, and is accompanied by an exponentially closing gap to the ground state, which renders quantum annealing intractable~\cite{Jorg2010}. Subsequent work has analyzed this model from the point of view of mean-field theory~\cite{Bapst2012}, entanglement in quantum phase transitions~\cite{Filippone2011}, and a variety of approaches to tame the exponential complexity for efficient quantum annealing and optimization~\cite{Kong2017, Matsuura2017}. In previous work we studied quantum simulations of $p$-spin models using tools of measurement-based feedback control~\cite{Munoz-Arias2020}. 

Our characterization of the nonlinear dynamics and classical/quantum chaos of the kicked $p$-spin family is structured in a similar fashion as the original kicked top paper~\cite{Haake1987}, in order to emphasize the similarities/differences between the kicked top and its generalizations. For the classical system, in the limit $N_s \to \infty$, borrowing from foundational results in the theory of area preserving maps~\cite{Meyer1970,Mackay1983,Henon1969,Simo1981} we characterize and classify the structural changes and instabilities, appearing far from the emergence of chaos, induced by bifurcations. Explicit computation of the largest Lyapunov exponent provides a characterization of the transition to global chaos, and the local structural aspects of the emergence of chaotic regions are assessed by estimating their surface areas. Quantum chaos is studied via \textit{kinematic} and \textit{dynamic} signatures. In the former case we focus on the statistics and localization properties of eigenphases and eigenvectors of the Floquet operator, respectively. In the latter case we study the growth of the out-of-time-order correlator. Our analysis generalizes the work of Haake on the quantum kicked top in the light of modern developments in quantum chaos.

The remainder of the manuscript is organized as follows. In Sec. \ref{sec:p_spin_model} we introduce the Hamiltonian for the kicked $p$-spin model and derive the stroboscopic map that describes the evolution in the classical limit. In Sec. \ref{sec:ndyna_p_spin} we analyze the classical nonlinear dynamics by means of studying fixed points and their stability and the largest Lyapunov exponent during the transition to global chaos. In Sec. \ref{sec:quantum_chaos} we characterize the quantum chaotic properties of the stroboscopic Floquet dynamics via kinematic signatures (including level spacing statistics and localization of the Floquet eigenstates) and dynamical indicators like the growth rate of the out-of-time-order correlator. Finally in Sec. \ref{sec:final_remarks} we summarize, conclude and give an overview of future research directions.

\section{The kicked $p$-spin model}
\label{sec:p_spin_model}
We study the delta-kicked version of the $p$-spin model, Eq. (1), governed by the Hamiltonian~\footnote{In our delta-kicked Hamiltonian in Eq. (\ref{eqn:kicked_p_spin_hamil}) we have dropped a minus sign compared to the effective Hamiltonian obtained from the Trotterization of the $p$-spin evolution. This is done in order to be faithful with the conventions in Haake's original work~\cite{Haake1987}, as in the present work we aim to stress the differences between the kicked top and its generalizations. Notice however that for the present study, of dynamical character, the choice of sign does not alter the observed phenomenology. It does change the character of the ground state phase diagram, which will be important in the context of analog quantum simulation of $p$-spin models, study that will be address in a future work.} 
\begin{equation}
\label{eqn:kicked_p_spin_hamil}
 \hat{H}_{\delta-p} (t)= \frac{\alpha}{\tau}\hat{J}_y + \frac{ k}{pJ^{p-1}}\hat{J}_z^p\sum_{n=-\infty}^\infty\delta(t-n\tau),
\end{equation}
where $\alpha$ is the precession angle, $\tau$ the time interval of free precession, and $k$ the strength of the nonlinear kick.  The time evolution operator under this Hamiltonian is the Floquet map
\begin{equation}
 \label{eqn:floquet_kicked_p_spin}
 \hat{U}_{p} = \mathcal{T}\left\{
 e^{-i\int_{0}^t dt'\hat{H}_{\delta-p}(t')}\right\}=
e^{-i\frac{k}{pJ^{p-1}}\hat{J}_z^p}e^{-i\alpha\hat{J}_y},
\end{equation} 
(here and throughout $\hbar =1$). Choosing $\alpha=B \tau$ and $ k= \Omega \tau$, this Floquet map is the Trotterized version of the unitary evolution generated by Eq. (\ref{eqn:intro_p_spin_hamil}). As the magnitude of the spin $J=N_s/2$ is conserved, the quantum dynamics take place in the $N_s+1$ dimensional symmetric irreducible subspace. In the classical limit $N_s \rightarrow \infty$ the mean spin executes motion on the surface of a sphere, described by a rotation of the spin about the $y$-axis by angle $\alpha$ followed by a nonlinear ``twist" about the $z$-axis. This twist can be understood as a rotation around the $z$-axis by an angle proportional to the $p-1$ power of the $z$-projection of the spin, inducing nonlinear dynamics with strength $k$.

The Heisenberg evolution of the collective spin is defined by the map $\hat{\mathbf{J}}' = \hat{U}_{p}^\dagger \hat{\mathbf{J}} \hat{U}_{p}$, with components
\begin{widetext}
\begin{subequations}
\label{eqn:collective_ope_evolved}
\begin{align}
 \hat{J}_x' &= \frac{1}{2}\left[ \left(\cos(\alpha)\hat{J}_x + 
\sin(\alpha)\hat{J}_z + i\hat{J}_y\right)e^{i \mathcal{Q}_{+}(k, \alpha, p)} + 
\left(\cos(\alpha)\hat{J}_x + \sin(\alpha)\hat{J}_z 
- i\hat{J}_y\right)e^{i \mathcal{Q}_{-}(k, \alpha, p)}\right], \\
 \hat{J}_y' &= \frac{1}{2i}\left[ \left(\cos(\alpha)\hat{J}_x + 
\sin(\alpha)\hat{J}_z + i\hat{J}_y\right)e^{i \mathcal{Q}_{+}(k, \alpha, p)} - 
\left(\cos(\alpha)\hat{J}_x + \sin(\alpha)\hat{J}_z 
- i\hat{J}_y\right)e^{i \mathcal{Q}_{-}(k, \alpha, p)}\right], \\
\hat{J}_z' &= -\sin(\alpha)\hat{J}_x + \cos(\alpha)\hat{J}_z,
\end{align}
\end{subequations}
\end{widetext}
where the arguments of the exponentials are given by
\begin{equation}
 \label{eqn:argument_heisenberg_equs}
 \mathcal{Q}_{\pm}(k, \alpha, p) =\frac{k}{pJ^{p-1}} \sum_{a=1}^{p} (\pm 1)^a 
\mathcal{J}_a(\hat{J}_x, \hat{J}_z;p,\alpha), 
\end{equation}
with 
\begin{equation}
\mathcal{J}_a(\hat{J}_x, \hat{J}_z;p,\alpha) = \binom{p}{a} \left( 
\cos(\alpha)\hat{J}_z - \sin(\alpha)\hat{J}_x \right)^{p-a}.
\end{equation}
Notice that for general $p$, a single evolution step couples the components of collective spin operators to a polynomial in these components of degree $p-1$. As a consequence, evolution under $\hat{U}_{p}$ leads to high complexity and rapidly takes an initially localized state, e.g., a spin coherent state, into a highly nonclassical spin state. Further details on the derivation of these equations of motion are presented in Appendix \ref{app:eqns_motion}.

Taking the proper limit allows us to define classical variables and obtain the classical nonlinear dynamical map when $J\to\infty$. In the standard way, we take the expectation value of the evolved operators in Eq. 
(\ref{eqn:collective_ope_evolved}) and neglect all correlations, i.e. $\langle\hat{A}\hat{B}
\rangle = \langle\hat{A} \rangle \langle\hat{B} \rangle$, with $\hat{A}$, $\hat{B}$ two Hermitian operators. Then, we introduce the classical unit vector $\bm{X}=\langle\hat{\mathbf{J}} \rangle/J$, and take the limit $J\to\infty$. The resulting stroboscopic map of the classical coordinates of  $\mathbf{X}=(X,Y,Z)$ on the unit sphere is given by
\begin{widetext}
 \begin{subequations}
 \label{eqn:classical_kicked_p_spin}
 \begin{align}
  X_{m+1} &= \cos\left(k (\cos(\alpha)Z_m - \sin(\alpha)X_m)^{p-1} 
\right)(\cos(\alpha)X_m + \sin(\alpha)Z_m) - \sin\left(k (\cos(\alpha)Z_m - \sin(\alpha)X_m)^{p-1} 
\right)Y_m, \\
 Y_{m+1} &= \sin\left(k (\cos(\alpha)Z_m - \sin(\alpha)X_m)^{p-1} 
\right)(\cos(\alpha)X_m + \sin(\alpha)Z_m) + \cos\left(k (\cos(\alpha)Z_m - \sin(\alpha)X_m)^{p-1} 
\right)Y_m, \\
 Z_{m+1} &= -\sin(\alpha)X_m +\cos(\alpha)Z_m,
 \end{align}
 \end{subequations}
\end{widetext}
with the respective inverse map given by
\begin{widetext}
\begin{subequations}
\label{eqn:inverse_classical_kicked_p_spin}
\begin{align}
   X_m &= \cos(\alpha)\cos(kZ_{m+1}^{p-1})X_{m+1} 
+\cos(\alpha)\sin(kZ_{m+1}^{p-1})Y_{m+1} - \sin(\alpha)Z_{m+1}, \\
  Y_m &= -\sin(kZ_{m+1}^{p-1})X_{m+1} + \cos(kZ_{m+1}^{p-1})Y_{m+1}, \\
  Z_m &= \sin(\alpha)\cos(kZ_{m+1}^{p-1})X_{m+1} + 
\sin(\alpha)\sin(kZ_{m+1}^{p-1})Y_{m+1} + \cos(\alpha)Z_{m+1}.
\end{align}
\end{subequations}
\end{widetext}
We will refer to a single application of the stroboscopic classical Floquet map in Eq. (\ref{eqn:classical_kicked_p_spin}) as $F[\bm{X}_m]$ and the respective inverse map in Eq. (\ref{eqn:inverse_classical_kicked_p_spin}) as $F^{-1}[\bm{X}_m]$.

The classical nonlinear dynamics arise from the mean-field approximation in the thermodynamic limit~\footnote{The classical phase space is restricted to the surface of the unit sphere, and hence one can also write the map in Eq. (\ref{eqn:classical_kicked_p_spin}) in terms of the angular variables of spherical coordinates $(\theta,  \phi)$,  where these are the polar and azimuthal angle, respectively.}. This is achieved by replacing the interaction term in Eq. (\ref{eqn:kicked_p_spin_hamil}) with its mean field approximation, $\hat{J}_z^p \to p\langle \hat{J}_z \rangle^{p-1}\hat{J}_z$. The resulting effective Hamiltonian yields an evolution operator composed of two components: a linear rotation by $\alpha$ and rotation depending on the current state.  The latter is ``nonlinear" in that the angle is proportional to the average of the $p-1$ power of the $z$-component. Note, given our choice of coordinates in Eq. (\ref{eqn:kicked_p_spin_hamil}), for any choice of $\alpha$, trajectories undergo Larmor precession around the $y$-axis. We thus refer to the points $(0,\pm1,0)$ as ``poles" and the great circle in the $x$-$z$ plane as the ``equator."

\section{Nonlinear dynamics of a classical kicked $p$-spin}
\label{sec:ndyna_p_spin}
In order to better identify and understand the general properties of the kicked $p$-spin models, we first summarize Haake's analysis of the nonlinear dynamics of the classical kicked top ($p=2$)~\cite{Haake1987}. The classical kicked top has doubly reversible dynamics under the appropriate choice of time reversal symmetry (given below), parity symmetry, and an additional symmetry of the iterated map $F^2$ when $\alpha=\pi/2$. The two fixed points on the poles of the sphere $(0,\pm 1, 0)$ bifurcate from elliptic (stable) to hyperbolic (unstable) at $k = 2$, leading to the onset of a cascade of period doubling bifurcations and a transition from regular to mixed phased space, before leading to global chaos. Additionally when $\alpha=\pi/2$ the period-$4$ orbit on the equator (defined below) changes from stable to unstable at $k=\pi$.

In the remainder of this section we extend this analysis to the whole family of $p$-spin models. In order to illustrate the differences as well as similarities between models with $p=2$ and $p>2$, we will often compare the three models with $p=2,3,4$. We stress that this choice does not restrict the generality of our findings, as those three test cases exhaust the kicked $p$-spin phenomenology (see \cite{Bapst2012} for a similar discussion with the $p$-spin family). In fact, all the models with odd $p>2$ exhibit the same phenomenology as $p=3$ and all the models with even $p>2$ exhibit that of $p=4$. We pay close attention to the value of $\alpha=\pi/2$ as it allows us to directly contrast models with $p>2$ with Hakee's kicked top results. However, as we will see models with $p>2$ exhibit a rich and intricate behavior in the range $\alpha\in[0,\pi]$, which we fully characterize as well.

\subsection{Symmetries}
\label{subsec:symmetries}
Symmetries of the map $F[\mathbf{X}]$ can be found with the help of the following two transformations
\begin{equation}
\label{eqn:t_trans}
 T[\mathbf{X}] = (-\cos(\alpha)X -\sin(\alpha)Z, Y, -\sin(\alpha)X + 
\cos(\alpha)Z),
\end{equation}
and
\begin{equation}
\label{eqn:t_tilde_trans}
\tilde{T}[\mathbf{X}] = (\cos(\alpha)X +\sin(\alpha)Z, Y, \sin(\alpha)X - \cos(\alpha)Z),
\end{equation}
which are both involutions, \textit{i.e.} $T^2 = \tilde{T}^2 = 1$ and have determinants $\det(T) = \det(\tilde{T}) = -1$. These transformations allow us 
to introduce time reversal operations of the stroboscopic evolution. One can 
easily check that $F$ and $F^{-1}$ satisfy 
\begin{equation}
\label{eqn:t_reversal}
 TFT \equiv T[F[T[\mathbf{X}]]] = F^{-1}[\mathbf{X}],
\end{equation}
and 
\begin{equation}
\label{eqn:t_tilde_reversal}
 \tilde{T}F\tilde{T} \equiv \tilde{T}[F[\tilde{T}[\mathbf{X}]]] = F^{-1}[\mathbf{X}],
\end{equation}
when $p$ is even, indicating the map has double reversible dynamics. However, for odd values of $p$ only Eq. (\ref{eqn:t_reversal}) is satisfied, hence only the $T$ transformation yields a proper time reversal operation. The major 
consequence of this time reversal is that the images of $n$-periodic orbits of 
$F$ under $T$ (and $\tilde{T}$ for even $p$'s) are also $n$-periodic orbits, 
where it may happen that the orbit is its own image~\cite{Haake1987}. 

\begin{figure*}[ht]
 \centering{\includegraphics[width=0.98\textwidth]
  {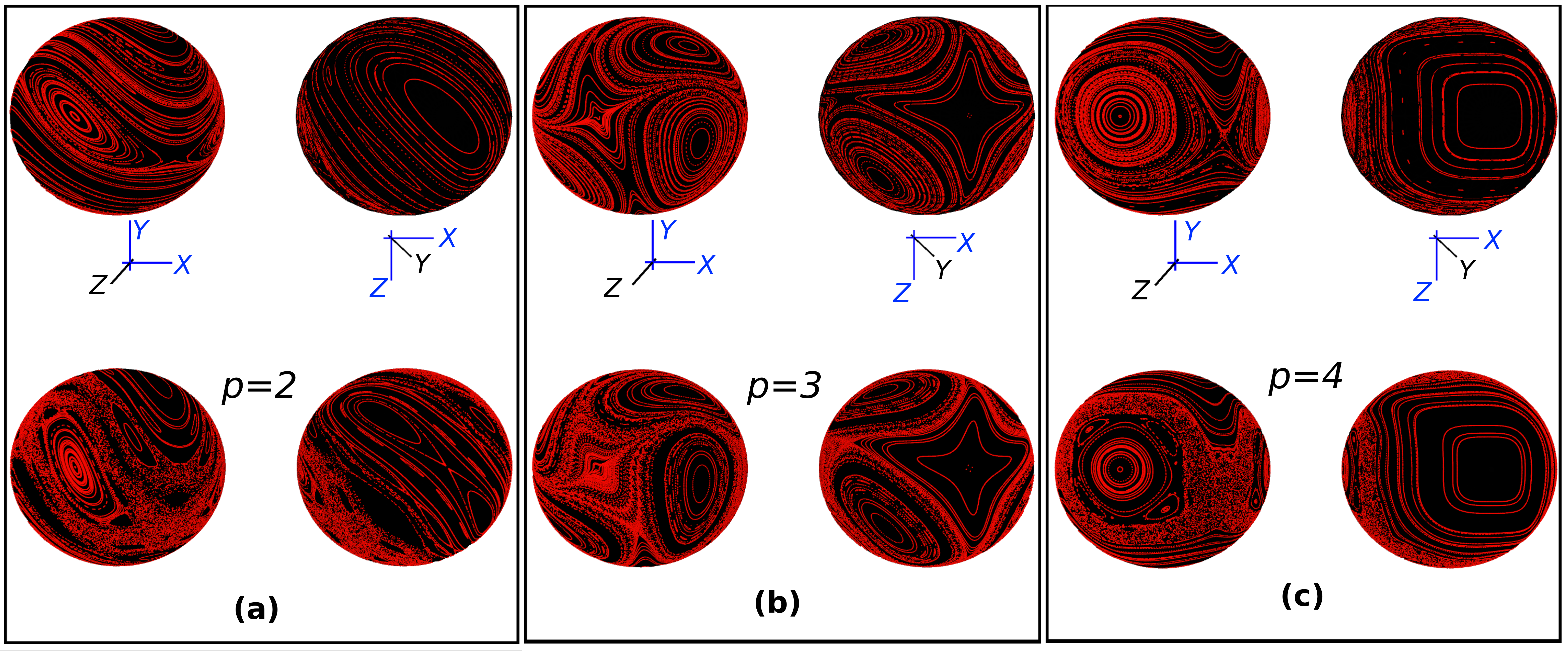}}
\caption{Phase space portraits for different values 
of the parameter $p$. (a) Case $p=2$ (kicked top), with
$k = 1.6$ (top), and $k = 2.3$ (bottom). (b) Case $p = 3$ 
with $k = 0.8$ (top), and $k = 1.2$ (bottom), (c) Case $p = 4$ with 
$k = 1.5$ (top), and $k = 2.2$ (bottom). In each of the panels we oriented the sphere on the left such that we look straight at the positive $z$ (along the equator), and the sphere on the right such that we look at the positive $y$ (north pole). In this way one of the regular islands on the poles and one of 
the resonant islands on the period-$4$ orbit along the equator are visible. A schematic of the Cartesian directions is also included to guide the eye. All cases correspond to $\alpha = \pi/2$.}\label{fig:all_phase_spaces}
\end{figure*}

Using Eq. (\ref{eqn:t_reversal}) and Eq. (\ref{eqn:t_tilde_reversal}) we define a family of symmetry curves on the unit sphere composed of orbits invariant under the application of any of the involutions 
$\mathcal{I}[\mathbf{X}_m]$ where $\mathcal{I} = T, \tilde{T}, TF, FT, ...$. The $T$ 
and $\tilde{T}$ invariant curves are given by the great circles satisfying 
\begin{eqnarray}
 \sin(\alpha)X - (\cos(\alpha) - 1)Z &=& 0, \\
 \sin(\alpha)X - (\cos(\alpha) + 1)Z &=& 0,
\end{eqnarray}
respectively. In general the invariant curves for the higher involutions, 
$\mathcal{I}$, have fairly complicated shapes. 

If an orbit is invariant under an involutions $\mathcal{I}$, the structural changes it might undergo are constrained, since the resulting orbit must still respect this invariance. For instance, if the periodic orbit is of even/odd period then it must have an even/odd number of points on the corresponding symmetry line of $\mathcal{I}$. Other consequences of time reversal by $T$ and $\tilde{T}$, and invariance under $\mathcal{I}$ are explored in~\cite{Haake1987}.

For the case $k=0$, the phase space is filled by regular orbits describing Larmor precession  around the $y$-axis which  deform as $k$ increases. Rotations around the precession axis then provide information about the symmetries of our map. Particularly, for even values of $p$, the map $F$ is invariant under $\pi$-rotations around the $y$-axis,
\begin{equation}
\label{eqn:y_symmetry}
R_y(\pi)F = FR_y(\pi), 
\end{equation}
where $R_y(\pi)[\mathbf{X}_m] = (-X, Y, -Z)$. To understand this fact we notice that the rotation $R_y(\pi)$ can be constructed 
as $R_y(\pi) = T\tilde{T} = \tilde{T}T$. Thus, invariance under $R_y(\pi)$ 
immediately implies time reversal under both $T$ and $\tilde{T}$. Conversely, the absence of time reversal under either $T$ or $\tilde{T}$ implies no invariance under $R_y(\pi)$. Thus, the maps for even $p$ have the feature that the image under $R_y(\pi)$ of every $n$-periodic orbit of $F$ is also an $n$-periodic orbit.

Finally, when specializing for $\alpha = \pi/2$ and even $p$'s, the map $F$ has an additional symmetry. To see this we use the following identity
\begin{equation}
\label{eqn:additional_symme}
 FR_x(\pi) = R_x(\pi)FR_y(\pi),
\end{equation}
where $R_X(\pi)[\mathbf{X}_m] = (X,-Y,-Z)$, is a rotation around the $x$-axis by an angle of $\pi$. Using Eq. (\ref{eqn:additional_symme}) it is easy to show that the iterated map $F^2$ is invariant under $R_x(\pi)$. 

With these symmetries in mind we can give an informed description of the phase portraits of the kicked $p$-spin models. In Figs. \ref{fig:all_phase_spaces}(a-c) we display characteristic phase portraits for the cases of $p=2,3,4$ and $\alpha=\pi/2$. In each of the three panels the spheres on the right show the regular island on the pole $(0,1,0)$ and the spheres on the left show one of the islands in a period-$4$ orbit along the equator (see description below). For $p=3$, Fig. \ref{fig:all_phase_spaces}b, the starry shape of the regular island on the pole is a consequence of the absence of the additional symmetry under rotations around the $y$-axis. 

\subsection{Fixed points}
\label{subsec:fixed_points}
The precession axis determines two fixed points of $F$, the poles $(0, \pm1, 0)$. Additional ones can be found by solving the equation $F[\bm{X}_{m}] = 
\bm{X}_m$. Writing all the components in terms of the $Z$ coordinate, we find that new fixed points appear when
\begin{eqnarray}
 X_m &=& -\cot(\alpha/2)Z_m, \nonumber \\  
 Y_m &=& \cot\left( \frac{kZ_m^{p-1}}{2} \right)\tan\left(\alpha/2\right)Z_m, \\
 \mathcal{F}(Z_m; k,\alpha,p) &=& 0, \nonumber
\end{eqnarray}
where $\mathcal{F}(Z_m;k,\alpha,p)$ is given by
\begin{multline}
 \mathcal{F}(Z_m;k,\alpha,p) = \\
  Z_m^2 - \frac{1}{\cot^2\left( \frac{kZ_m^{p-1}}{2}\right)\tan^2(\alpha/2) + 
\csc^2(\alpha/2)}.
\end{multline}
When we specialize to the case $\alpha = \pi/2$, writing the 
expressions in terms of the $X$ coordinate, new fixed points appear if
\begin{eqnarray}
\label{eqn:fixed_point_conditions}
 Z_m = -X_m, \enspace Y_m &=& (-1)^p\cot\left(\frac{kX_m^{p-1}}{2}\right), \\ 
\enspace \mathcal{F}(X_m; k, p) &=& 0, \nonumber
\end{eqnarray}
where $\mathcal{F}(X_m;k,p)$ is given by
\begin{equation}
\mathcal{F}(X_m; k, p) = \frac{\sin^2\left(\frac{kX_m^{p-1}}{2} \right)}{1 
+ \sin^2\left(\frac{kX_m^{p-1}}{2}\right)} - X_m^2,
\end{equation}
which recovers the kicked top result when $p=2$~\cite{Haake1987}. The solutions of $\mathcal{F}(Z_m; k, \alpha, p)=0$ are invariant under $Z_m \to -Z_m$, and thus any nontrivial solution gives two new fixed points. We can then focus on solutions for positive values of $Z$, where $-Z$ provides a valid solution as well. Let us study the solutions of Eq. (\ref{eqn:fixed_point_conditions}), \textit{i.e} fixing $\alpha=\pi/2$. For $p=2$ the first nontrivial fixed point appears at $k=2$. For $p\ge3$ solutions for positive $X$ come in pairs, which means every solution gives four new fixed points. In particular, for $p=3$ the first nontrivial solutions appear at $k\sim4.7$, for $p=4$ they appear at $k\sim7.5$. We observe then that new fixed points for the models with $p>2$ appear at fairly large values of $k$, for which chaotic region of considerable size have already developed, as we will see in Sec. \ref{subsec:trasition_hamil_chaos}. This indicates that, for these cases, the emergence of new fixed points does not influence the transition to chaos. This point will be further explored next via the stability analysis of various fixed points of the map $F$.

\subsection{Stability}
\label{subsec:stability}
The stability of a fixed point or orbit is investigated using the eigevalues of the tangent map (Jacobi matrix), $\mathbf{M}(\mathbf{X}_m) = \frac{\partial \mathbf{X}_{m+1}}{\partial\mathbf{X}_m}$ of $F$, evaluated at the fixed point or along the orbit~\cite{Reichl,Schuster1995}. 

For the family of models under study, the condition $|\bm{X}_m|^2=1$ guarantees that one of the eigenvalues of $\mathbf{M}(\bm{X}_m)$ is always one. Therefore stability analysis reduces to that of a two dimensional area preserving map~\cite{Mackay1993}. Area preservation implies $\det(\mathbf{M}(\bm{X_m})) = 1$, thus one has that the other two eigenvalues, $\mathcal{M}=(\mathcal{M}_1,\mathcal{M}_2)$, of $\mathbf{M}$ behave in one of three ways: 
\begin{enumerate}[(i)]
    \item If the eigenvalues $\mathcal{M}$ of $\mathbf{M}$ form a complex conjugated pair and live on the unit circle, satisfying $|{\rm Tr}(\mathbf{M})|<2$, the fixed point is elliptic and known to be stable as a consequence of Moser's twist theorem~\cite{Moser1962} (excluding the situation when $\mathcal{M}$ is the $l$-th root of unity). 
    \item If the eigenvalues $\mathcal{M}$ of $\mathbf{M}$ form a reciprocal real pair and live on the real line, satisfying $|{\rm Tr}(\mathbf{M})|>2$, the fixed point is hyperbolic and unstable. 
    \item If the eigenvalues of $\mathbf{M}$ are real and degenerate, both equal to either $1$ or $-1$, satisfying $|{\rm Tr}(\mathbf{M})|=2$, the fixed point is parabolic. Determining its stability, i.e., whether or not the fixed point is surrounded by closed invariant curves, requires a case-by-case study (see~\cite{Simo1981,Aharonov1990} for some early works).
\end{enumerate}
A negative value of the trace indicates an inversion hyperbolic/parabolic point~\cite{Mackay1983,Reichl}. The above classification characterize the shape of trajectories in the vicinity of a fixed point or orbit. The effective eccentricity, $e_{\rm eff} = \frac{1}{2}|{\rm Tr}(\mathbf{M})|$, connects the stability classification and the different conic sections.


In this context, a parabolic point is the hallmark of a bifurcation process~\cite{Meyer1970}. One eigenvalue equal to $1$ implies isolation and persistence of the fixed point are not guaranteed~\footnote{meaning that additional fixed points could exist arbitrarily close to the original one and the original fixed point might not be robust to small perturbations (see for instance~\cite{Mackay1983})}. In particular, if $\mathcal{M} = 1$ one observes a tangent bifurcation, \textit{i.e} change in stability, and if $\mathcal{M}=-1$ one observes a period doubling bifurcation~\cite{Meyer1970,Mackay1983}. 

The above stability classification covers period-$l$ orbits of $F$ as well, \textit{i.e} fixed points of the the map $F^l[\bm{X}_m]$. A parabolic point of $F^l[\bm{X}_m]$ with $\mathcal{M} = 1$ corresponds with an elliptic fixed point of $F$ with $\mathcal{M}_i$ equal to the $l$-th root of $1$, indicating a $1$ to $l$ bifurcation~\cite{Meyer1970}. The aforementioned types of bifurcations constitute a classification of these processes in area preserving maps~\cite{Meyer1970,Mackay1993}, and are dubbed generic. Nongeneric bifurcations might exists (see Sec. 1.2.4.7 of~\cite{Mackay1993}). For instance, when additional symmetry constraints are imposed on the orbits of $F$, as it is the case in doubly reversible maps (see Sec. \ref{subsec:symmetries}).

Parabolic points in conjunction with the symmetries of the map provide a large amount of information regarding the structures that one might observe in phase space (see the example in~\footnote{Considering $p=2$ with $\alpha=\pi/2$ and the fixed points on the poles, one has that ${\rm Tr}(\mathbf{M}) = \pm k$ for $(0,\pm1,0)$. Therefore, at $k=2$ the poles are parabolic fixed points. The north pole undergoes a tangent bifurcation, the south pole undergoes a period doubling bifurcation. Additionally, at $\alpha=\pi/2$, $F^2$ is invariant under $R_x(\pi)$, which forces the north pole to undergo a period doubling bifurcation as well. This is one of the main results of Haake~\cite{Haake1987}, and a good example of the importance of parabolic fixed points.\label{fn_example}}). For the current study they will play an crucial role in the behavior of the models with $p>2$, as we will see below. 

We split the stability analysis of the $p$-spin models in two cases. First, the case $\alpha=\pi/2$, where the main structures in phase space are the regular regions around the poles and a period-$4$ orbit on the equator. Second the case of models with $\alpha\ne\pi/2$ where phase space is dominated by the regular regions around the poles.

\subsubsection{Stability of models with $\alpha=\pi/2$}
Using the eigenvalues of the tangent map when $\alpha=\pi/2$, a fixed point of $F$ is stable when the following inequality is satisfied,
\begin{equation}
 \label{eqn:stability_condition}
 \left| (-1)^p(p-1)kX^{p-2}Y + \cos\left(kX^{p-1}\right) -1 \right|<2.
\end{equation}
The cases of $\mathcal{M}$ equal to the $l$-th root of $1$ should be treated separately, as they indicate bifurcation processes. In the case of $p=2$, Eq. (\ref{eqn:stability_condition}) reduces to $|kY +\cos(kX)-1|<2$ as obtained by Haake~\cite{Haake1987}. 

Consider now the fixed points on the poles. For $p=2$, by virtue of Eq. (\ref{eqn:stability_condition}) these points are stable only if $k<2$. At $k=2$ the appearance of new fixed points, as dictated by Eq. (\ref{eqn:fixed_point_conditions}), together with the change in stability, indicate a bifurcation processes (see left spheres on Fig. \ref{fig:all_phase_spaces}a). At larger values of $k$ further period doubling bifurcations occur, leading to a cascade of these bifurcations, as investigated by Haake~\cite{Haake1987}. 

For the models with $p>2$, the left hand side of Eq. (\ref{eqn:stability_condition}) evaluated on the poles yields zero regardless of the value of $k$. We observe closed invariant curves surrounding the poles (see right spheres in Fig. \ref{fig:all_phase_spaces}b,c), hinting at the poles being stable. However at $\alpha=\pi/2$, the eigenvalues are $\mathcal{M}=\pm i$, the fourth root of unity. Therefore the poles undergo a 1-to-4 bifurcation as a function of $\alpha$ (details of which will be given in the next subsection). The local stability of the poles at this particular value of $\alpha$ is studied by constructing the $2$D area preserving map describing dynamics in the vicinity of the poles (see Appendix \ref{app:stability_general} for details). This map satisfies the conditions of the theorem in~\cite{Aharonov1990}, therefore the parabolic point at the origin is guaranteed to be surrounded by closed invariant curves. More specifically, the local area preserving map coincides with those in example $1$ and $2$ in~\cite{Aharonov1990} for even and odd $p$'s, respectively. This confirms our initial observations and allow us to conclude that the regular islands around the poles are stable for all values of $k$.  

The stability features of the poles outlined above represent a major distinction between the models with $p=2$ and $p>2$, for the special case of $\alpha=\pi/2$. In contrast with the cascade of period doubling in the model with $p=2$, in the models with $p>2$ we expect to find regular islands around the poles which survive even at large values of the kicking strength $k$, gradually reducing their size. This has defining consequences for the crossover mechanism to global chaos as we will see in Sec. \ref{subsec:trasition_hamil_chaos}.

Let us now study the period-$4$ orbit on the equator. This orbit is given by $\bm{X}_1\to \bm{X}_2\to \bm{X}_3 \to \bm{X}_4 \to \bm{X}_1$,
were $\bm{X}_1=(1,0,0)$, $\bm{X}_2=(0,0,1)$, $\bm{X}_3=(-1,0,0)$, 
$\bm{X}_4=(0,0,-1)$. The tangent map of this orbit has the form
$ \mathbf{M}_{4{\rm p}}= 
\mathbf{M}(\bm{X}_4)\mathbf{M}(\bm{X}_3)\mathbf{M}(\bm{X}_2)\mathbf{M} (\bm{X}_1)$.
In the case $p=2$ the orbit is stable if $(2\cos(k) + k\sin(k))^2 < 4$, which is not satisfied for the first time when $k=\pi$. For the case $p>2$ the relevant $2\times2$ subblock of $\mathbf{M}_{4\rm{p}}$ takes the form
\begin{multline}
 \label{eqn:orbin_4_tangent_map}
 \mathbf{M}_{4{\rm p}}^{(2\times2)} =\\ 
 \begin{pmatrix}
 \cos^2(k)-(-1)^p\sin^2(k) && \cos(k)\sin(k)(1+(-1)^p)\\
 -\cos(k)\sin(k)(1+(-1)^p) && \cos^2(k) - (-1)^p\sin^2(k)
 \end{pmatrix},
\end{multline}
with eigenvalues
\begin{equation}
 \mathcal{M}_{4{\rm p}}^{(\pm)} = \left(\cos(k) \mp 
i\sin(k)\right)\left(\cos(k) \mp i(-1)^p\sin(k)\right).
\end{equation}
If $p$ is odd, then $\mathcal{M}_{4{\rm 
p}}^{(\pm)} = 1$ and thus $F^4$ has a parabolic point. Local stability analysis indicates the points on the orbit are not stable, meaning that the neighborhood of points on the orbit is not composed of closed curves (see Appendix \ref{app:stability_general}). The vicinity of the orbit is populated by trajectories which belong to either the north or south hemispheres, orbiting around the corresponding pole. Thus, trajectories shear along the equator which divides the counter rotating flow between the two hemispheres (see Fig. \ref{fig:all_phase_spaces}b for a view of the phase space around the parabolic fixed point).

For this period-4 orbit, if $p$ is even, the two eigenvalues are given by
\begin{equation}
\mathcal{M}^{(\pm)}_{4{\rm p}} = e^{\mp i2k}.
\end{equation}
Thus, the period-$4$ orbit is composed of elliptic (stable) fixed points, except at the discrete values $k=s\frac{\pi}{2}$ with $s=1,2,3,...$, for which it becomes parabolic and bifurcation processes take place. When $s$ is odd $\mathcal{M}^{(\pm)}_{4{\rm p}}=-1$, indicating the period-$8$ orbit constructed as two cycles of the period-$4$ orbit bifurcates, and each of the points on the original period-$4$ orbit undergoes a 1-to-4 bifurcation. When $s$ is even, $\mathcal{M}^{(\pm)}_{4{\rm p}}=1$, and each of the points on the original period-$4$ orbit undergoes a $1$ to $2$ bifurcation.

The stability of the period-$4$ orbit on the equator allows us to make a distinction between models with odd and even values of $p$. For the former, the orbit is parabolic and always unstable. For the latter it is stable (elliptic), except for a discrete set of values at which it bifurcates. Both cases stand in contrast with the model with $p=2$ where the bifurcation processes change the stability of the orbit. The long-lived regularity of trajectories in the vicinity of the poles and the stability of trajectories near the equator have important consequences for the way in which models for $p>2$ crossover to global chaos, in contrast to that of the model with $p=2$. We will see this in detail in Sec. \ref{subsec:trasition_hamil_chaos}.

\subsubsection{Stability of models with $\alpha\ne\pi/2$}
For the models with $p>2$, the eigenvalues at the poles are $\mathcal{M}_j=e^{\pm i \alpha}$. Therefore, the poles undergo a $1$ to $l$-bifurcation as $\alpha$ is varied in $[0,\pi]$, with bifurcation points at $\alpha=\alpha_{\rm b}=2\pi q/l$, with $q,l$ relative primes, $q<l$ and $l>2$. 

For our kicked $p$-spin models all of these bifurcations are generic, meaning that they correspond to the classification in~\cite{Mackay1993,Meyer1970}. There is, however, one exception. Models with even $p$ are double reversible, therefore the involution $C=T\tilde{T}$ (and $\tilde{C}=\tilde{T}T$) commutes with the map $F$, that is $CF = FC$~\footnote{This is nothing but a restatement of invariance under $R_y(\pi)$.}. For these models, the poles are a fixed point of $C$ as well; they are strongly symmetric orbits (see Sec. 1.2.4.7 of~\cite{Mackay1993}). Therefore, any orbit emerging as a result of the bifurcation process must satisfy the symmetry imposed by $C$, \textit{i.e} orbit points lie on the symmetry lines of $C$. This implies that when $l$ is even the bifurcation is generic, but when $l$ is odd the bifurcation is double, since the orbit should have an even number of points in order to satisfy the symmetry imposed by $C$. Thus we observe the emergence of two period-$l$ orbits which look essentially identical to a single period-$2l$ orbit emerging from a $1$ to $2l$ bifurcation.

Bifurcation processes provides additional insights into the distinction between models with odd and even $p$ for $p>2$. When $p$ is odd, dynamics in the vicinity of north and south poles is described by the same $2$D are preserving map, and bifurcations on both poles take place for $\alpha>\alpha_{\rm b}$. On the other hand, when $p$ is even, dynamics in the vicinity of the poles is described by the same $2$D area preserving map only under the trivial change $k\to-k$. This indicates that north/south poles bifurcate on opposite sides of $\alpha_{\rm b}$ (an example of this is given in Appendix \ref{app:stability_general}).

As an example we consider the two lower order bifurcations, taking place at values of $q=1$, $l=3,4$, \textit{i.e.} $\alpha_{b}=\pi/2,2\pi/3$, corresponding to a $1$ to $4$ and $1$ to $3$ bifurcations, respectively. When $\alpha_{\rm b}=\pi/2$, for odd values of $p$ the bifurcation takes place at both north and south poles in the direction of $\alpha > \pi/2$. For even values of $p$ we see the bifurcation in the north pole in the direction of $\alpha > \pi/2$, and in the south pole in the direction of $\alpha < \pi/2$. Furthermore, the period-$4$ orbit appearing as a result of the bifurcation process is composed of unstable points, and it ceases to exists at $\alpha\sim2$ for $p$ odd, and $\alpha\sim2$, $\alpha\sim1$ for the north and south poles, respectively, in the case of even $p$'s. 

\begin{figure}[!t]
 \centering{\includegraphics[width=0.45\textwidth]{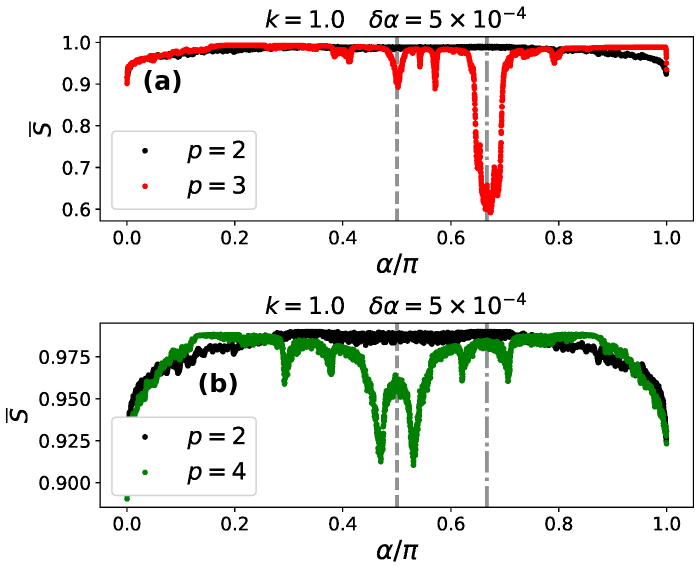}}
\caption{Averaged phase space similarity $\overline{\mathcal{S}}$ as a function of $\alpha$. The panels display the cases (a) $p=3$ and (b) $p=4$; in both cases we included the curve for $p=2$ for comparison purposes. The vertical lines indicate the values of $\alpha = 2\pi/4$ (dashed), and $\alpha = 2\pi/3$ (dashed-dotted), respectively, which mark the position of the two most prominent bifurcations. The other parameters are: $k=1$, $\delta\alpha = 5\times10^{-4}$, $n_{\rm tot} = 1500$ initial conditions and $N = 200$ kicks.}
\label{fig:simi_bifu}
\end{figure}

Consider now $\alpha_{\rm b}=2\pi/3$. For models with odd values of $p$ the new orbit emerges, in both north and south poles, when $\alpha>2\pi/3$. For models with even values of $p$ the new orbit emerges, in the north pole, when $\alpha>2\pi/3$, and in the south pole when $\alpha<2\pi/3$. In the latter case the bifurcation is double; we observe two period-$3$ orbits emerging from the pole, looking structurally the same as a period-$6$ orbit. Importantly, phase space is structurally the same in the vicinity of $\alpha=2\pi/3$ and in the vicinity of $\alpha=2\pi/6$, where a generic $1$ to $6$ bifurcation takes place. Therefore, any consequence of the stability of dynamics around the poles will display a symmetric character between these two points (see, for instance, Fig. \ref{fig:lyaps_all}c and Fig. \ref{fig:areas_all}c). Additionally, at this bifurcation point, for models with odd $p$, the poles have an unstable character, as was described by Simó in~\cite{Simo1981}. This will have a defining consequence on the early emergence of large chaotic seas, as we will study in Sec. \ref{subsec:trasition_hamil_chaos}.

\subsubsection{Identification of the most prominent bifurcations for models with $p>2$}
In the previous subsection we focused on the bifurcations taking place at $\alpha = 2\pi/3$ and $\alpha=2\pi/4$. For the models with $p>2$, $p=3,4$ as studied here, these two bifurcations are the most prominent/important ones. We define the importance of a bifurcation by the magnitude of global structural changes it generates in phase space. The degree of global structural changes can be quantified by the similarity/dissimilarity of two phase space portraits constructed starting with the same set of initial conditions and with parameters that are only infinitesimally different. Thus, if one phase space portrait corresponds to parameters $(\alpha, k)$, the second one corresponds to parameters $(\alpha', k') = (\alpha +\delta\alpha, k + \delta k)$ with $\delta\alpha, \delta k \ll 1$.

We consider a similarity/dissimilarity quantifier $\mathcal{S}$ based on the Pearson correlation coefficient~\cite{Lee1988}, first introduced in \cite{Munoz-Arias2020}. We review its explicit construction in Appendix \ref{app:simi_quant}. As we are interested purely on global structural changes induced by bifurcation processes, we will fix $k=1$, value for which the chaotic instability is not present yet, and will study $\mathcal{S}$ as a function of $\alpha$. In Fig. \ref{fig:simi_bifu}a,b we present the results of averaging $\mathcal{S}$ over the generated phase space portraits, \textit{i.e} a fixed set of initial conditions chosen uniformly over the unit sphere, for the systems with $p=3,4$ and include the $p=2$ curve for comparison purposes.

In this setting $\overline{\mathcal{S}} = 1$ indicates two phase space portraits which are identical and $\overline{\mathcal{S}} = 0$ indicates two phase space portraits which are completely different. Intermediate values indicate phase space portraits having a subset of trajectories which undergo a structural change and hence are dissimilar. In both Fig. \ref{fig:simi_bifu}a,b the vertical lines indicate $\alpha = 2\pi/3, 2\pi/4$, respectively. Notice that the most prominent dips of $\overline{\mathcal{S}}$ appear around these two positions, leading us to the conclusion that the two more prominent bifurcations in systems with $p>2$ take place at $\alpha = 2\pi/3, 2\pi/4$. We will see that these strong structural changes will have influence in the early emergence of chaotic trajectories.

\subsection{The transition to Hamiltonian chaos}
\label{subsec:trasition_hamil_chaos}
The transition to chaos in perturbed Hamiltonian systems with few degrees of freedom is well understood~\cite{Lichtenberg1992,Reichl,Wimberger}. For a small enough perturbation almost all invariant tori remain unchanged, as dictated by the KAM theorem~\cite{Wimberger,Reichl,Schuster1995}, with the exception of small chaotic regions appearing in the vicinity of unstable manifolds~\cite{Zaslavsky1991}. At larger perturbation strengths some invariant tori are destroyed, giving birth to chains of regular regions and new unstable manifolds, providing new ground for the chaotic region to expand. Area preserving mappings of the Poincare surface of section display this same behavior~\cite{Mackay1993}, with the emergence of chains of regular regions dictated by the Poincare-Birkoff theorem~\cite{Birkhoff1935,Arnold1964,Benettin1978}.

In the case of a two dimensional phase space, the chaotic region is clamped in between the regular regions, and generally the emergence and growth of chaotic regions adheres strictly to the mechanism described above. However, some Hamiltonian systems exhibit period doubling cascades~\footnote{transition to chaos via a cascade of period doubling bifurcations was observed initially in dissipative systems~\cite{Feigenbaum1979,Sander2012}, for instance the Logistic map} in conjunction with the destruction of KAM tori, and therefore the transition from regular to global chaotic motion is enhanced (see~\cite{Reichl}, Appendix G of~\cite{Schuster1995} and~\cite{Bountis1981,Greene1981}). In fact, a period doubling bifurcation is the last instability to occur before the neighborhood of the fixed point becomes completely chaotic~\cite{Mackay1983}.

In his pioneering work~\cite{Haake1987} Haake showed the existence of a period doubling cascade in the kicked top ($p=2$), which is interwoven with the destruction of KAM tori. From our stability analysis, it follows that none of the models with $p>2$ exhibit period doubling bifurcations, and in fact the bifurcations present on these models correspond to $l$-cycle bifurcations with $l>2$. Therefore, for the special case of $\alpha = \pi/2$, where the period doubling cascade occurs for $p=2$, we expect the kicked top to transition faster than any other model to the global chaos regime. On the other hand, for values of $\alpha \ne \pi/2$ we expect to encounter a different situation, as the presence of the $l$-cycle bifurcations influences the emergence of chaotic regions in the models with $p>2$. In the following we study this transition in detail, by characterizing the behavior of the largest Lyapunov exponent and the surface area of the chaotic sea.

\subsubsection{Largest Lyapunov exponent}
Chaotic behavior is identified with a positive value of the largest Lyapunov exponent, indicating that nearby initial conditions diverge exponentially fast, \textit{i.e.}, knowledge of the initial state is lost exponentially fast~\cite{Kolmogorov1958,Latora1999,Boffetta2002}

\begin{figure}[!t]
 \centering{\includegraphics[width=0.48\textwidth]{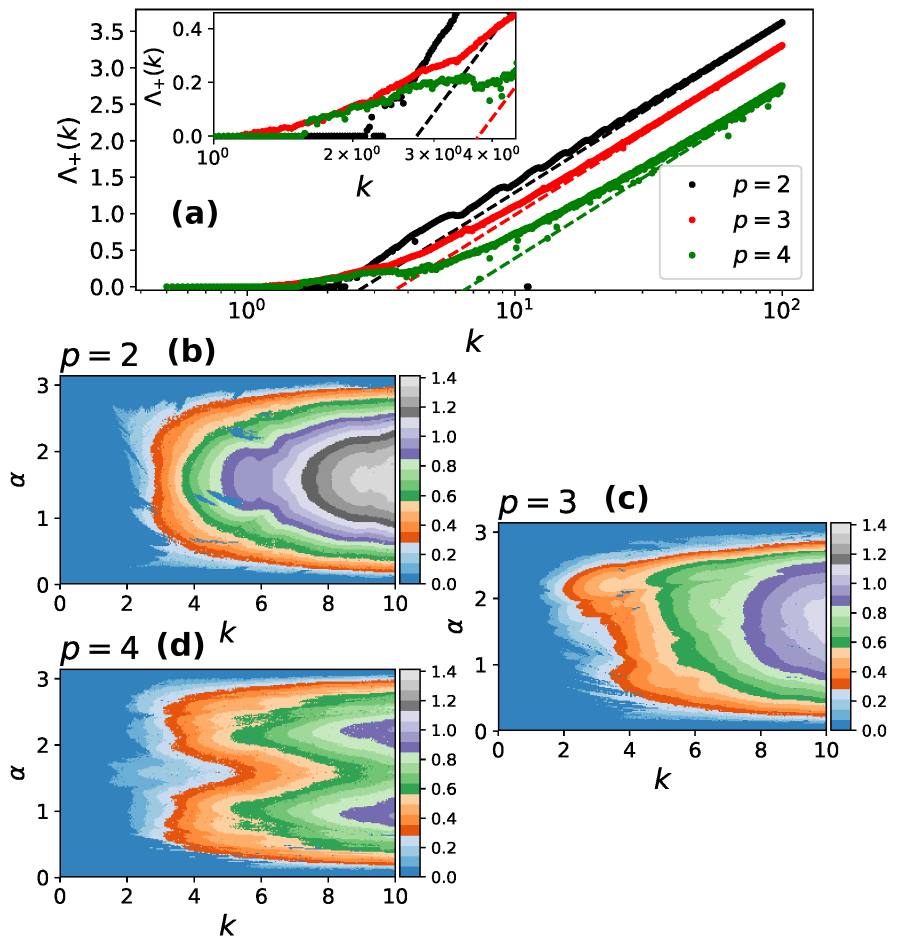}}
\caption{(a) Largest Lyapunov exponent of the kicked 
$p$-spin map in Eq. (\ref{eqn:classical_kicked_p_spin}) as a function of $k$ and for the special case of $\alpha=\pi/2$. We show the cases of $p=2$ (black), $p=3$ (red) and $p=4$ 
(green). Chaos emerges first for models with $p>2$ due to either the instability or bifurcations of the period-$4$ orbit along the equator. For $p=2$ it takes the first period doubling bifurcation, $k=2$, before chaos can appear, then the transition to strong chaotic trajectories (dashed black line) happens faster than for any other $p$. The inset shows a zoom into the parameter range $k\in[1.0, 4.5]$. (b,c,d) Largest Lyapunov exponent as a function of $\alpha$ and $k$, for the models with $p=2,3,4$, respectively. For the model with $p=2$ the dominant behavior of $\Lambda_+$ occurs at $\alpha\sim\pi/2$. For the models with $p>2$ and odd values ($p=3$ in (c)), dominant behavior of $\Lambda_+$ takes place around $\alpha\sim2\pi/3$. For the models with $p>2$ and even values ($p=4$ in (d)), dominant behavior of $\Lambda_+$ appears at $\alpha\sim \pi/2 \pm \pi/6$. Values corresponding to the $1$-to-$3$ bifurcation processes of the poles (see Sec. \ref{sec:ndyna_p_spin}).}
\label{fig:lyaps_all}
\end{figure}

When considering a map like the one in Eq. (\ref{eqn:classical_kicked_p_spin}), using Oseledets ergodic theorem~\cite{Ose1968,Eckmann1985} one can compute the largest Lyapunov exponent, $\Lambda_{+}$, via 
\begin{equation}
 \label{eqn:oseledec}
 \Lambda_{+}(\alpha, k, p) = \lim_{N\to\infty}[\lambda_{+}(\alpha, k, p)]^{1/2N},
\end{equation}
where $N$ is the number of time steps, $\lambda_{+}$ is the largest eigenvalue of the matrix  
$\prod_{m=1}^N\mathbf{M}^T(\bm{X}_m)\mathbf{M}(\bm{X}_m)$ and $\mathbf{M}(\bm{X}_m)$ is the tangent map introduced before. We can gain some insight on the chaotic behavior of the kicked $p$-spin models by computing an estimate of $\Lambda_{+}$ in the limit of strongly chaotic trajectories, $k \gg 1$. This estimate for the model with $p=2$ was first obtained in~\cite{Constantoudis1997}. For models with a general value of $p$, we show in Appendix \ref{app:big_k_lyap} that the largest Lyapunov exponent can be approximated by
\begin{equation}
\label{eqn:big_k_lyap}
\Lambda_{+}(\alpha, k, p) = \ln\left[(p-1)\sin(\alpha)k\right] - (p-1),
\end{equation}
Several observations follow from the form of Eq. (\ref{eqn:big_k_lyap}). First, strong global chaos behaves similarly in all the models, regardless of the value of $p$, since $\Lambda_+\sim\ln(k)$. Second, the value of $k$ at which the limit of strong chaotic trajectories is reached is exponential in the size of $p$. Third, the periodicity of $\Lambda_{+}(\alpha, k, p)$ with $\alpha$ implies that chaotic dynamics cannot develop when $\alpha$ is an integer multiple of $\pi$, since $\Lambda_+=0$. At these values of $\alpha$, the precession will map the system to itself or to its $y$-image. Finally, in the limit of large kicking strengths, Eq. (\ref{eqn:big_k_lyap}) has a maximum at $\alpha = \pi/2$, indicating that the system will exhibit the strongest chaotic limit at this value of $\alpha$. 

Here we completely characterize $\Lambda_+$, including the case of weak chaos, by numerically calculating Eq. (\ref{eqn:oseledec}). We use a method based on QR decomposition~\cite{Benettin1976,Benettin1980,Geist1990} and compute $\Lambda_{+}$ for values of $k\in[0, 100]$ and $\alpha=\pi/2$. Results for the models with $p=2,3,4$, with $N$ up to $10^6$ steps, are shown as dots in Fig. \ref{fig:lyaps_all}a. Note that the models with $p>2$ already have a nonzero Lyapunov exponent at values of $k\sim1$. In the case of the model with $p=3$, red dots in Fig. \ref{fig:lyaps_all}a, we know that the instability of the parabolic points on the period-$4$ orbit along the equator guarantees the existence of small regions of chaotic trajectories in the vicinity of the orbit whose size grows continuously as the kicking strength increases. For the model with $p=4$, blue dots in Fig. \ref{fig:lyaps_all}a, the exponent becomes positive for the first time around $k\sim\pi/2$, when the period-$4$ orbit on the equator bifurcates for the first time (see inset in Fig. \ref{fig:lyaps_all}a). 

In contrast, the exponent for the model with $p=2$ remains zero up to $k>2$, when the first period doubling bifurcation takes place. Once the period doubling bifurcations begin, the model with $p=2$ approaches the limit of strong chaotic trajectories (dashed black line in Fig. \ref{fig:lyaps_all}a) faster than the models with $p>2$. In fact, for $p=2$, already for small values of $k$, the estimate in Eq. (\ref{eqn:big_k_lyap}) is a good approximation to $\Lambda_+$. After the onset of chaos, the system rapidly approaches the limit of strongly chaotic trajectories. However, it does not capture the small oscillations appearing at intermediate values of $k$, which where studied and characterized in~\cite{Constantoudis1997}. On the other hand, for larger values of $p$, larger kicking strengths are required to push the system into the strong chaotic trajectories regime, as noted from Eq. (\ref{eqn:big_k_lyap}).

In summary, for the case of $\alpha=\pi/2$ two important features stand out. On the one hand, chaos is an early phenomenon in models with $p>2$, either due to the instability of the period-$4$ orbit on the equator or its bifurcations. However, at larger values of $k$ this process slows down due to the everlasting stability of the fixed points at the poles. On the other hand, the model with $p=2$ exhibits a cascade of period doubling bifurcations which brings phase space to global chaos faster than any other model. This is due to the fact that a period doubling bifurcation is the last one to take place before the vicinity of the fixed point becomes completely chaotic~\cite{Reichl,Mackay1983}.

We conclude the study of the largest Lyapunov exponent $\Lambda_+$ with a numerical exploration of its behavior as a function of both model parameters $(k,\alpha)$, in the ranges $\alpha\in[0,\pi]$ and $k\in[0,10]$. Numerical results are shown in Figs. \ref{fig:lyaps_all}b,c,d. For the model with $p=2$ the behavior of $\Lambda_+$ is dominated by the case of $\alpha=\pi/2$ (see Fig. \ref{fig:lyaps_all}b) as we described above. However, for models with $p>2$ this is not the case. If $p$ is odd the transition to chaos occurs first in the region $\pi/2<\alpha<2\pi/3$, where both poles undergo $1$ to $4$ and $1$ to $3$ bifurcations. In particular, chaos appears fairly early when $\alpha\sim 2\pi/3$ (see Fig. \ref{fig:lyaps_all}c) since at this bifurcation point, the poles have an unstable character and a small value of $k\sim1$ is enough to generate a chaotic sea of considerable size; we will expand on this in the next subsection. For the models with even values of $p$ the main features of $\Lambda_+$ appear symmetrically around $\alpha = \pi/2 \pm \pi/6$ (see Fig. \ref{fig:lyaps_all}d) values for which bifurcations similar to those taking place in the models with odd $p$'s appear. However, as we saw in Sec. \ref{sec:ndyna_p_spin}, here the north/south pole undergoes the bifurcations to the right/left of $\alpha = \pi/2$, respectively.

\subsubsection{Behavior of the chaotic sea surface area}
\begin{figure}[!t]
 \centering{\includegraphics[width=0.48\textwidth]{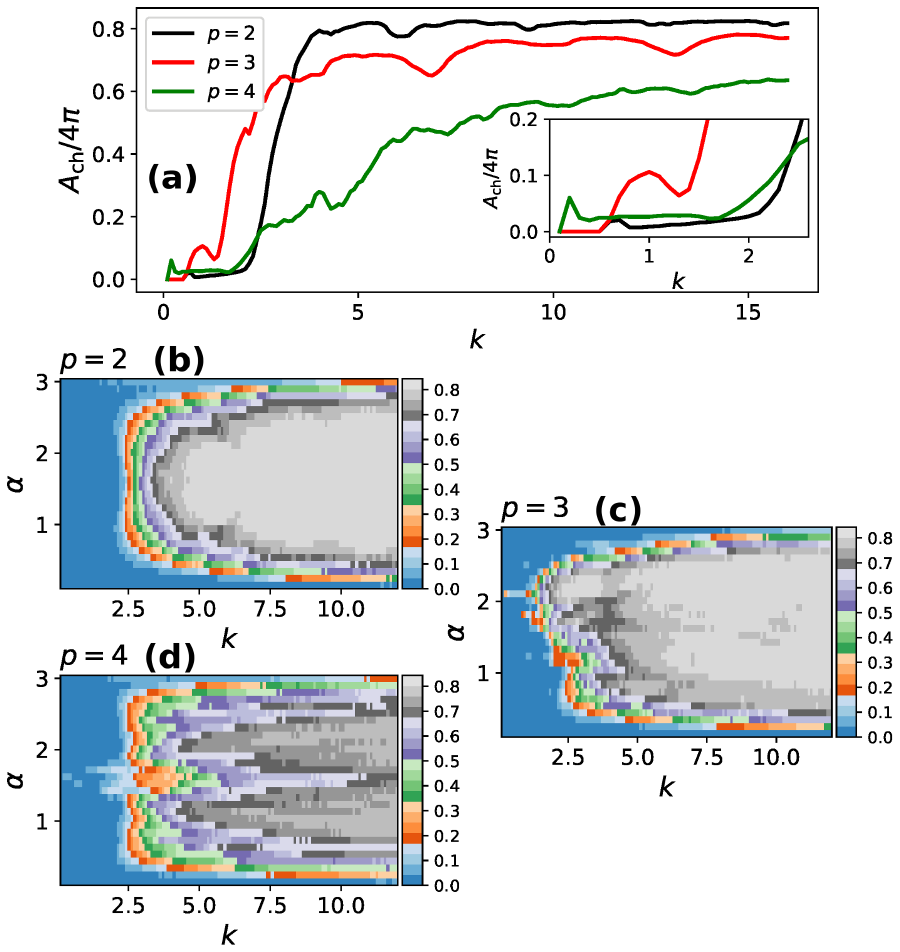}}
\caption{(a) Surface of the chaotic sea $A_{\rm ch}$ as 
a function of the nonlinear parameter $k$ for the special case of $\alpha = \pi/2$. In the models with $p>2$ (red and green), instability and/or bifurcations of the orbit on the equator lead to a small chaotic region at fairly small values of $k$. For $p=2$, $A_{\rm ch}$ grows exponentially after the first period doubling bifurcation at $k=2$. The inset shows a zoom into the region with $k\in[0, 2.5]$. (b,c,d) Surface area of the chaotic region as a function of both $k$ and $\alpha$ for the models with $p=2,3,4$, respectively. The fastest growth of $A_{\rm ch}$ for $p=2$ occurs at $\alpha=\pi/2$, for $p=3$ occurs at $\alpha=2\pi/3$ as the poles are unstable, for $p=4$ it occurs at $\alpha=\pi/2 \pm \pi/6$, as the bifurcation processes on the poles occur for the north/south poles symmetrically with respect to $\alpha=\pi/2$. All the curves and points on the heat maps were obtained by averaging the result of the calculation explained in Appendix \ref{app:numerical_chaotic_area} over values of $t_{\rm max} = 120, 121,..., 140$ and using $d_{\rm min} = 6\time10^{-2}$.}
\label{fig:areas_all}
\end{figure}

The study of the largest Lyapunov exponent provided a distinction between the models with $p=2$ and $p>2$, which we connected to the stability/instability of the main regular regions of their corresponding phase space. However, $\Lambda_{+}$ is a global measure and does not provide explicit information of the shapes and sizes of regular and chaotic regions. To complement our previous observations we study the behavior of the size of the chaotic region as a function of the model parameters. 

The surface area of the chaotic sea, denoted here $A_{\rm ch}$, can be estimated following a Metropolis sampling-like algorithm, as presented in the Appendix of~\cite{Fortes2019}. The key idea behind this method is the concept of recurrence times~\cite{Anishchenko2013}. In short, given some set of $n_{\rm tot}$ initial conditions uniformly distributed on the manifold of interest, we count how many have not returned sufficiently close to the initial neighborhood after some finite time $t_{\rm max}$. Given the surface area of the phase space manifold, this number gives a good approximation to the portion that is occupied by a chaotic region. Further details on the method and our choice of parameters are given in Appendix \ref{app:numerical_chaotic_area}. 

Using this Metropolis-like method we numerically study the behavior of the surface area of the chaotic region as a function of $k$ and $\alpha$, and pay special attention to the case $\alpha=\pi/2$. Results for $A_{\rm ch}(k)$ in this latter case are shown in Fig. \ref{fig:areas_all}a. For values of $k<2$, the area of the chaotic region for $p=3,4$ is always larger than that of $p=2$. In particular, for the model with $p=3$ we know a chaotic sea develops in the vicinity of the period-$4$ orbit along the equator, as it is composed of unstable points. 

For the model with $p=2$ (black line in Fig. \ref{fig:areas_all}), after $k=2$, $A_{\rm ch}$ grows exponentially fast, as a consequence of the period doubling cascade, already covering the whole sphere at $k\approx 3.5$, in agreement with our observations steaming from the study of the largest Lyapunov exponent. Notice that the models with $p>2$ cannot follow this exponential growth of $A_{\rm ch}$ for this large range of values of $k$, since the chaotic sea is constrained between stable regions, either the poles (odd $p$) or the poles and equator (even $p$), and they remain stable for all values of $k$, only gradually reducing its size. 

The behavior of $A_{\rm ch}(k,\alpha)$ for the models with $p=2,3,4$, in the ranges $k\in[0,12]$, $\alpha\in[0,\pi]$, are shown in Fig. \ref{fig:areas_all}b,c,d, respectively. In agreement with our observations for the largest Lyapunov exponent, the behavior of $A_{\rm ch}$ for the model with $p=2$ is dominated by the bifurcation processes taking place as a function of $k$ when $\alpha=\pi/2$. In fact, as a function of $k$, $A_{\rm ch}(k,\alpha)$ reaches the saturation value when $\alpha=\pi/2$ faster than for any other value of $\alpha$ (see Fig. \ref{fig:areas_all}b). 

In the case of models with $p>2$ and odd, the dominant behavior of $A_{\rm ch}(k,\alpha)$ takes place around $\alpha = 2\pi/3$, value at which a $1$ to $3$ bifurcation occurs. Furthermore the unstable character of the poles lead to an early emergence of a considerable sized chaotic region, $A_{\rm ch}\approx 30\%$ around the point $(k,\alpha) \sim (1,2\pi/3)$ in Fig. \ref{fig:areas_all}c. For models with $p>2$ and even, we do not observe a chaotic sea for values of $k<2.5$, except when $\alpha=\pi/2$, where the bifurcations of the period-$4$ orbit a long the equator create a narrow chaotic region. 

Finally, we note that the behavior of $A_{\rm ch}$ as a function of both $k$ and $\alpha$ is in direct correspondence with that of $\Lambda_+$, as can be seen by comparing Figs. \ref{fig:lyaps_all}b,c,d with Figs.  \ref{fig:areas_all}b,c,d. With this two quantities we have complete information regarding the sizes of chaotic regions in phase space and the strength of the local instability of trajectories inside these chaotic seas.

\section{Quantum chaos of a kicked $p$-spin}
\label{sec:quantum_chaos}
In this section we characterize the quantum chaotic features of the kicked $p$-spin model, informed by the previous analysis of the classical nonlinear dynamics. 

Signatures of quantum chaos arise from two different points of view.  On the one hand, signatures of chaos are found in properties of the eigenvalues and eigenvectors of the Hamiltonian or Floquet operator driving the dynamics~\cite{Haake2001a}. We refer to these as \emph{kinematic} signatures. In their study, the system symmetries play a central role. On the other hand, quantum chaos can be characterized via \emph{dynamical} signatures appearing in the time evolution of the states or observables~\cite{Emerson2002, Srednicki1999, Torres-Herrera2017, Zhuang2013, Zurek1995}. These include the dynamical generation of entanglement~\cite{Trail2008, Kumari2019,Wang2004,Arul2001,Kubotani2006}, ``hypersensitivity'' to perturbations~\cite{Schack1996}, tripartite mutual information~\cite{Seshadri2018}, the easiness/hardness of reconstructing an initial state via tomographic protocols~\cite{Madhok2016, Madhok2014}, to name few. More recently, the use of high order correlation functions, in particular the out-of-time-order correlator (OTOC)~\cite{Larkin1969,Maldacena2016}, 
a four point correlation between two observables with vanishing commutator at the initial time, has received attention given the relationship between chaos and information scrambling~\cite{Sachdev1993,Swingle2018,Riddell2019,Landsman2019,Pappalardi2018}.

As a first step we study the symmetries of the Floquet map $\hat{U}_{p}$ in Eq. (\ref{eqn:floquet_kicked_p_spin}). The map $F$ in Eq. (\ref{eqn:classical_kicked_p_spin}) is the classical limit of this quantum map, and therefore we expect that each of the symmetries of $F$ should be manifested as a symmetry of $\hat{U}_p$. First we investigate how symmetry under $R_y(\pi)$ (or the lack of it) is manifested in the quantum system. It follows from Eq. (\ref{eqn:y_symmetry}) that $\hat{U}_{p}$ is invariant under $R_y(\pi) = e^{-i\pi\hat{J}_y}$ for even values of $p$. Thus, the Floquet eigenvectors come with two different parities according to how 
they transform under $R_{y}(\pi)$, and thus a block diagonal 
representation for $\hat{U}_{p}$ can be constructed. Time reversal 
is obtained from the two appropriate 
anti-unitary operators $\hat{T}$ and $\hat{\tilde{T}}$, which yield the 
doubly reversible character of the quantum evolution for even values of $p$, 
with the composition rules for $T, \tilde{T}$ and $R_{y}(\pi)$ described in Sec. \ref{subsec:symmetries}. Similar to the classical case, the broken rotational symmetry around  $y$ for odd values of $p$ implies that only $\hat{T}$ is a proper time reversal operator for the dynamics of those models. Additionally, for $\alpha=\pi/2$, $\hat{U}_{p}^2$ is invariant under $\pi$ rotations around the $x$-axis when $p$ is even, as this symmetry requires invariance under $R_{y}(\pi)$. In correspondence with the family of involutions $\mathcal{I}$ introduced in Sec. \ref{subsec:symmetries}, one can construct operators $\hat{\mathcal{I}}$, which provide a way of identifying additional symmetries.

\subsection{Diagnosing quantum chaos: the kinematic view}
\label{subsec:qchaos_indicators}
\begin{figure}[!t]
 \centering{\includegraphics[width=0.48\textwidth]{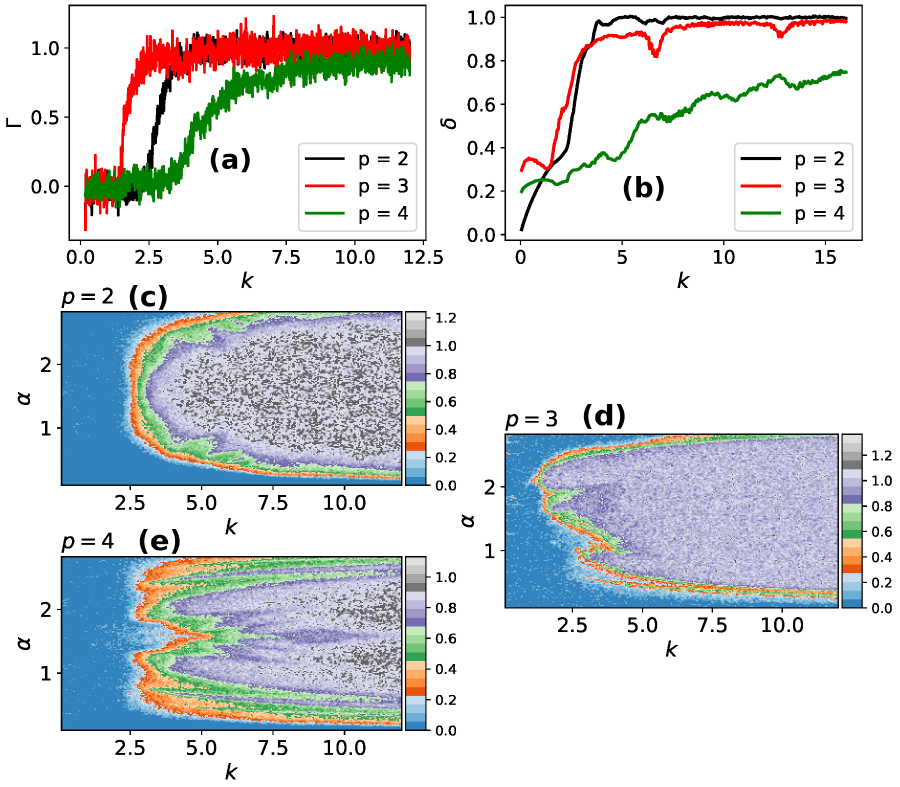}}
\caption{(a) Normalized averaged 
adjacent ratios, $\Gamma(k)$, of the eigenphases of 
$\hat{U}_{p}$ \textcolor{blue}{for the} special case of $\alpha=\pi/2$. Eigenphases show a nontrivial degree of correlation for the model with $p=3$ at small values of $k\sim1.5$. The model with $p=2$ only shows a nontrivial value after $k>2$. Both behaviors in concordance with the observations for $A_{\rm ch}$ in Fig. \ref{fig:areas_all}a. In order to acquire good statistics we use $N_s=2048$. (b) Averaged \textcolor{blue}{PR}, $\delta$, as a function of $k$ for the special case of $\alpha = \pi/2$. The models with $p>2$ show some degree of Floquet vectors delocalization at small $k$'s, Floquet vectors retain some degree of the instability of classical trajectories. We use $N = 1024$. (c,d,e) Normalized average adjacent ratio of eigenphases as a function of $k$ and $\alpha$, in the ranges $k\in[0,12]$, $\alpha\in[0,\pi]$, for the models with $p=2,3,4$, respectively. $\Gamma(k,\alpha)$ exhibits the same behavior as the classical measures $\Lambda(k,\alpha)$, $A_{\rm ch}(k,\alpha)$, indicating that the eigenphases carry information of the stability/instability and different bifurcation processes taking place in the classical model. (c-e) Numerical results with $N=1024$ for the models with even value of $p$ were obtained by combining the statistics of the parity-symmetric and the parity-antisymmetric subblocks of $\hat{U}_{p}$.}
\label{fig:ratios_ipr}
\end{figure}
Different kinematic signatures have been proposed to quantify the chaoticity of quantum systems~\cite{Stockmann2006a,Haake2001a}. Among these, the statistics of the level spacing of eigenphases
$\{\mu_j\}_{j=1,..,N_{s}+1}$ of $\hat{U}_{p}$, 
is widely accepted as an indicator of the transition from regularity to chaos, in particular for systems with a chaotic classical counterpart~\cite{Berry1977,Bohigas1984}. We consider the statistics of ratios of level spacings between two adjacent eigenphases, as introduced in~\cite{Oganesyan2007}, to quantify the degree of repulsion between eigenphases. A simple test of the degree of regularity of the spectrum is provided by computation of the average adjacent spacing ratio~\cite{Atas2013}, defined as  
\begin{equation}
 \label{eqn:adjacent_ratio}
 \overline{r} = \frac{1}{N_s+1}\sum_{j=1}^{N_{s}+1}r_j, \quad r_j = 
\frac{{\rm min}(d_j, d_{j+1})}{{\rm max}(d_j, d_{j+1})},
\end{equation}
where $d_j = \mu_{j+1} - \mu_j$ is the eigenphase spacing and $N_s = 2J$.
The regular regime is characterized by the absence of correlations between the eigenphases, in which case the statistics of the spacings $\{d_j\}$ follow that of a Poisson distribution, with an average adjacent eigenphase spacing ratio given by $\overline{r}_{\rm POS}\approx0.39$~\cite{Atas2013}. On the other hand, chaos is associated with the presence of strong correlations between the eigenphases (after removing additional symmetries in $\hat{U}_{p}$, for instance, parity symmetry for even $p$ values). In this case, the eigenphase spacing follows the statistics of the circular orthogonal ensemble (COE) of random matrices, where time-reversal symmetry is the only remaining symmetry. The average adjacent spacing ratio $\overline{r}_{\rm COE}$ for this ensemble has a value of $\overline{r}_{\rm COE}\approx0.530(1)$ \cite{Atas2013}. Given these two limiting values for the mean adjacent ratio, we define the following normalized indicator 
\begin{equation}
 \label{eqn:normalized_adjacent_ratios}
 \Gamma = \frac{\overline{r} - \overline{r}_{\rm POS}}{\overline{r}_{\rm COE} - 
\overline{r}_{\rm POS}},
\end{equation}
where now a value of $\Gamma\sim0$ indicates a regular regime of in the 
quantum kicked $p$-spin Floquet operator, and a value of $\Gamma\sim1$ signals the chaotic regime.

We numerically study the behavior or $\Gamma$ as a function of $k$ and $\alpha$. For a fixed value of $\alpha=\pi/2$ and for the models with $p=2,3,4$ results  are shown in Fig. \ref{fig:ratios_ipr}a, corresponding to the black, red and green lines, respectively. The model with $p=3$ presents a value of $\Gamma$ which deviates from the Poisson value when $k\sim1$, in agreement with our observations for the classical model where instability of some regions of phase space gave birth to small chaotic seas at similar values of $k$. We then see that the eigenphase repulsion encodes information about the instability present in this model. For the model with $p=2$, we see a nonzero value of $\Gamma$ only for $k>2$, in agreement with the existence of a classical mixed phase space due to the emergence of chaotic regions after the period doubling bifurcation of the corresponding classical model. For large enough kicking strength all models saturate to the random matrix prediction (regardless of the value of $p$), giving evidence of the fully chaotic character of the spectral statistics.

As we observed in our analysis of the classical nonlinear dynamics, when $\alpha\ne\pi/2$ there are rich and intricate phase space structures for the models with $p>2$. To explore their manifestations in the quantum map, we numerically compute the normalized averaged adjacent ratio, $\Gamma(k,\alpha)$ in the ranges $k\in[0,12]$ and $\alpha\in[0,\pi]$. Results are shown in Fig. \ref{fig:ratios_ipr}c,d,e. We observe similar behavior to that of the classical indicators $\Lambda_+(k,\alpha)$ and $A_{\rm ch}(k,\alpha)$ presented in Fig. \ref{fig:lyaps_all} and Fig. \ref{fig:areas_all}, respectively. 

For the model with $p=2$, the behavior of $\Gamma(k,\alpha)$ is dominated by the case of $\alpha=\pi/2$, and values of $\alpha\ne\pi/2$ lead to a wider ranges of $k$ for which the eigenphases do not display strong repulsion (see the blue regions in Fig. \ref{fig:ratios_ipr}a). For models with $p>2$ and odd, as $p=3$ in Fig. \ref{fig:ratios_ipr}d, the spectrum exhibits strong eigenphase repulsion, almost saturating the random matrix prediction, at small values of $k$ when $\alpha\sim2\pi/3$, values at which the classical model has an unstable bifurcation point. For models with $p>2$ and even, as $p=4$ in Fig. \ref{fig:ratios_ipr}e, $\Gamma(k,\alpha)$ is symmetric with respect to $\alpha=\pi/2$, and it displays the strongest eigenphase repulsion around $\alpha\sim\pi/2\pm\pi/6$. Around those two values, it saturates the random matrix prediction only for $k \gtrsim 4$, thus approaching the chaotic regime slower than the other models. This is a direct consequence of the high regularity and stability of the corresponding classical model.

Another useful kinematic signature of quantum chaos is the participation ratio (PR) associated with the Floquet eigenstates. Generally the PR is defined as the inverse of the second moment of the distribution elements
\begin{equation}
 \label{eqn:ipr}
 {\rm PR}(|\psi\rangle) = \left(\sum_{l=1}^{N_s +1}|\langle\psi|\phi_l\rangle|^4\right)^{-1},
\end{equation}
where $|\psi\rangle$ is an arbitrary state and 
$\{|\phi_l\rangle\}_{l=1,..,N_s+1}$ is a reference basis set. In our case it corresponds to the eigenbasis of $\hat{J}_y$, which defines the precession axis and thus the canonical direction for our $p$-spin. The PR measures how localized or delocalized the state $|\psi\rangle$ is in the 
reference basis. Thus, we can use the PR to construct a 
measure of localization of the Floquet eigenbasis $\{|\mu_l\rangle\}_{l=1,..,N_s+1}$ by taking the averge PR of the Floquet states in the reference basis. We then define
\begin{equation}
\label{eqn:simi_floquet}
 \delta = 
\frac{1}{\delta_{\rm COE}(N_s +1)} \sum_{l'=1}^{N_s +1}{\rm 
PR}(|{\mu}_{l'}\rangle),
\end{equation}
where $\delta_{\rm COE} \sim \frac{N_s+1}{3}$ is the value of the PR averaged over the COE  ensemble (see methods in~\cite{Sieberer2019} and \cite{Gubin2012}), and 
$\{|\mu_l\rangle\}_{l=1,..,N_s +1}$ are the eigenvectors of 
the Floquet operator $\hat{U}_{p}$. Under this definition $\delta\sim \frac{1}{\delta_{\rm COE}}$ indicates strong localization of the Floquet eigenvectors, associated with the regular regime, and $\delta\sim1$ indicates highly delocalized Floquet eigenvectors which are generically associated with the chaotic regime.

Numerical results for $\delta(k)$ in the case of $\alpha=\pi/2$ are shown in Fig. \ref{fig:ratios_ipr}b, with $p=2,3,4$ corresponding to the black, red, and green lines, respectively. From the average localization of the Floquet states in the basis of $\hat{J}_y$ we recognize a similar behavior to that of the surface area of the chaotic sea, $A_{\rm ch}$ in Sec. \ref{subsec:trasition_hamil_chaos}. For small values of $k$, Floquet states for $p=3, 4$ show nonzero average delocalization, indicating that the Floquet states retain some of the unstable character of trajectories in the corresponding classical model. As we increase $k$, $\delta$ increases for all values of $p$ eventually saturating the random matrix prediction. However in the case of $p=2$, $\delta(k)$ grows faster than any other models, saturating the random matrix prediction first. We highlight how the kinematic signatures studied here are in excellent correspondence with our observations on stability and transition to chaos in the family of classical models~\footnote{This was expected, since in the semiclassical limit $N_s\gg1$ Floquet states will have a strong correspondence with classical phase space trajectories, as can be seen by looking at their phase space representation (for instance using the Husimi $Q$-function) and therefore they will inherit properties of the classical system, that we observe in the kinematic signatures.}.

\subsection{Early time Lyapunov growth of the OTOC}
\label{subsec:early_time_otoc}

The out-of-time-order correlator (OTOC) is a temporal correlation function measuring the growth in time of the overlap between two observables that initially commute. It was initially introduced as a probe of nonlinear behavior in the mean-field theory of superconductivity~\cite{Larkin1969}, later rediscovered and popularized due to its importance in the study of information 
scrambling~\cite{Maldacena2016,Riddell2019,Landsman2019,Pappalardi2018} in nonequilibrium many-body quantum systems and its relation with the classical Lyapunov exponent. In this context, the OTOC is given by
\begin{equation}
\label{eqn:fotoc}
f(t) = {\rm tr}\left(\rho_0\hat{V}^\dagger\hat{W}^\dagger(t)\hat{V}\hat{W}(t)\right),
\end{equation}
with $\rho_0$ a reference initial state, $\hat{V}$, $\hat{W}$ two operators of interest which commute at the initial time, \textit{i.e} $[\hat{V}, \hat{W}(0)] = 0$, and $\hat{W}(t)$ denotes the Heisenberg evolution of $\hat{W}(0)=\hat{W}$ up to some finite time $t$.

A related quantity of interest is the operator growth of the commutator between $\hat{V}$ and $\hat{W}(t)$, since it provides information on the speed at which the available degrees of freedom are occupied in time. The growth of the square commutator is quantified by
\begin{equation}
 \label{eqn:otoc}
 C(t) = {\rm tr}\left(\rho_0[\hat{W}(t), \hat{V}]^\dagger[\hat{W}(t), \hat{V}]\right),
\end{equation}
where $\hat{V}$, $\hat{W}$ are as in Eq. (\ref{eqn:fotoc}). The exact form of $f(t)$ and $C(t)$ will depend on the choice of operators and reference state. For the latter, if one considers a thermal state the growth rate and saturation value of $C(t)$ strongly depends on the temperature~\cite{Jalabert2018,Hashimoto2017}. Here our interest is to study the growth of the commutator purely due to operator growth, and so we choose $\rho_0 = \frac{1}{\mathcal{D}}\mathbf{I}$, the infinite temperature state, where expectation values are given by $\langle \hat{B} \rangle = 
\frac{1}{\mathcal{D}}{\rm tr}(\hat{B})$. Furthermore we take the operators 
$\hat{V}$ and $\hat{W}$ to be Hermitian. Under these conditions, Eq. (\ref{eqn:otoc}) takes the form 
\begin{equation}
 \label{eqn:otoc_hermitian}
 C(t) = \frac{2}{\mathcal{D}}\left( {\rm tr}(\hat{V}^2\hat{W}^2(t)) - 
{\rm Re}\left[f(t)\right]\right),
\end{equation}
where $\mathcal{D}$ is the dimension of the Hilbert space.
\begin{figure}[!t]
 \centering{\includegraphics[width=0.48\textwidth]{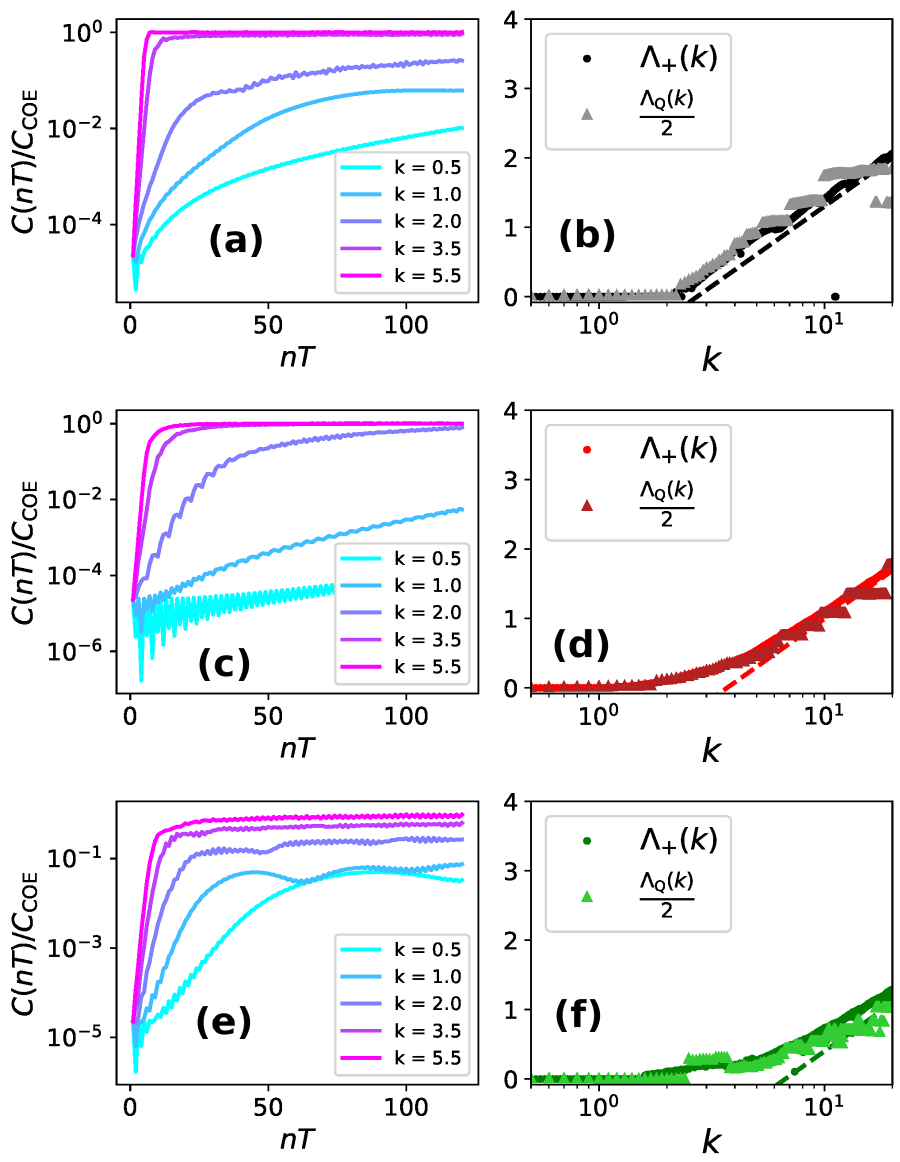}}
\caption{(a,c,e) Short time evolution of the OTOC for different values 
of $k$ with $N=512$. (b) Lyapunov exponent as obtained from the short time growth 
rate of the OTOC (triangles), numerically from the classical map (dots), 
and analytically from Eq. (\ref{eqn:big_k_lyap}) (dashed line). 
(d) Lyaounov exponent as obtained from the short time growth 
rate of the OTOC (triangles), numerically from the classical map (dots), and analytically from Eq. (\ref{eqn:big_k_lyap}) (dashed line). 
(f) Lyaounov exponent as obtained from the short time growth 
rate of the OTOC (triangles), numerically from the classical map (dots), 
and analytically from Eq. (\ref{eqn:big_k_lyap}) (dashed line). From top to 
bottom we show, $p=2$ (a,b), $p=3$ (c,d), and $p=4$ 
(e,f).}\label{fig:otoc_lyap}
\end{figure}
The square commutator as defined in Eq. (\ref{eqn:otoc}) typically exhibits two different behaviors, at short and long times. The short-time behavior is characterized by a monotonic growth, which has been reported to follow different functional forms~\cite{Dora2017,Kukuljan2017,Riddell2019,Fortes2019}, especially in generic many-body systems. Furthermore it has been conjectured that the initial growth rate saturates and is always
bounded~\cite{Maldacena2016}. For quantum systems with chaotic classical counterparts, it has been shown in several models that the growth rate of $C(t)$ at early times is exponential and characterized by the classical Lyapunov exponent or by a factor proportional to it~\cite{Rozenbaum2017,Garcia-Mata2018,Chavez-Carlos2019}. A discussion of the origin of this phenomena in the semiclassical regime for systems of collective spin variables was recently given in~\cite{Lerose2020}, and for a generic bosonic mode in~\cite{Yan2020}. 

We point out that in some cases, quantum systems with integrable classical counterparts can also lead to ``scrambling" in the sense of an exponentially increasing $C(t)$ at short times. This behavior is typically attributed to the presence of saddle points in the classical dynamics~\cite{Xu2020,Kidd2021}. Due to this fact, the long-time behavior of $C(t)$ has been proposed as a complementary probe for quantum chaos~\cite{Fortes2019,Kidd2021}, since for chaotic systems $C(t)$ is expected to present oscillations of exponentially vanishing amplitude. For the case of the kicked $p$-spin  models, the exponential growth of $C(t)$ can be safely attributed to chaos, for $p>2$, and we see in the classical analysis that there are no saddle points. This conclusion holds for the case of $p=2$ and $\alpha = \pi/2$, studied in Fig.~\ref{fig:otoc_lyap}, as the first saddle point appears at $k=2$, value at which a nonnegligible chaotic sea is already present in phase space.

We now turn our attention to the short time regime of $C(t)$ for the dynamics of the kicked $p$-spin models. In particular we look at the square commutator with the choice of operators $\hat{V}=\hat{W}=\hat{J}_z$ and thus $\hat{W}(t)=\hat{W}(nT)=\hat{U}_{p}^{\dagger n}\hat{J}_z\hat{U}_{p}^n$, operators which are accessible in state of the art proposals for measuring OTOC's~\cite{Swingle2016}. In Fig. 
\ref{fig:otoc_lyap}a,c,e we present the early time evolution of $C(nT)/C_{\rm COE}$ for $p=2,3,4$, respectively. The normalization factor $C_{\rm COE}$ is obtained by replacing $\hat{U}_p$ in Eq. (\ref{eqn:otoc}) by a random unitary from the COE ensemble (see methods in~\cite{Sieberer2019} for further details). Notice how the exponential growth is already visible 
at $k\sim1.5$ for $p>2$ (green and orange lines in Fig. 
\ref{fig:otoc_lyap}c,e). On the other hand, once $C(nT)/C_{\rm 
COE}$ grows exponentially, the rate of growth is larger for $p=2$ (see red and purple lines in Fig. \ref{fig:otoc_lyap}a,c,e). These two aspects are in direct agreement with the behavior of $\Lambda_{+}$ in Fig. \ref{fig:lyaps_all}a. 

Finally, by a linear fit of the section that grows exponentially, we extracted the quantum Lyapunov exponent $\Lambda_{\rm Q}$, shown as light dots in Fig. \ref{fig:otoc_lyap}b,d,f. From this fit and direct comparison with the largest Lypaunov exponent, $\Lambda_{+}$ in Fig. (\ref{fig:lyaps_all}), we found $\Lambda_{\rm Q} \approx 2\Lambda_{+}$. This result expands those in~\cite{Garcia-Mata2018, Rozenbaum2017}, providing evidence of the early-time Lyapunov growth of the OTOC for a system whose dynamics is constrained to a compact phase space, here, the unit sphere. This is in agreement with the recent result of Lerose and Pappalardi who, using a quantum generalization of the Oseledets ergodic theorem in the semiclassical limit~\cite{Lerose2020},  provided an explicit construction that connected the OTOC and other dynamical signatures such as entanglement entropy, with the classical Lyapunov exponent and Kolmogorov-Sinai entropy.

\section{Summary and outlook}
\label{sec:final_remarks}
We studied the Floquet dynamics of a family of Ising $p$-spin models subject to time-periodic delta kicks. These models can be regarded as the generalization of the paradigmatic quantum kicked top, which is recovered for $p=2$. We fully characterized the classical nonlinear dynamics of these models by studying its symmetries, fixed points, stability, bifurcations, and the emergence of chaos. This analysis allowed us to draw several distinctions between the models with different $p$'s. With this foundation, we characterized the quantum chaotic features of the kicked $p$-spin models via both kinematic (eigenvalues and eigenvectors) and dynamical indicators (OTOCs).  We saw how the classical dynamics informed the emergence of quantum chaos in the limit of large spins.

The generalization of the kicked top for $p>2$ showed new phenomena arising from the decoupling of the effects of the two different dynamical processes: precession and nonlinear kicking, characterized by the parameters $\alpha$ and $k$, respectively. In other words, in the case of $p=2$, structural changes of phase space as well as the transition to global chaos are dependent on both $\alpha$ and $k$. Here, the most prominent parameter regime takes place at $\alpha = \pi/2$, value at which a cascade of period doubling bifurcations accelerates the transition to global chaos. On the other hand, for $p>2$, structural changes of phase space are strictly dictated by $\alpha$, and the transition to global chaos is dictated only by $k$. A further distinction within the models with $p>2$, is given by the nature of the structural changes, in particular bifurcations. When $p>2$ and even, some bifurcations are double as it is required to satisfy the symmetries imposed by the double reversibility of the models, whereas for $p>2$ and odd, all bifurcation processes are generic~\cite{Meyer1970}.

We illustrate many of the studied phenomena with the models with $p=3,4$. The observed phenomena is exhaustive and covers the whole family of models, where one might need larger values of $k$ with increasing $p$ in order to observe chaotic regions of considerable size. This is in agreement with the instability of the ferromagnetic phase of the $p$-spin models with increasing $p$~\cite{Bapst2012}.

 To characterize quantum chaos, we studied the normalized mean adjacent ratio of level spacings of the eigenphases of the Floquet operator $\hat{U}_p$, and the averaged inverse participation ratio of its eigenvectors. The behavior of these two quantities was seen to be in direct correspondence with that of the classical Lyapunov exponent and the area of the chaotic region in phase space, respectively. Finally, we studied the short time growth of the OTOC. We showed numerically that the growth rate is dictated by twice the classical Lyapunov exponent, $2\Lambda_{+}$, providing further evidence to the connection of the the OTOC with the classical Lyapunov exponent~\cite{Maldacena2016}, for a system whose evolution lies on the unit sphere.


In the present work we studied the kicked $p$-spin models as Hamiltonian dynamical systems. As mentioned in the introduction to the present work, $p$-spin models are of importance in some areas of quantum information processing. In the context of quantum simulation it is now known that such kicked system will naturally arise in an analog quantum simulator where you have restrictions on the allowed ``native gates'' which can be implemented. The effects of chaos in such simulator were studied in~\cite{Sieberer2019} for the case $p=2$. We have shown that the $p$-spin models with $p>2$ display a richer behavior beyond the case of $p=2$, and that chaotic instability is not the only instability playing an important role in these models. Given the complete characterization of these family of models provided in this work, we will extend its application \textcolor{blue}{to} analog simulation in future research.

Furthermore, $p$-spin models are important toy models in adiabatic quantum computing. Given the recently studied connection between discretized adiabatic evolution and certain variational optimization schemes such as quantum approximate optimization algorithm (QAOA)~\cite{Crooks2018,Zhou2020}, the relation, if any, between the instabilities of the kicked dynamics and the performance and efficiency of QAOA in $p$-spin models is an interesting future directions. The phenomenology of $p$-spin models can also be investigated with other types of analog quantum simulators, for instance programmable quantum processors~\cite{Lysne2020}. In that situation, the relation between observed simulation errors, native imperfections, and nonlinear dynamical effects of the simulator model is a research avenue currently under investigation.

\section*{Acknowledgments}
The authors are grateful to Tameem Albash for his insights into the phenomenology of $p$-spin models and Karthik Chinni for helpful discussions on deterministic chaos. This work was supported by NSF Grants No. PHY-1606989, No. PHY-1630114, No. PHY-1820758 and Quantum Leap Challenge Institutes program, Award No. 2016244. This material is based upon work supported by the U.S. Department of Energy, Office of Science, National Quantum Information Science Research Centers, Quantum Systems Accelerator (QSA).

\appendix 
\section{Computation of the Heisenberg equations of motion}
\label{app:eqns_motion}
In this appendix we present the main steps behind the derivation of the stroboscopic Heisenberg equations of motion for the collective operators in Eq. (\ref{eqn:collective_ope_evolved}). 

Consider first the evolution of $\hat{J}_z$, given our choice axis in the $p$-spin Hamiltonian the only nontrivial evolution is generated by the precession unitary. The Heisenberg evolution of $\hat{J}_z$ is then a rotation around the $y$-axis by an angle $\alpha$. The equations for $\hat{J}_x$ and $\hat{J}_y$ can be constructed from the evolution equations of $\hat{J}_{\pm}$. The Heisenberg evolution of the latter are computed exploiting the commutation relations between spin ladder operators and $\hat{J}_z$. A single step of the stroboscopic evolution of the Ladder operators is given by
\begin{equation}
 \hat{J}_{\pm}' = 
e^{i\alpha\hat{J}_y}e^{\frac{ik}{pJ^{p-1}}\hat{J}_z^p}\hat{J}_\pm 
e^{\frac{-ik}{pJ^{p-1}}\hat{J}_z^p}e^{-i\alpha\hat{J}_y}.
\end{equation}
To deal with the unitary involving $\hat{J}_z^p$ we apply the 
Baker-Campbell-Haussdorf formula and get 
\begin{equation}
\label{eqn:p_term}
 e^{\frac{ik}{pJ^{p-1}}\hat{J}_z^p}\hat{J}_\pm 
e^{\frac{-ik}{pJ^{p-1}}\hat{J}_z^p} = \hat{J}_\pm + \sum_{n=1}^\infty 
\frac{1}{n!}\left(\frac{ik}{pJ^{p-1}} \right)^n \left[\hat{J}_z^p, 
\hat{J}_\pm\right]^n,
\end{equation}
where the notation $[\enspace, \enspace]^n$ indicates nested applications of 
the commutator. Noticing that the commutator $[\hat{J}_z^p, \hat{J}_\pm]$ can 
be written as
\begin{eqnarray}
 [\hat{J}_z^p, \hat{J}_\pm] &=& \pm \sum_{a=1}^{p} \hat{J}_z^{p-a}\hat{J}_\pm 
\hat{J}_z^{a-1}, \\
 &=& \pm \hat{J}_\pm\left(\sum_{a=1}^{p}(\pm1)^{a+1}\binom{p}{a} 
\hat{J}_z^{p-a}\right),
\label{eqn:bch_p_term}
\end{eqnarray}
where to go from the first line to the second line we introduced the 
commutation relation $[\hat{J}_z, \hat{J}_\pm] = \pm\hat{J}_\pm$, a total of $(p-a)$-times and move $\hat{J}_\pm$ all the way to the left. 
After substituting Eq. (\ref{eqn:bch_p_term}) into Eq. (\ref{eqn:p_term}) one easily recognizes that the Baker-Campbell-Haussdorf series is nothing but the series expansion of the exponential of the operator sum in Eq. (\ref{eqn:bch_p_term}), and we write
\begin{equation}
 e^{\frac{ik}{pJ^{p-1}}\hat{J}_z^p}\hat{J}_\pm 
e^{\frac{-ik}{pJ^{p-1}}\hat{J}_z^p} = 
\hat{J}_{\pm}e^{\frac{ik}{pJ^{p-1}}\sum_{a=1}^{p}(\pm1)^a 
\binom{p}{a}\hat{J}_z^{p-a}}.
\end{equation}
Now we can easily apply the rotation part of the Floquet operator, and a single step of the stroboscopic evolution of the ladder operators takes the form
\begin{equation}
 \hat{J}_\pm' = \left(\cos(\alpha)\hat{J}_x + \sin(\alpha)\hat{J}_z \pm 
i\hat{J}_y \right)e^{\mathcal{Q}_{\pm}(k, \alpha, p)},
\end{equation}
where the functions $\mathcal{Q}_{\pm}(k, \alpha, p)$ was defined in Eq. (\ref{eqn:argument_heisenberg_equs}) of the main text. From this last expression the equations of motion for $\hat{J}_{x,y}'$ in Eq. 
(\ref{eqn:collective_ope_evolved}) follow.

\section{Details of some stability results}
\label{app:stability_general}
\subsection{Explicit form of the tangent ma}

For all the results presented in the main text and this appendix we have used the tangent map of the inverse classical stroboscopic evolution in Eq. (\ref{eqn:classical_kicked_p_spin}). Explicitly it has the matrix form
\begin{widetext}
\begin{multline}
\label{eqn:tangent_map_p_spin}
 \mathbf{M}^{(p)}(\bm{X}_m) = \\\begin{pmatrix}
 \cos(\alpha)\cos(kZ_m^{p-1}) && \cos(\alpha)\sin(kZ_m^{p-1}) && 
\mathcal{C}(Z_m;k, \alpha, p)\cos(\alpha)\left(-\sin(kZ_m^{p-1})X_m + 
\cos(kZ_m^{p-1})Y_m \right) - \sin(\alpha)\\
 -\sin(kZ_m^{p-1}) && \cos(kZ_m^{p-1}) && -\mathcal{C}(Z_m;k, \alpha, p)\left( 
\cos(kZ_m^{p-1})X_m + \sin(kZ_m^{p-1})Y_m \right) \\
\sin(\alpha)\cos(kZ_m^{p-1}) && \sin(\alpha)\sin(KZ_m^{p-1}) && 
\mathcal{C}(Z_m;k, \alpha, p)\sin(\alpha)\left(-\sin(kZ_m^{p-1})X_m + 
\cos(kZ^{p-1})Y_m \right) 
+ \cos(\alpha)
 \end{pmatrix},
\end{multline}
\end{widetext}
where $\mathcal{C}(Z_m;k, \alpha, p) = (p-1)kZ_m^{p-2}$.

\subsection{Stability of fixed points for arbitrary $\alpha$ and $k$}
In the case of the map in Eq. (\ref{eqn:classical_kicked_p_spin}) for arbitrary values of $k$ and $\alpha$, a general expression for the stability of a fixed point, \textit{i.e} $\bm{X}_m$ such that $F[\bm{X}_m] = \bm{X}_m$, is written by noticing that the tangent map, evaluated at $\bm{X}_m$ has characteristic polynomial of the form
\begin{multline}
\label{eqn:poly_chara_fixed}
\mathcal{M}^3 - G_1(\bm{X}_m;k,\alpha,p)\mathcal{M}^2 + G_2(\bm{X}_m;k,\alpha,p)\mathcal{M} \\ 
+ G_3(\bm{X}_m;k,\alpha,p) = 0,
\end{multline}
where $\mathcal{M}$ are the eigenvalues of the tangent map, and the coefficients $G_{i}(\bm{X}_m;k,\alpha,p)$ with $i=1,2,3$ are given by 
\begin{widetext}
\begin{subequations}
\begin{align}
G_1(\bm{X}_m;k,\alpha,p) &= \mathcal{C}(Z_m;k,\alpha,p)\sin(\alpha)Y_m + \cos(\alpha) + \cos(k Z_m^{p-1})\cos(\alpha) + \cos(k Z_m^{p-1}), \\
G_2(\bm{X}_m;k,\alpha,p) &= -\mathcal{C}(Z_m;k,\alpha,p)\sin(\alpha)\cos(\alpha)\cos(k Z_m^{p-1})Y_m - G_1(\bm{X}_m;k,\alpha,p), \\ 
G_3(\bm{X}_m;k,\alpha,p) &= - \mathcal{C}(Z_m;k,\alpha,p)\left( \cos(k Z_m^{p-1}) - \sin(k Z_m^{p-1}) \right)\cos(\alpha)\sin(k Z_m^{p-1})(1-\cos(\alpha))Z_m- 1.
\end{align}
\end{subequations}
\end{widetext}
Dynamics is constrained to the unit sphere, $|\bm{X}_m|^2 = 1$, thus one of the eigenvalues of $\mathbf{M}(\bm{X}_m)$ is always $1$. We can then write a factorization for the characteristic polynomial in Eq. (\ref{eqn:poly_chara_fixed}) as
\begin{equation}
(\mathcal{M}-1)(B_1\mathcal{M}^2 + B_2\mathcal{M} + B_3) = 0
\end{equation}
where the coefficients $B_i$ with $i=1,2,3$ are functions of $\bm{X}_m$ with parameters $k,\alpha$ and $p$. From this last expression and Eq. (\ref{eqn:poly_chara_fixed}) we identify, $B_1 = 1$, $B_3 = -G_3(\bm{X}_m;k,\alpha,p)$ and $B_2 = 1 - G_1(\bm{X}_m;k,\alpha,p)$. Given these coefficients the other two eigenvalues of $\mathbf{M}$ have the forms $-\frac{B_2}{2} \pm \frac{1}{2}\sqrt{B_2^2 - 4B_3}$, thus the fixed point under study is stable if
\begin{equation}
\label{eqn:stability_general}
B_2^2 - 4B_3 < 0.    
\end{equation}
As a sanity check consider the case of $\alpha=\pi/2$ studied in the main text. For this value of $\alpha$ the coefficients $B_2 \to 1 - (p-1)k Z_m^{p-2}Y_m - \cos(k Z_m^{p-1})$ and $B_3 \to 1$, after which Eq. (\ref{eqn:stability_general}) takes the form 
\begin{equation}
\left( 1 - (p-1)k Z_m^{p-2}Y_m - \cos(k Z_m^{p-1}) \right)^2  - 4 < 0,  
\end{equation}
which recovers the expression given in the main text since, given a fixed point, $X_m = -Z_m$ when $\alpha=\pi/2$. 

\subsection{Stability of the fixed points at the poles}
We study now the stability for a general value of $\alpha$, and the bifurcation processes highlighted in Sec. \ref{sec:ndyna_p_spin} for the fixed points on the poles.

Consider first the model with $p=2$, for the fixed points on the poles we have, $\mathcal{C}(Z_m;k, \alpha, p) \to k$, $\cos(k Z_m)\to 1$, and the coefficients $B_2 \to \mp k\sin(\alpha) - 2\cos(\alpha)$, $B_3 \to 1$, giving the stability condition 
\begin{equation}
\left(2\cos(\alpha) \pm k\sin(\alpha) \right)^2 < 4,
\end{equation}
which reduces to the inequality $k^2<4$ when $\alpha=\pi/2$ as expected. 

In the case of models with $p>2$, for the fixed points at the poles we have, $\mathcal{C}(Z_m;k, \alpha, p) \to 0$, $\cos(k Z_m^{p-1}) \to 1$ and the coefficients $B_2 \to -2\cos(\alpha)$, $B_3 \to 1$, giving the stability condition $\cos^2(\alpha)<1$. Which is satisfied for all $\alpha$ except at the discrete set of values $\alpha = r\pi$ with $r$ an integer. At these particular values the two nontrivial eigenvalues of $\mathbf{M}$ are equal to $\mathcal{M} = \frac{-B_2}{2} = \cos(r\alpha) = \pm1$ depending on the parity of $r$, thus poles are parabolic points. These values of $\alpha$ lead to trivial dynamics. Every point gets mapped to itself after either one or two applications of $F$. We can conclude then, on the stability of the fixed points at the poles for the models with $p>2$ for all values of $k$ and almost all values of $\alpha$. With the only exceptions being given by elliptic points with eigenvalues equal to the $l$-th root of one, as they signal bifurcation points. 

The eigenvalues of $\mathbf{M}$ at the poles as function of $\alpha$ are
\begin{equation}
\mathcal{M} = e^{\pm i\alpha},
\end{equation}
they are roots of $1$ when $\alpha=\alpha_{\rm b}=2\pi q/l$ with $q$ and $l$ relative primes, $q<l$ and $l>2$. We investigate these bifurcations by restricting dynamics to the local neighborhood around the poles.

Consider, for instance, the north pole $(0,1,0)$ and construct the area preserving map for points in its vicinity. This is achieved by taking $Y_m=1$, $X_m=\delta X$ and $Z_m = \delta Z$, with $\delta X,\enspace\delta Z \ll 1$. Expanding Eq. (\ref{eqn:inverse_classical_kicked_p_spin}) to leading order, and noticing that by placing the origin at $(0,1,0)$, the $x$-direction requires a reflection to be oriented in the appropriate fashion, thus we apply the additional transformation $\delta X \to -\delta X$. After these steps we obtain
\begin{subequations}
\label{eqn:at_north_pole}
\begin{align}
\delta X' &= \sin(\alpha) \delta Z  + \cos(\alpha) \left(\delta X -k\delta Z^{p-1}\right), \\
\delta Z' &= \cos(\alpha) \delta Z - \sin(\alpha) \left(\delta X - k \delta Z^{p-1}\right),
\end{align}
\end{subequations}
which is a generalization of the paradigmatic quadratic map initially studied by Michel Henon in~\cite{Henon1969}. 

In the particular case of $p=3$, we recover the general form of the quadratic map~\cite{Henon1969}. Importantly for us, Henon studied the periodic orbits of this map up to period-$4$. He found that there are no period-$2$ orbits. There are two period-$3$ orbits which appear at $\alpha = \cos^{-1}(1-\sqrt{2})$, one composed of unstable points and one composed of stable points up to $\alpha = 2\pi/3$, value at which it changes stability. There are two period-$4$ orbits which appear at $\alpha = \pi/2$. One of them is composed of unstable points, the other one of stable points up to $\alpha = \cos^{-1}(-0.10336015)$, value at which it changes stability. The existence of the period-$3$ orbits is not the result of a $1$ to $3$ bifurcation, as this one is expected to occur at $\alpha = 2\pi/3$, value at which both orbits already exists. On the other hand, the period-$4$ orbits are indeed the result of a bifurcation process and they emerge from the origin at $\alpha=\pi/2$.

The positions of all the points in the period-$3$ and period-$4$ orbits move away from the origin as a function of $\alpha$, therefore these periodic orbits only exists during a, sometimes, more restricted range of $\alpha$'s as the one presented in~\cite{Henon1969}. The identification with the Henon quadratic map is only exact when $p=3$, however the observed phenomenology is similar for all odd values of $p$, where larger values of $k$ are required in order to observe the emergence of these orbits. 

The models with even $p$'s display a different, yet qualitatively similar, phenomenology. In the case of $p=4$ the local area preserving map is cubic. To illustrate these points, and some of the remarks made in Sec. \ref{subsec:stability}, we study the bifurcation at $\alpha=\pi/2$.

First we address the question of whether the fixed point is surrounded by closed invariant curves. Evaluating Eq. \ref{eqn:at_north_pole} at $\alpha = \pi/2$ and taking the second iteration of the resulting map we obtain
\begin{subequations}
\label{eqn:second_map_poles}
\begin{align}
\delta X' &= -(\delta X - k \delta Z^{p-1}), \\
\delta Z' &= -\delta Z - (-1)^{p-1} k (\delta X - k\delta Z^{p-1})^{p-1}.
\end{align}
\end{subequations}
For models with even $p$, Eq. (\ref{eqn:second_map_poles}) satisfies the conditions of the main theorem in~\cite{Aharonov1990}. In fact, it is equivalent to the area preserving map considered in example $1$ in~\cite{Aharonov1990}. Hence, the fixed point at the origin is surrounded by close invariant curves. Similarly, when $p$ is odd the map in Eq. (\ref{eqn:second_map_poles}) is equivalent to the one investigated in example $2$ of~\cite{Aharonov1990}, thus we are guaranteed to have close invariant curves surrounding the fixed point.

The bifurcation process can be studied by considering the area preserving map in Eq. (\ref{eqn:at_north_pole}) with $\alpha = \pi/2 + \gamma$ and $\gamma \ll 1$. Then taking the fourth iterate of the resulting map one finds 
\begin{subequations}
\label{eqn:map_for_bifu}
\begin{align}
\delta X' &= \delta X + (6\gamma^2 - 1)k \delta Z^{p-1} + 4\gamma \delta Z - 6\gamma^2 \delta X, \\ 
\delta Z' &= \delta Z - 4\gamma \delta X + 4\gamma k \delta Z^{p-1} - 6 \gamma^2 \delta Z
\end{align}
\end{subequations}
where we have kept terms up to order $\mathcal{O}(\delta Z^{p-1})$ and $\mathcal{O}(\gamma^2)$. New fixed points of this map are 
\begin{equation}
\label{eqn:bifu_fixed_points}
\delta Z = \left( \frac{4\gamma + 9\gamma^3}{k} \right)^{\frac{1}{p-2}},\quad \delta X = \gamma k \delta Z^{p-1} - \frac{3}{2}\gamma^2 \delta Z.
\end{equation}
We obtain Eq. (\ref{eqn:at_north_pole}) as the local dynamical description around the north pole, however, for odd $p$'s, it also describes local dynamics around the south pole. Therefore Eq. (\ref{eqn:bifu_fixed_points}) gives the bifurcation of the south pole as well. 

For even $p$'s, local dynamics around the south pole is given by Eq. (\ref{eqn:at_north_pole}) only after taking $k\to-k$, thus the bifurcation takes place only when $\gamma<0$, only then Eq. (\ref{eqn:bifu_fixed_points}) yields real values.
\begin{figure}[!t]
\centering{\includegraphics[width=0.48\textwidth]{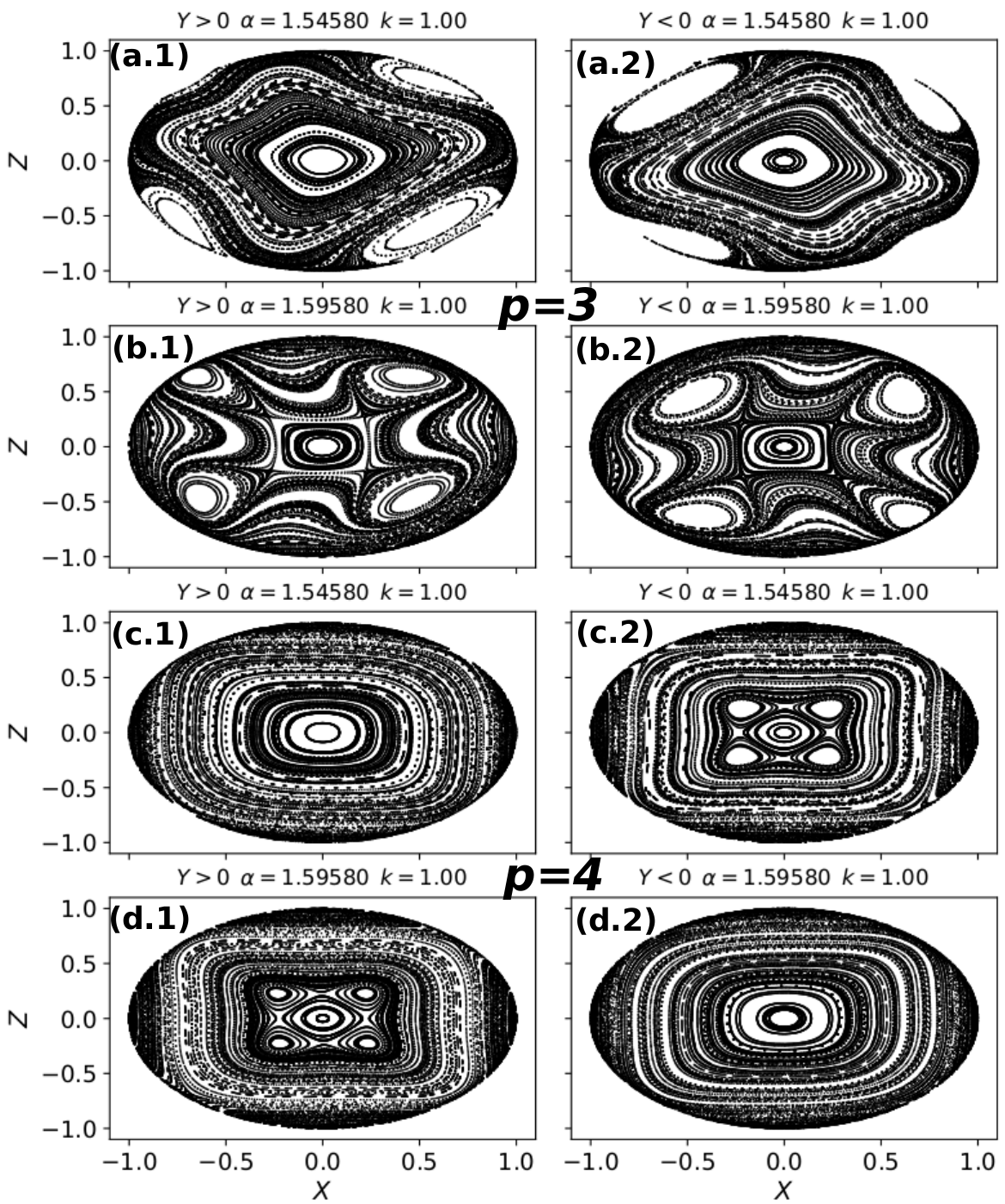}}
\caption{Snapshots of phase space right before and after the 1-to-4 bifurcation studied in this appendix. Left/right columns show the north/south pole projected onto the $x$-$z$.  (a,b) Bifurcation in the model with $p=3$, the bifurcation takes place at $\alpha>\pi/2$ in both north and south poles. (c,d) Bifurcation in the model with $p=4$. The north pole bifurcates for $\alpha>\pi/2$, and the south pole for $\alpha<\pi/2$. The parameters of the displayed phase portraits are: $\alpha=\pi/2 \pm 0.025$, $k=1.0$.}\label{fig:ps_bifu}
\end{figure}
Notice that this asymmetry in the direction of the bifurcation is allowed since the invariance of $F^2[\bm{X}_m]$ under $R_x(\pi)$ only exists at $\alpha=\pi/2$. We display projections of both hemispheres of phase spaces showing the 1-to-4 bifurcation, for the exemplary models with $p=3,4$ in Fig. \ref{fig:ps_bifu}.

Regardless of the parity of $p$, the emergent fixed points are unstable. We see this evaluating the trace of the tangent map of Eq. (\ref{eqn:map_for_bifu}) at the new fixed points,
\begin{equation}
{\rm Tr}(\mathbf{M}) = 2\left[1 + 2\gamma^2\left((p-1)(4 - 15\gamma^2) - 4 \right)\right],
\end{equation}
where we have kept terms up to order $\mathcal{O}(\gamma^2)$, and computed the trace by writing its square in terms of the determinant. In the worst case, given by $p=3$, the trace is always larger than $2$ provided $\gamma^2<\frac{2}{15}$. However, we observe that at those values of $\alpha$ the period-$4$ orbits do not exists anymore, as their positions do not comply with neither the locality condition nor with $\bm{X}_m^2= 1$. We conclude then, the period-$4$ orbit emerging as a consequence of the 1-to-4 bifurcation of the poles is composed of unstable points.

\subsection{Stability of the period-$4$ orbit on the equator at $\alpha=\pi/2$}
We saw that the period-$4$ orbit along the equator is composed of parabolic points for all the models with odd value of $p$. To investigate the stability of this orbit, we construct the area preserving map describing motion of points in the vicinity of the orbit. This is achieved by considering small increments on the two directions perpendicular to each of the points on the orbit, then concatenating the resulting four area preserving maps. Additionally we consider $\alpha = \pi/2 + \gamma$ with $\gamma \ll 1$, and write $\cos(\alpha)\sim-\gamma$, $\sin(\alpha)\sim 1-\gamma^2$. Going over these steps, and writing the orbit as in Sec. \ref{sec:ndyna_p_spin}, beginning and ending at $(1,0,0)$ we find
\begin{subequations}
\label{eqn:APM_orbit_equator}
\begin{align}
\delta X' &\approx 1, \\
\delta Y' &= (-1)^p\sin(k)\gamma(2-\gamma^2) \nonumber \\
&+ (1 + (-1)^p)\sin(k)\cos(k)(\gamma + (1-\gamma^2)\delta Z)\nonumber \\
&+ \left(\cos^2(k) - (-1)^p\sin^2(k) \right)(\delta Y - k\delta Z^{p-1}) \nonumber \\ 
&+ k\cos(k)\mathcal{W}, \\ 
\delta Z' &= \gamma + \gamma(2-\gamma^2)(1-\gamma^2)\cos(k) \nonumber\\ 
&+ \left(\cos^2(k) + (-1)^{p-1}\sin^2(k)\right)(1-\gamma^2)(\gamma + (1-\gamma^2) \delta Z) \nonumber \\ 
&+ (1-2\gamma^2)\cos^2(k)(\gamma + (1-\gamma^2)\delta Z)\nonumber \\ 
&- (1-(-1)^{p-1})(1-\gamma^2)\cos(k)\sin(k)(\delta Y - k\delta Z^{p-1})\nonumber \\ 
&- (1-2\gamma^2)\cos(k)\sin(k)(\delta Y - k \delta Z^{p-1})\nonumber \\ 
&+ (-1)^{p-1}(1-\gamma^2)k\sin(k)\mathcal{W},
\end{align}
\end{subequations}
where $\mathcal{W} = \left(\delta Y - k \delta Z^{p-1} - \gamma\right)^{p-1}$. When considering odd values of $p$ Eq. (\ref{eqn:APM_orbit_equator}) reduces to 
\begin{subequations}
\label{eqn:APM_orbit_equator_odd}
\begin{align}
\delta Y' &= \delta Y - k\delta Z^{p-1} + k\cos(k)\mathcal{W} -\sin(k)\gamma(2-\gamma^2), \\
\delta Z' &= \gamma + \gamma(2-\gamma^2)(1-\gamma^2)\cos(k) \nonumber \\ 
&+ (1-\gamma^2)(\gamma+(1-\gamma^2)\delta Z) \nonumber \\ 
&+ (1-2\gamma^2)\cos^2(k)(\gamma + (1-\gamma^2)\delta Z) \nonumber \\ 
&-(1-2\gamma^2)\cos(k)\sin(k)(\delta Y - k \delta Z^{p-1})\nonumber \\ 
&- (1-\gamma^2)k\sin(k)\mathcal{W}.
\end{align}
\end{subequations}
With this last expression we can compute the tangent map, keeping up to terms of order $\mathcal{O}(\delta Z^{p-1})$ and $\mathcal{O}(\gamma^2)$, at $(1,0,0)$. Its trace is given by 
\begin{equation}
\label{eqn:stability_orbit_odd}
{\rm Tr}(\mathbf{M}) \approx 2\left[1 + \frac{1}{8}\left(2\gamma^2 - (1-2\gamma^2)(1-\gamma^2)\cos^2(k)\right)^2\right].    
\end{equation}
\begin{figure}[!t]
\centering{\includegraphics[width=0.48\textwidth]{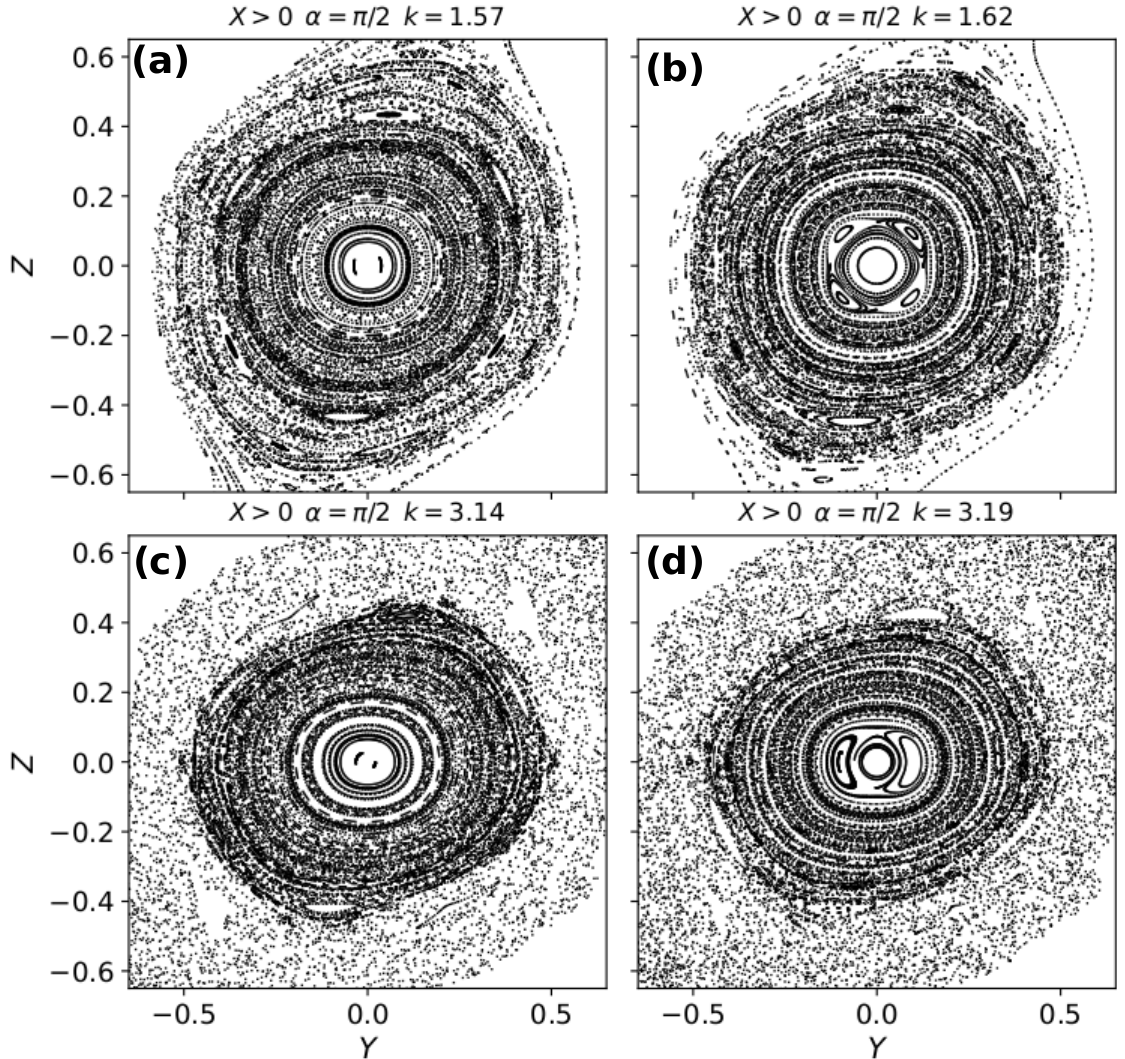}}
\caption{(a,b) Bifurcation of the period-$8$ orbit for $p=4$ and $\alpha = \pi/2$ constructed as two cycles of the period-$4$ orbit on the equator. The parameters are $k=\pi/2$ (a) and $k=\pi/2+0.05$ (b). (c,d) Bifurcation of the period-$4$ orbit. The parameters are $k=\pi$ (c) and $k=\pi+0.05$ (d). }\label{fig:equator_bifu}
\end{figure}
Which is always larger than $2$. Furthermore, the periodic orbit is only well define at $\alpha=\pi/2$, thus $(1,0,0)$ is a fixed point of the map in Eq. (\ref{eqn:APM_orbit_equator}) when $\gamma\to0$. In this limit Eq. (\ref{eqn:stability_orbit_odd}) gives $2\left[1+\frac{\cos^4(k)}{8}\right]$, which is always larger than $2$, confirming the observations made in Sec. \ref{sec:ndyna_p_spin}, trajectories in the vicinity of the periodic orbit do not form closed curves. In fact, the equator is a region where trajectories belonging to opposite hemispheres shear, leading to the instability of the parabolic points forming the period-$4$ orbit.

Finally, for models with even value of $p$ the orbit is composed of elliptic fixed points, except when $k$ is a multiple of $\pi/2$. These values of $k$ signal bifurcations of either the period-$8$ orbit, form by two cycles of the period-$4$ orbit (for instance at $k=\pi/2$), or the period-$4$ orbit (for instance at $k=\pi$). We present two snapshots of these bifurcation processes for the model with $p=4$ in Fig. \ref{fig:equator_bifu}, where we show projections of phase space on the $y$-$z$ plane, with the origin at $(1,0,0)$. Fig. \ref{fig:equator_bifu}a,b show the $4$ to $16$ bifurcation taking place at $k=\pi/2$ and Fig. \ref{fig:equator_bifu}c,d show the $4$ to $8$ bifurcation taking place at $k=\pi$. 

\section{Construction of the similarity/dissimilarity quantifier}
\label{app:simi_quant}
Let $\{\bm{X}_l\}_{l=1,..,n_{\rm tot}}$ and $\{\bm{X}'_l\}_{l=1,..,n_{\rm tot}}$ be two sets with $n_{\rm tot}$ trajectories defining the phase space portraits of the two parameter sets $(\alpha, k)$ and $(\alpha', k') = (\alpha +\delta\alpha, k + \delta k)$. Each phase space portrait obtained from the same set of $n_{\rm tot}$ initial conditions chosen uniformly on the unit sphere. Each trajectory is generated up to the same final time $N$.

Consider a trajectory on each set, say $\bm{X}_{k}$ and $\bm{X}'_k$, belonging to the same initial condition, we quantify their similarity by the product of the Pearson correlation coefficients~\cite{Lee1988} of their three Cartesian components extended in time, 
\begin{equation}
\label{eqn:simi}
 \mathcal{S}(\bm{X}_k, \bm{X}'_k) = {\rm cor}(\tilde{X}_k, \tilde{X}'_k){\rm cor}(\tilde{Y}_k, \tilde{Y}'_k){\rm cor}(\tilde{Z}_k, \tilde{Z}'_k),
\end{equation}
where $\tilde{X}_k = (X_k^{(1)}, X_k^{(2)},...,X_k^{(N)})$. The Pearson correlation coefficient is given by 
\begin{equation}
 \label{eqn:pearson}
 {\rm cor}(A,B) = \frac{{\rm cov(A, B)}}{\sqrt{{\rm var}(A){\rm var}(b)}},
\end{equation}
with ${\rm cov}(A,B)$ the covariance between vectors $A$ and $B$ of same length, and ${\rm var}(A)$ the variance of vector $A$. Notice that Eq. (\ref{eqn:pearson}) gives $1$ for perfect correlation between $A$ and $B$ and $0$ in absence of correlations.

We construct the similarity/dissimilarity quantifier between phase space portraits by taking the average of $\mathcal{S}$ over the $n_{\rm tot}$ initial conditions
\begin{multline}
 \label{eqn:average_simi}
 \overline{\mathcal{S}} = \frac{1}{n_{\rm 
tot}}\sum_{k=1}^{n_{\rm tot}}\mathcal{S} \left( \{\bm{X}_k\}, 
\{\bm{X}'_k\} \right) \\
= \frac{1}{n_{\rm tot}}\sum_{k=1}^{n_{\rm tot}}{\rm cor}(\tilde{X}_k, \tilde{X}'_k){\rm 
cor}(\tilde{Y}_k, \tilde{Y}'_k){\rm cor}(\tilde{Z}_k, \tilde{Z}'_k).
\end{multline}
For this quantity a value of $\overline{\mathcal{S}} = 1$ tells 
that the two phase spaces are identical, and $\overline{\mathcal{S}} 
= 0$ tells the two phase spaces are completely different.

\section{Lyapunov exponent in the limit of strongly chaotic trajectories}
\label{app:big_k_lyap}
In this appendix we provide the derivation of the analytic expression for the Largest Lyapunov exponent in the limit of strongly chaotic trajectories, Eq. (\ref{eqn:big_k_lyap}) in the main text.

For strongly chaotic trajectories~\cite{Chirikov1979,Constantoudis1997} the 
largest Lyapunov exponent is given by 
\begin{equation}
\label{eqn:exponent_strong_chaos}
 \Lambda_{+}(\alpha, k) = \lim_{N\to\infty} \frac{1}{N}\sum_{m=1}^N 
\ln|\mathcal{M}_{+}(\bm{X}_m)|,
\end{equation}
where $\mathcal{M}_{+}(\bm{X}_m)$ is the largest eigenvalue of the tangent map in Eq. 
(\ref{eqn:tangent_map_p_spin}). Using 
the ergodic hypothesis we change the time average in Eq. 
(\ref{eqn:exponent_strong_chaos}) for a phase space average (average over the unit sphere). Then Eq. (\ref{eqn:exponent_strong_chaos}) takes the form
\begin{equation}
 \label{eqn:largest_on_sphere}
 \Lambda_{+}(\alpha, k) = \frac{1}{4\pi}\int_{-1}^1 dZ 
\int_{0}^{2\pi}d\phi \ln|\mathcal{M}_{+}(\bm{X}_m)|,
\end{equation}
where $Z = \cos(\theta)$ and $(\theta, \phi)$ represent the same direction on 
the unit sphere as $\bm{X}$ but in angular variables. In the limit of $k \gg 1$ we can approximate $\mathcal{M}_{+}$ by
\begin{equation}
\label{eqn:largest_k_big}
 \mathcal{M}_{+}^{(p)}(\bm{X}_m) 
\approx (p-1)k\sin(\alpha)Z^{p-2}\sqrt{1-Z^2}\sin(\phi),
\end{equation}
obtained by writing Eq. (\ref{eqn:tangent_map_p_spin}) in angular variables and keeping terms to first order in $k$. Substituting Eq. (\ref{eqn:largest_k_big}) into Eq. (\ref{eqn:largest_on_sphere}) and computing the integral we obtain the expression in Eq. (\ref{eqn:big_k_lyap}) of the main text.

\begin{figure}[!t]
 \centering{\includegraphics[width=0.48\textwidth]{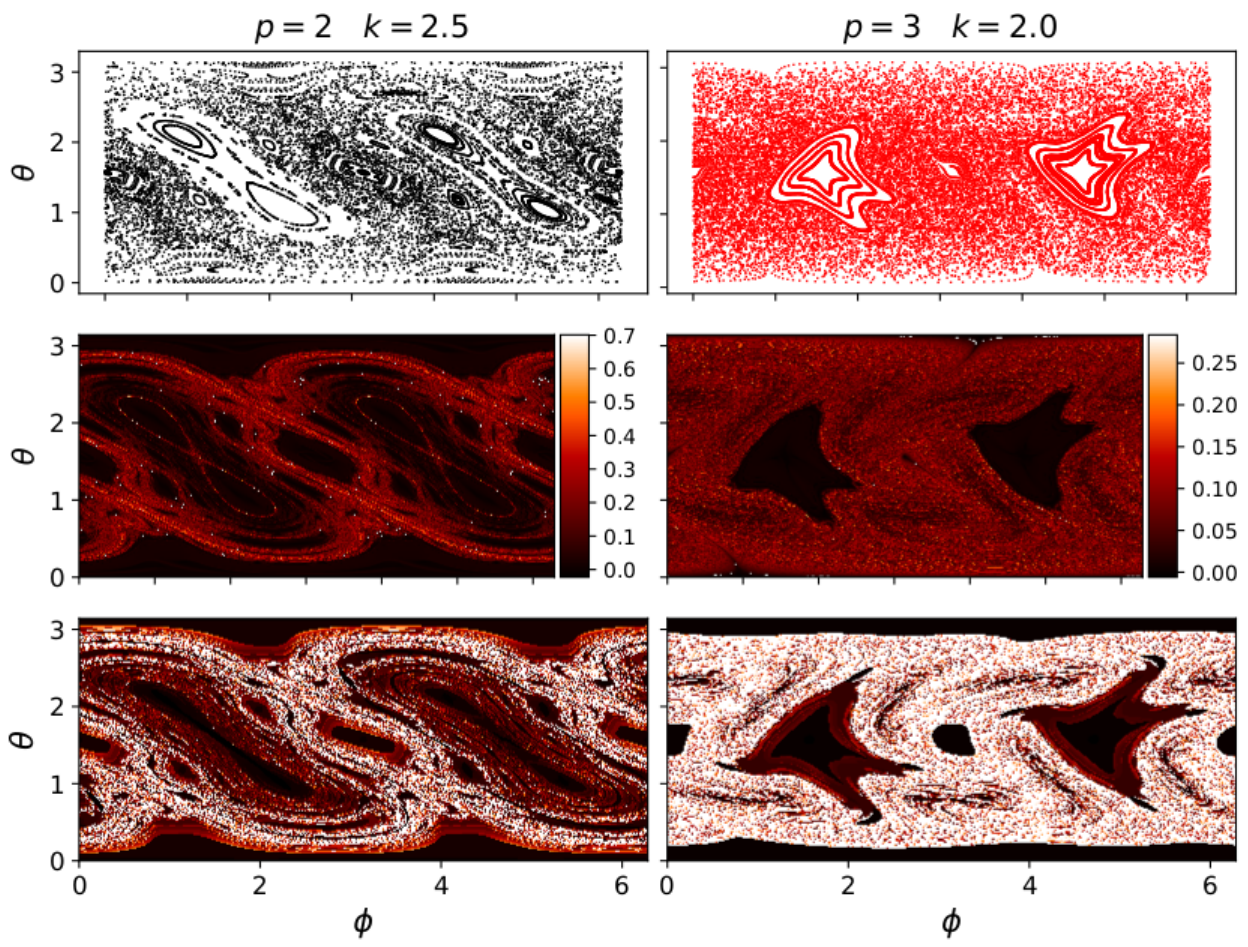}}
\caption{(top) Mixed phase phase spaces of the kicked $p$-spin model. 
From left to right: models with $p=2,3$ and $k=2,2.5$. 
(center) Phase portraits as in (top) colored with the value of the local Lyapunov exponent, we used $10^4$ initial conditions distributed approximately uniformly on the unit sphere. 
(bottom) Values of the recurrence time for each of the $N_{\rm tot}$ initial conditions used in the $A_{\rm ch}$ calculation. Notice how well the algorithm identifies the chaotic regions (white) for the two values of $p$ 
shown. Compare the white regions with the chaotic ones in the 
(center) and (top) panels.}
\label{fig:chaos_areas_test}
\end{figure}
\section{Numerical computation of the surface area of the chaotic region}
\label{app:numerical_chaotic_area}
In this appendix we provide further details on the method used for the estimation of the surface are of the chaotic region.

An estimate of the surface area of the chaotic region, $A_{\rm ch}$, can be constructed using the concept Poincar\'e 
recurrence times~\cite{Anishchenko2013}. Given some initial condition, when the dynamics is regular time evolution will bring the system arbitrarily close to the initial condition after a short time, meaning that the system usually displays some degree of periodicity. On 
the other hand, when the dynamics is chaotic these ``recurrence" times can be exponentially large. Thus, we can construct an estimate of the area of the chaotic region by setting a truncation time $t_{\rm max}$ and a distance $d_{\rm min}$ defining a small local neighborhood around the initial condition, and counting the number $n_{t_{\rm max}}$ of initial conditions which have not returned inside this neighborhood after $t_{\rm max}$ time steps. 

Recalling that the surface area of the unit sphere is $4\pi$, in this approach we can write the are of the chaotic region as 
\begin{equation}
 \label{eqn:area_chaotic_region}
 A_{\rm ch} = 4\pi\frac{n_{t_{\rm max}}}{n_{\rm tot}},
\end{equation}
and the area of the regular region is then given by $A_{\rm reg} = 
4\pi-A_{\rm ch}$. In all our numerical experiments we use a grid of $n_{\rm tot} = 10^4$ initial conditions, evenly spaced on the unit sphere. This grid, on the sphere, can only be constructed to an approximate degree, we use the Fibonacci algorithm (see for instance~\cite{Keinert2015}), which is known to give fairly accurate results. To avoid fluctuations in our counting of initial conditions, we construct $n_{t_{\rm max}}$ as an average over $20$ different values of $t_{\rm max}$. 

In order to check the accuracy of our implementation of the above described method, we compare pictures of phase space (in Mercator projection), with figures of the same phase spaces colored according to the value of the local Lyapunov exponent, and figures of the same phase spaces colored according to the values of the returning time obtained with our implementation. These phase portraits are shown in Fig. \ref{fig:chaos_areas_test}, where the left column corresponds to results for the model with $p=2$, $k=2.5$, $\alpha=\pi/2$ and the right column to the model with $p=3$, $k=2.0$, $\alpha = \pi/2$. We observe an excellent agreement between the chaotic region identified via the Metropolis-like sampling (withe region in bottom panels of fig. \ref{fig:chaos_areas_test}), and the region displaying a nonzero value of the local Lyapunov (red region on center panels of Fog. \ref{fig:chaos_areas_test}). Therefore we verify the that our implementation of the method accurately identifies the chaotic region in phase space.

\bibliography{kicked_p_spin}

\begin{thebibliography}{122}%
\makeatletter
\providecommand \@ifxundefined [1]{%
 \@ifx{#1\undefined}
}%
\providecommand \@ifnum [1]{%
 \ifnum #1\expandafter \@firstoftwo
 \else \expandafter \@secondoftwo
 \fi
}%
\providecommand \@ifx [1]{%
 \ifx #1\expandafter \@firstoftwo
 \else \expandafter \@secondoftwo
 \fi
}%
\providecommand \natexlab [1]{#1}%
\providecommand \enquote  [1]{``#1''}%
\providecommand \bibnamefont  [1]{#1}%
\providecommand \bibfnamefont [1]{#1}%
\providecommand \citenamefont [1]{#1}%
\providecommand \href@noop [0]{\@secondoftwo}%
\providecommand \href [0]{\begingroup \@sanitize@url \@href}%
\providecommand \@href[1]{\@@startlink{#1}\@@href}%
\providecommand \@@href[1]{\endgroup#1\@@endlink}%
\providecommand \@sanitize@url [0]{\catcode `\\12\catcode `\$12\catcode
  `\&12\catcode `\#12\catcode `\^12\catcode `\_12\catcode `\%12\relax}%
\providecommand \@@startlink[1]{}%
\providecommand \@@endlink[0]{}%
\providecommand \url  [0]{\begingroup\@sanitize@url \@url }%
\providecommand \@url [1]{\endgroup\@href {#1}{\urlprefix }}%
\providecommand \urlprefix  [0]{URL }%
\providecommand \Eprint [0]{\href }%
\providecommand \doibase [0]{http://dx.doi.org/}%
\providecommand \selectlanguage [0]{\@gobble}%
\providecommand \bibinfo  [0]{\@secondoftwo}%
\providecommand \bibfield  [0]{\@secondoftwo}%
\providecommand \translation [1]{[#1]}%
\providecommand \BibitemOpen [0]{}%
\providecommand \bibitemStop [0]{}%
\providecommand \bibitemNoStop [0]{.\EOS\space}%
\providecommand \EOS [0]{\spacefactor3000\relax}%
\providecommand \BibitemShut  [1]{\csname bibitem#1\endcsname}%
\let\auto@bib@innerbib\@empty
\bibitem [{\citenamefont {Haake}\ \emph {et~al.}(1987)\citenamefont {Haake},
  \citenamefont {Ku{\'{s}}},\ and\ \citenamefont {Scharf}}]{Haake1987}%
  \BibitemOpen
  \bibfield  {author} {\bibinfo {author} {\bibfnamefont {F.}~\bibnamefont
  {Haake}}, \bibinfo {author} {\bibfnamefont {M.}~\bibnamefont {Ku{\'{s}}}}, \
  and\ \bibinfo {author} {\bibfnamefont {R.}~\bibnamefont {Scharf}},\
  }\bibfield  {title} {\enquote {\bibinfo {title} {{Classical and quantum chaos
  for a kicked top}},}\ }\href {\doibase 10.1007/BF01303727} {\bibfield
  {journal} {\bibinfo  {journal} {Zeitschrift f{\"{u}}r Physik B Condensed
  Matter}\ }\textbf {\bibinfo {volume} {65}},\ \bibinfo {pages} {381--395}
  (\bibinfo {year} {1987})}\BibitemShut {NoStop}%
\bibitem [{\citenamefont {Nakahara}(2013)}]{Nakahara2013}%
  \BibitemOpen
  \bibfield  {author} {\bibinfo {author} {\bibfnamefont {Mikio}\ \bibnamefont
  {Nakahara}},\ }\href@noop {} {\emph {\bibinfo {title} {Lectures on quantum
  computing, thermodynamics and statistical physics}}},\ Vol.~\bibinfo {volume}
  {8}\ (\bibinfo  {publisher} {World Scientific},\ \bibinfo {year}
  {2013})\BibitemShut {NoStop}%
\bibitem [{\citenamefont {Lucas}(2014)}]{Lucas2014}%
  \BibitemOpen
  \bibfield  {author} {\bibinfo {author} {\bibfnamefont {Andrew}\ \bibnamefont
  {Lucas}},\ }\bibfield  {title} {\enquote {\bibinfo {title} {Ising
  formulations of many np problems},}\ }\href@noop {} {\bibfield  {journal}
  {\bibinfo  {journal} {Frontiers in Physics}\ }\textbf {\bibinfo {volume}
  {2}},\ \bibinfo {pages} {5} (\bibinfo {year} {2014})}\BibitemShut {NoStop}%
\bibitem [{\citenamefont {Albash}\ and\ \citenamefont
  {Lidar}(2018)}]{Albash2018}%
  \BibitemOpen
  \bibfield  {author} {\bibinfo {author} {\bibfnamefont {Tameem}\ \bibnamefont
  {Albash}}\ and\ \bibinfo {author} {\bibfnamefont {Daniel~A.}\ \bibnamefont
  {Lidar}},\ }\bibfield  {title} {\enquote {\bibinfo {title} {Adiabatic quantum
  computation},}\ }\href {\doibase 10.1103/RevModPhys.90.015002} {\bibfield
  {journal} {\bibinfo  {journal} {Rev. Mod. Phys.}\ }\textbf {\bibinfo {volume}
  {90}},\ \bibinfo {pages} {015002} (\bibinfo {year} {2018})}\BibitemShut
  {NoStop}%
\bibitem [{\citenamefont {Biamonte}\ \emph {et~al.}(2017)\citenamefont
  {Biamonte}, \citenamefont {Wittek}, \citenamefont {Pancotti}, \citenamefont
  {Rebentrost}, \citenamefont {Wiebe},\ and\ \citenamefont
  {Lloyd}}]{Biamonte2017}%
  \BibitemOpen
  \bibfield  {author} {\bibinfo {author} {\bibfnamefont {Jacob}\ \bibnamefont
  {Biamonte}}, \bibinfo {author} {\bibfnamefont {Peter}\ \bibnamefont
  {Wittek}}, \bibinfo {author} {\bibfnamefont {Nicola}\ \bibnamefont
  {Pancotti}}, \bibinfo {author} {\bibfnamefont {Patrick}\ \bibnamefont
  {Rebentrost}}, \bibinfo {author} {\bibfnamefont {Nathan}\ \bibnamefont
  {Wiebe}}, \ and\ \bibinfo {author} {\bibfnamefont {Seth}\ \bibnamefont
  {Lloyd}},\ }\bibfield  {title} {\enquote {\bibinfo {title} {Quantum machine
  learning},}\ }\href@noop {} {\bibfield  {journal} {\bibinfo  {journal}
  {Nature}\ }\textbf {\bibinfo {volume} {549}},\ \bibinfo {pages} {195}
  (\bibinfo {year} {2017})}\BibitemShut {NoStop}%
\bibitem [{\citenamefont {Kirkpatrick}\ and\ \citenamefont
  {Thirumalai}(1987)}]{Kirkpatrick1987}%
  \BibitemOpen
  \bibfield  {author} {\bibinfo {author} {\bibfnamefont {T.~R.}\ \bibnamefont
  {Kirkpatrick}}\ and\ \bibinfo {author} {\bibfnamefont {D.}~\bibnamefont
  {Thirumalai}},\ }\bibfield  {title} {\enquote {\bibinfo {title}
  {p-spin-interaction spin-glass models: Connections with the structural glass
  problem},}\ }\href {\doibase 10.1103/PhysRevB.36.5388} {\bibfield  {journal}
  {\bibinfo  {journal} {Phys. Rev. B}\ }\textbf {\bibinfo {volume} {36}},\
  \bibinfo {pages} {5388--5397} (\bibinfo {year} {1987})}\BibitemShut {NoStop}%
\bibitem [{\citenamefont {Sachdev}(2011)}]{Sachdev2011}%
  \BibitemOpen
  \bibfield  {author} {\bibinfo {author} {\bibfnamefont {Subir}\ \bibnamefont
  {Sachdev}},\ }\href {\doibase 10.1017/CBO9780511973765} {\emph {\bibinfo
  {title} {Quantum Phase Transitions}}},\ \bibinfo {edition} {2nd}\ ed.\
  (\bibinfo  {publisher} {Cambridge University Press},\ \bibinfo {year}
  {2011})\BibitemShut {NoStop}%
\bibitem [{\citenamefont {Peng}\ \emph {et~al.}(2005)\citenamefont {Peng},
  \citenamefont {Du},\ and\ \citenamefont {Suter}}]{Peng2005}%
  \BibitemOpen
  \bibfield  {author} {\bibinfo {author} {\bibfnamefont {Xinhua}\ \bibnamefont
  {Peng}}, \bibinfo {author} {\bibfnamefont {Jiangfeng}\ \bibnamefont {Du}}, \
  and\ \bibinfo {author} {\bibfnamefont {Dieter}\ \bibnamefont {Suter}},\
  }\bibfield  {title} {\enquote {\bibinfo {title} {Quantum phase transition of
  ground-state entanglement in a heisenberg spin chain simulated in an nmr
  quantum computer},}\ }\href@noop {} {\bibfield  {journal} {\bibinfo
  {journal} {Physical review A}\ }\textbf {\bibinfo {volume} {71}},\ \bibinfo
  {pages} {012307} (\bibinfo {year} {2005})}\BibitemShut {NoStop}%
\bibitem [{\citenamefont {Filippone}\ \emph {et~al.}(2011)\citenamefont
  {Filippone}, \citenamefont {Dusuel},\ and\ \citenamefont
  {Vidal}}]{Filippone2011}%
  \BibitemOpen
  \bibfield  {author} {\bibinfo {author} {\bibfnamefont {Michele}\ \bibnamefont
  {Filippone}}, \bibinfo {author} {\bibfnamefont {S\'ebastien}\ \bibnamefont
  {Dusuel}}, \ and\ \bibinfo {author} {\bibfnamefont {Julien}\ \bibnamefont
  {Vidal}},\ }\bibfield  {title} {\enquote {\bibinfo {title} {Quantum phase
  transitions in fully connected spin models: An entanglement perspective},}\
  }\href {\doibase 10.1103/PhysRevA.83.022327} {\bibfield  {journal} {\bibinfo
  {journal} {Phys. Rev. A}\ }\textbf {\bibinfo {volume} {83}},\ \bibinfo
  {pages} {022327} (\bibinfo {year} {2011})}\BibitemShut {NoStop}%
\bibitem [{\citenamefont {Zhang}\ \emph {et~al.}(2017)\citenamefont {Zhang},
  \citenamefont {Pagano}, \citenamefont {Hess}, \citenamefont {Kyprianidis},
  \citenamefont {Becker}, \citenamefont {Kaplan}, \citenamefont {Gorshkov},
  \citenamefont {Gong},\ and\ \citenamefont {Monroe}}]{Zhang2017}%
  \BibitemOpen
  \bibfield  {author} {\bibinfo {author} {\bibfnamefont {Jiehang}\ \bibnamefont
  {Zhang}}, \bibinfo {author} {\bibfnamefont {Guido}\ \bibnamefont {Pagano}},
  \bibinfo {author} {\bibfnamefont {Paul~W}\ \bibnamefont {Hess}}, \bibinfo
  {author} {\bibfnamefont {Antonis}\ \bibnamefont {Kyprianidis}}, \bibinfo
  {author} {\bibfnamefont {Patrick}\ \bibnamefont {Becker}}, \bibinfo {author}
  {\bibfnamefont {Harvey}\ \bibnamefont {Kaplan}}, \bibinfo {author}
  {\bibfnamefont {Alexey~V}\ \bibnamefont {Gorshkov}}, \bibinfo {author}
  {\bibfnamefont {Z-X}\ \bibnamefont {Gong}}, \ and\ \bibinfo {author}
  {\bibfnamefont {Christopher}\ \bibnamefont {Monroe}},\ }\bibfield  {title}
  {\enquote {\bibinfo {title} {Observation of a many-body dynamical phase
  transition with a 53-qubit quantum simulator},}\ }\href@noop {} {\bibfield
  {journal} {\bibinfo  {journal} {Nature}\ }\textbf {\bibinfo {volume} {551}},\
  \bibinfo {pages} {601--604} (\bibinfo {year} {2017})}\BibitemShut {NoStop}%
\bibitem [{\citenamefont {Jurcevic}\ \emph {et~al.}(2017)\citenamefont
  {Jurcevic}, \citenamefont {Shen}, \citenamefont {Hauke}, \citenamefont
  {Maier}, \citenamefont {Brydges}, \citenamefont {Hempel}, \citenamefont
  {Lanyon}, \citenamefont {Heyl}, \citenamefont {Blatt},\ and\ \citenamefont
  {Roos}}]{Jurcevic2017}%
  \BibitemOpen
  \bibfield  {author} {\bibinfo {author} {\bibfnamefont {P}~\bibnamefont
  {Jurcevic}}, \bibinfo {author} {\bibfnamefont {H}~\bibnamefont {Shen}},
  \bibinfo {author} {\bibfnamefont {P}~\bibnamefont {Hauke}}, \bibinfo {author}
  {\bibfnamefont {C}~\bibnamefont {Maier}}, \bibinfo {author} {\bibfnamefont
  {T}~\bibnamefont {Brydges}}, \bibinfo {author} {\bibfnamefont
  {C}~\bibnamefont {Hempel}}, \bibinfo {author} {\bibfnamefont
  {BP}~\bibnamefont {Lanyon}}, \bibinfo {author} {\bibfnamefont {Markus}\
  \bibnamefont {Heyl}}, \bibinfo {author} {\bibfnamefont {R}~\bibnamefont
  {Blatt}}, \ and\ \bibinfo {author} {\bibfnamefont {CF}~\bibnamefont {Roos}},\
  }\bibfield  {title} {\enquote {\bibinfo {title} {Direct observation of
  dynamical quantum phase transitions in an interacting many-body system},}\
  }\href@noop {} {\bibfield  {journal} {\bibinfo  {journal} {Physical review
  letters}\ }\textbf {\bibinfo {volume} {119}},\ \bibinfo {pages} {080501}
  (\bibinfo {year} {2017})}\BibitemShut {NoStop}%
\bibitem [{\citenamefont {\ifmmode \check{Z}\else
  \v{Z}\fi{}unkovi\ifmmode~\check{c}\else \v{c}\fi{}}\ \emph
  {et~al.}(2018)\citenamefont {\ifmmode \check{Z}\else
  \v{Z}\fi{}unkovi\ifmmode~\check{c}\else \v{c}\fi{}}, \citenamefont {Heyl},
  \citenamefont {Knap},\ and\ \citenamefont {Silva}}]{Zunkovic2018}%
  \BibitemOpen
  \bibfield  {author} {\bibinfo {author} {\bibfnamefont {Bojan}\ \bibnamefont
  {\ifmmode \check{Z}\else \v{Z}\fi{}unkovi\ifmmode~\check{c}\else
  \v{c}\fi{}}}, \bibinfo {author} {\bibfnamefont {Markus}\ \bibnamefont
  {Heyl}}, \bibinfo {author} {\bibfnamefont {Michael}\ \bibnamefont {Knap}}, \
  and\ \bibinfo {author} {\bibfnamefont {Alessandro}\ \bibnamefont {Silva}},\
  }\bibfield  {title} {\enquote {\bibinfo {title} {Dynamical quantum phase
  transitions in spin chains with long-range interactions: Merging different
  concepts of nonequilibrium criticality},}\ }\href {\doibase
  10.1103/PhysRevLett.120.130601} {\bibfield  {journal} {\bibinfo  {journal}
  {Phys. Rev. Lett.}\ }\textbf {\bibinfo {volume} {120}},\ \bibinfo {pages}
  {130601} (\bibinfo {year} {2018})}\BibitemShut {NoStop}%
\bibitem [{\citenamefont {Heyl}\ \emph {et~al.}(2013)\citenamefont {Heyl},
  \citenamefont {Polkovnikov},\ and\ \citenamefont {Kehrein}}]{Heyl2013}%
  \BibitemOpen
  \bibfield  {author} {\bibinfo {author} {\bibfnamefont {M.}~\bibnamefont
  {Heyl}}, \bibinfo {author} {\bibfnamefont {A.}~\bibnamefont {Polkovnikov}}, \
  and\ \bibinfo {author} {\bibfnamefont {S.}~\bibnamefont {Kehrein}},\
  }\bibfield  {title} {\enquote {\bibinfo {title} {Dynamical quantum phase
  transitions in the transverse-field ising model},}\ }\href {\doibase
  10.1103/PhysRevLett.110.135704} {\bibfield  {journal} {\bibinfo  {journal}
  {Phys. Rev. Lett.}\ }\textbf {\bibinfo {volume} {110}},\ \bibinfo {pages}
  {135704} (\bibinfo {year} {2013})}\BibitemShut {NoStop}%
\bibitem [{\citenamefont {Gubin}\ and\ \citenamefont
  {F.~Santos}(2012)}]{Gubin2012}%
  \BibitemOpen
  \bibfield  {author} {\bibinfo {author} {\bibfnamefont {Aviva}\ \bibnamefont
  {Gubin}}\ and\ \bibinfo {author} {\bibfnamefont {Lea}\ \bibnamefont
  {F.~Santos}},\ }\bibfield  {title} {\enquote {\bibinfo {title} {Quantum
  chaos: An introduction via chains of interacting spins 1/2},}\ }\href
  {\doibase 10.1119/1.3671068} {\bibfield  {journal} {\bibinfo  {journal}
  {American Journal of Physics}\ }\textbf {\bibinfo {volume} {80}},\ \bibinfo
  {pages} {246--251} (\bibinfo {year} {2012})},\ \Eprint
  {http://arxiv.org/abs/https://doi.org/10.1119/1.3671068}
  {https://doi.org/10.1119/1.3671068} \BibitemShut {NoStop}%
\bibitem [{\citenamefont {Kos}\ \emph {et~al.}(2018)\citenamefont {Kos},
  \citenamefont {Ljubotina},\ and\ \citenamefont {Prosen}}]{Kos2018}%
  \BibitemOpen
  \bibfield  {author} {\bibinfo {author} {\bibfnamefont {Pavel}\ \bibnamefont
  {Kos}}, \bibinfo {author} {\bibfnamefont {Marko}\ \bibnamefont {Ljubotina}},
  \ and\ \bibinfo {author} {\bibfnamefont {Toma\ifmmode
  \check{z}\else~\v{z}\fi{}}\ \bibnamefont {Prosen}},\ }\bibfield  {title}
  {\enquote {\bibinfo {title} {Many-body quantum chaos: Analytic connection to
  random matrix theory},}\ }\href {\doibase 10.1103/PhysRevX.8.021062}
  {\bibfield  {journal} {\bibinfo  {journal} {Phys. Rev. X}\ }\textbf {\bibinfo
  {volume} {8}},\ \bibinfo {pages} {021062} (\bibinfo {year}
  {2018})}\BibitemShut {NoStop}%
\bibitem [{\citenamefont {Gogolin}\ and\ \citenamefont
  {Eisert}(2016)}]{Gogolin2016}%
  \BibitemOpen
  \bibfield  {author} {\bibinfo {author} {\bibfnamefont {Christian}\
  \bibnamefont {Gogolin}}\ and\ \bibinfo {author} {\bibfnamefont {Jens}\
  \bibnamefont {Eisert}},\ }\bibfield  {title} {\enquote {\bibinfo {title}
  {{Equilibration, thermalisation, and the emergence of statistical mechanics
  in closed quantum systems}},}\ }\href {\doibase
  10.1088/0034-4885/79/5/056001} {\bibfield  {journal} {\bibinfo  {journal}
  {Reports on Progress in Physics}\ }\textbf {\bibinfo {volume} {79}},\
  \bibinfo {pages} {056001} (\bibinfo {year} {2016})}\BibitemShut {NoStop}%
\bibitem [{\citenamefont {D'Alessio}\ \emph {et~al.}(2016)\citenamefont
  {D'Alessio}, \citenamefont {Kafri}, \citenamefont {Polkovnikov},\ and\
  \citenamefont {Rigol}}]{Rigol2016}%
  \BibitemOpen
  \bibfield  {author} {\bibinfo {author} {\bibfnamefont {Luca}\ \bibnamefont
  {D'Alessio}}, \bibinfo {author} {\bibfnamefont {Yariv}\ \bibnamefont
  {Kafri}}, \bibinfo {author} {\bibfnamefont {Anatoli}\ \bibnamefont
  {Polkovnikov}}, \ and\ \bibinfo {author} {\bibfnamefont {Marcos}\
  \bibnamefont {Rigol}},\ }\bibfield  {title} {\enquote {\bibinfo {title}
  {{From quantum chaos and eigenstate thermalization to statistical mechanics
  and thermodynamics}},}\ }\href {\doibase 10.1080/00018732.2016.1198134}
  {\bibfield  {journal} {\bibinfo  {journal} {Advances in Physics}\ }\textbf
  {\bibinfo {volume} {65}},\ \bibinfo {pages} {239--362} (\bibinfo {year}
  {2016})}\BibitemShut {NoStop}%
\bibitem [{\citenamefont {Simon}\ \emph {et~al.}(2011)\citenamefont {Simon},
  \citenamefont {Bakr}, \citenamefont {Ma1}, \citenamefont {Tai}, \citenamefont
  {Preiss},\ and\ \citenamefont {Greiner}}]{Simon2011}%
  \BibitemOpen
  \bibfield  {author} {\bibinfo {author} {\bibfnamefont {Jonathan}\
  \bibnamefont {Simon}}, \bibinfo {author} {\bibfnamefont {Waseem~S.}\
  \bibnamefont {Bakr}}, \bibinfo {author} {\bibfnamefont {Ruichao}\
  \bibnamefont {Ma1}}, \bibinfo {author} {\bibfnamefont {M.~Eric}\ \bibnamefont
  {Tai}}, \bibinfo {author} {\bibfnamefont {Philipp~M.}\ \bibnamefont
  {Preiss}}, \ and\ \bibinfo {author} {\bibfnamefont {Markus}\ \bibnamefont
  {Greiner}},\ }\bibfield  {title} {\enquote {\bibinfo {title} {Quantum
  simulation of antiferromagnetic spin chains in an optical lattice},}\ }\href
  {https://doi.org/10.1038/nature09994} {\bibfield  {journal} {\bibinfo
  {journal} {Nature}\ }\textbf {\bibinfo {volume} {472}},\ \bibinfo {pages}
  {307–312} (\bibinfo {year} {2011})}\BibitemShut {NoStop}%
\bibitem [{\citenamefont {Zeiher}\ \emph {et~al.}(2017)\citenamefont {Zeiher},
  \citenamefont {Choi}, \citenamefont {Rubio-Abadal}, \citenamefont {Pohl},
  \citenamefont {van Bijnen}, \citenamefont {Bloch},\ and\ \citenamefont
  {Gross}}]{Zeiher2017}%
  \BibitemOpen
  \bibfield  {author} {\bibinfo {author} {\bibfnamefont {Johannes}\
  \bibnamefont {Zeiher}}, \bibinfo {author} {\bibfnamefont {Jae-yoon}\
  \bibnamefont {Choi}}, \bibinfo {author} {\bibfnamefont {Antonio}\
  \bibnamefont {Rubio-Abadal}}, \bibinfo {author} {\bibfnamefont {Thomas}\
  \bibnamefont {Pohl}}, \bibinfo {author} {\bibfnamefont {Rick}\ \bibnamefont
  {van Bijnen}}, \bibinfo {author} {\bibfnamefont {Immanuel}\ \bibnamefont
  {Bloch}}, \ and\ \bibinfo {author} {\bibfnamefont {Christian}\ \bibnamefont
  {Gross}},\ }\bibfield  {title} {\enquote {\bibinfo {title} {Coherent
  many-body spin dynamics in a long-range interacting ising chain},}\ }\href
  {\doibase 10.1103/PhysRevX.7.041063} {\bibfield  {journal} {\bibinfo
  {journal} {Phys. Rev. X}\ }\textbf {\bibinfo {volume} {7}},\ \bibinfo {pages}
  {041063} (\bibinfo {year} {2017})}\BibitemShut {NoStop}%
\bibitem [{\citenamefont {Blatt}\ and\ \citenamefont {Roos}(2012)}]{Blatt2012}%
  \BibitemOpen
  \bibfield  {author} {\bibinfo {author} {\bibfnamefont {Rainer}\ \bibnamefont
  {Blatt}}\ and\ \bibinfo {author} {\bibfnamefont {Christian~F}\ \bibnamefont
  {Roos}},\ }\bibfield  {title} {\enquote {\bibinfo {title} {Quantum
  simulations with trapped ions},}\ }\href@noop {} {\bibfield  {journal}
  {\bibinfo  {journal} {Nature Physics}\ }\textbf {\bibinfo {volume} {8}},\
  \bibinfo {pages} {277--284} (\bibinfo {year} {2012})}\BibitemShut {NoStop}%
\bibitem [{\citenamefont {Monroe}\ \emph {et~al.}(2019)\citenamefont {Monroe},
  \citenamefont {Campbell}, \citenamefont {Duan}, \citenamefont {Gong},
  \citenamefont {Gorshkov}, \citenamefont {Hess}, \citenamefont {Islam},
  \citenamefont {Kim}, \citenamefont {Pagano}, \citenamefont {Richerme} \emph
  {et~al.}}]{Monroe2019}%
  \BibitemOpen
  \bibfield  {author} {\bibinfo {author} {\bibfnamefont {C}~\bibnamefont
  {Monroe}}, \bibinfo {author} {\bibfnamefont {WC}~\bibnamefont {Campbell}},
  \bibinfo {author} {\bibfnamefont {L-M}\ \bibnamefont {Duan}}, \bibinfo
  {author} {\bibfnamefont {Z-X}\ \bibnamefont {Gong}}, \bibinfo {author}
  {\bibfnamefont {AV}~\bibnamefont {Gorshkov}}, \bibinfo {author}
  {\bibfnamefont {P}~\bibnamefont {Hess}}, \bibinfo {author} {\bibfnamefont
  {R}~\bibnamefont {Islam}}, \bibinfo {author} {\bibfnamefont {K}~\bibnamefont
  {Kim}}, \bibinfo {author} {\bibfnamefont {G}~\bibnamefont {Pagano}}, \bibinfo
  {author} {\bibfnamefont {P}~\bibnamefont {Richerme}},  \emph {et~al.},\
  }\bibfield  {title} {\enquote {\bibinfo {title} {Programmable quantum
  simulations of spin systems with trapped ions},}\ }\href@noop {} {\bibfield
  {journal} {\bibinfo  {journal} {arXiv preprint arXiv:1912.07845}\ } (\bibinfo
  {year} {2019})}\BibitemShut {NoStop}%
\bibitem [{\citenamefont {Scholl}\ \emph {et~al.}(2020)\citenamefont {Scholl},
  \citenamefont {Schuler}, \citenamefont {Williams}, \citenamefont
  {Eberharter}, \citenamefont {Barredo}, \citenamefont {Schymik}, \citenamefont
  {Lienhard}, \citenamefont {Henry}, \citenamefont {Lang}, \citenamefont
  {Lahaye}, \citenamefont {Läuchli},\ and\ \citenamefont
  {Browaeys}}]{Scholl2020}%
  \BibitemOpen
  \bibfield  {author} {\bibinfo {author} {\bibfnamefont {Pascal}\ \bibnamefont
  {Scholl}}, \bibinfo {author} {\bibfnamefont {Michael}\ \bibnamefont
  {Schuler}}, \bibinfo {author} {\bibfnamefont {Hannah~J.}\ \bibnamefont
  {Williams}}, \bibinfo {author} {\bibfnamefont {Alexander~A.}\ \bibnamefont
  {Eberharter}}, \bibinfo {author} {\bibfnamefont {Daniel}\ \bibnamefont
  {Barredo}}, \bibinfo {author} {\bibfnamefont {Kai-Niklas}\ \bibnamefont
  {Schymik}}, \bibinfo {author} {\bibfnamefont {Vincent}\ \bibnamefont
  {Lienhard}}, \bibinfo {author} {\bibfnamefont {Louis-Paul}\ \bibnamefont
  {Henry}}, \bibinfo {author} {\bibfnamefont {Thomas~C.}\ \bibnamefont {Lang}},
  \bibinfo {author} {\bibfnamefont {Thierry}\ \bibnamefont {Lahaye}}, \bibinfo
  {author} {\bibfnamefont {Andreas~M.}\ \bibnamefont {Läuchli}}, \ and\
  \bibinfo {author} {\bibfnamefont {Antoine}\ \bibnamefont {Browaeys}},\
  }\href@noop {} {\enquote {\bibinfo {title} {Programmable quantum simulation
  of 2d antiferromagnets with hundreds of rydberg atoms},}\ } (\bibinfo {year}
  {2020}),\ \Eprint {http://arxiv.org/abs/2012.12268} {arXiv:2012.12268
  [quant-ph]} \BibitemShut {NoStop}%
\bibitem [{\citenamefont {Ebadi}\ \emph {et~al.}(2020)\citenamefont {Ebadi},
  \citenamefont {Wang}, \citenamefont {Levine}, \citenamefont {Keesling},
  \citenamefont {Semeghini}, \citenamefont {Omran}, \citenamefont {Bluvstein},
  \citenamefont {Samajdar}, \citenamefont {Pichler}, \citenamefont {Ho},
  \citenamefont {Choi}, \citenamefont {Sachdev}, \citenamefont {Greiner},
  \citenamefont {Vuletic},\ and\ \citenamefont {Lukin}}]{Ebadi2020}%
  \BibitemOpen
  \bibfield  {author} {\bibinfo {author} {\bibfnamefont {Sepehr}\ \bibnamefont
  {Ebadi}}, \bibinfo {author} {\bibfnamefont {Tout~T.}\ \bibnamefont {Wang}},
  \bibinfo {author} {\bibfnamefont {Harry}\ \bibnamefont {Levine}}, \bibinfo
  {author} {\bibfnamefont {Alexander}\ \bibnamefont {Keesling}}, \bibinfo
  {author} {\bibfnamefont {Giulia}\ \bibnamefont {Semeghini}}, \bibinfo
  {author} {\bibfnamefont {Ahmed}\ \bibnamefont {Omran}}, \bibinfo {author}
  {\bibfnamefont {Dolev}\ \bibnamefont {Bluvstein}}, \bibinfo {author}
  {\bibfnamefont {Rhine}\ \bibnamefont {Samajdar}}, \bibinfo {author}
  {\bibfnamefont {Hannes}\ \bibnamefont {Pichler}}, \bibinfo {author}
  {\bibfnamefont {Wen~Wei}\ \bibnamefont {Ho}}, \bibinfo {author}
  {\bibfnamefont {Soonwon}\ \bibnamefont {Choi}}, \bibinfo {author}
  {\bibfnamefont {Subir}\ \bibnamefont {Sachdev}}, \bibinfo {author}
  {\bibfnamefont {Markus}\ \bibnamefont {Greiner}}, \bibinfo {author}
  {\bibfnamefont {Vladan}\ \bibnamefont {Vuletic}}, \ and\ \bibinfo {author}
  {\bibfnamefont {Mikhail~D.}\ \bibnamefont {Lukin}},\ }\href@noop {} {\enquote
  {\bibinfo {title} {Quantum phases of matter on a 256-atom programmable
  quantum simulator},}\ } (\bibinfo {year} {2020}),\ \Eprint
  {http://arxiv.org/abs/2012.12281} {arXiv:2012.12281 [quant-ph]} \BibitemShut
  {NoStop}%
\bibitem [{\citenamefont {Lloyd}(1996)}]{Lloyd1996}%
  \BibitemOpen
  \bibfield  {author} {\bibinfo {author} {\bibfnamefont {Seth}\ \bibnamefont
  {Lloyd}},\ }\bibfield  {title} {\enquote {\bibinfo {title} {Universal quantum
  simulators},}\ }\href@noop {} {\bibfield  {journal} {\bibinfo  {journal}
  {Science}\ }\textbf {\bibinfo {volume} {273}},\ \bibinfo {pages} {1073--1078}
  (\bibinfo {year} {1996})}\BibitemShut {NoStop}%
\bibitem [{\citenamefont {Heyl}\ \emph {et~al.}(2019)\citenamefont {Heyl},
  \citenamefont {Hauke},\ and\ \citenamefont {Zoller}}]{Heyl2019}%
  \BibitemOpen
  \bibfield  {author} {\bibinfo {author} {\bibfnamefont {Markus}\ \bibnamefont
  {Heyl}}, \bibinfo {author} {\bibfnamefont {Philipp}\ \bibnamefont {Hauke}}, \
  and\ \bibinfo {author} {\bibfnamefont {Peter}\ \bibnamefont {Zoller}},\
  }\bibfield  {title} {\enquote {\bibinfo {title} {Quantum localization bounds
  trotter errors in digital quantum simulation},}\ }\href@noop {} {\bibfield
  {journal} {\bibinfo  {journal} {Science advances}\ }\textbf {\bibinfo
  {volume} {5}},\ \bibinfo {pages} {eaau8342} (\bibinfo {year}
  {2019})}\BibitemShut {NoStop}%
\bibitem [{\citenamefont {Sieberer}\ \emph {et~al.}(2019)\citenamefont
  {Sieberer}, \citenamefont {Olsacher}, \citenamefont {Elben}, \citenamefont
  {Heyl}, \citenamefont {Hauke}, \citenamefont {Haake},\ and\ \citenamefont
  {Zoller}}]{Sieberer2019}%
  \BibitemOpen
  \bibfield  {author} {\bibinfo {author} {\bibfnamefont {Lukas~M.}\
  \bibnamefont {Sieberer}}, \bibinfo {author} {\bibfnamefont {Tobias}\
  \bibnamefont {Olsacher}}, \bibinfo {author} {\bibfnamefont {Andreas}\
  \bibnamefont {Elben}}, \bibinfo {author} {\bibfnamefont {Markus}\
  \bibnamefont {Heyl}}, \bibinfo {author} {\bibfnamefont {Philipp}\
  \bibnamefont {Hauke}}, \bibinfo {author} {\bibfnamefont {Fritz}\ \bibnamefont
  {Haake}}, \ and\ \bibinfo {author} {\bibfnamefont {Peter}\ \bibnamefont
  {Zoller}},\ }\bibfield  {title} {\enquote {\bibinfo {title} {{Digital quantum
  simulation, Trotter errors, and quantum chaos of the kicked top}},}\ }\href
  {\doibase 10.1038/s41534-019-0192-5} {\bibfield  {journal} {\bibinfo
  {journal} {npj Quantum Information}\ }\textbf {\bibinfo {volume} {5}},\
  \bibinfo {pages} {78} (\bibinfo {year} {2019})}\BibitemShut {NoStop}%
\bibitem [{\citenamefont {Lipkin}\ \emph {et~al.}(1965)\citenamefont {Lipkin},
  \citenamefont {Meshkov},\ and\ \citenamefont {Glick}}]{Lipkin1965}%
  \BibitemOpen
  \bibfield  {author} {\bibinfo {author} {\bibfnamefont {H.J.}\ \bibnamefont
  {Lipkin}}, \bibinfo {author} {\bibfnamefont {N.}~\bibnamefont {Meshkov}}, \
  and\ \bibinfo {author} {\bibfnamefont {A.J.}\ \bibnamefont {Glick}},\
  }\bibfield  {title} {\enquote {\bibinfo {title} {{Validity of many-body
  approximation methods for a solvable model: (I). Exact solutions and
  perturbation theory}},}\ }\href {\doibase 10.1016/0029-5582(65)90862-X}
  {\bibfield  {journal} {\bibinfo  {journal} {Nuclear Physics}\ }\textbf
  {\bibinfo {volume} {62}},\ \bibinfo {pages} {188--198} (\bibinfo {year}
  {1965})}\BibitemShut {NoStop}%
\bibitem [{\citenamefont {Schack}\ \emph {et~al.}(1994)\citenamefont {Schack},
  \citenamefont {D'Ariano},\ and\ \citenamefont {Caves}}]{Schack1994}%
  \BibitemOpen
  \bibfield  {author} {\bibinfo {author} {\bibfnamefont {R{\"{u}}diger}\
  \bibnamefont {Schack}}, \bibinfo {author} {\bibfnamefont {Giacomo~M.}\
  \bibnamefont {D'Ariano}}, \ and\ \bibinfo {author} {\bibfnamefont
  {Carlton~M.}\ \bibnamefont {Caves}},\ }\bibfield  {title} {\enquote {\bibinfo
  {title} {{Hypersensitivity to perturbation in the quantum kicked top}},}\
  }\href {\doibase 10.1103/PhysRevE.50.972} {\bibfield  {journal} {\bibinfo
  {journal} {Physical Review E}\ }\textbf {\bibinfo {volume} {50}},\ \bibinfo
  {pages} {972--987} (\bibinfo {year} {1994})}\BibitemShut {NoStop}%
\bibitem [{\citenamefont {Ghose}\ \emph {et~al.}(2008)\citenamefont {Ghose},
  \citenamefont {Stock}, \citenamefont {Jessen}, \citenamefont {Lal},\ and\
  \citenamefont {Silberfarb}}]{Ghose2008}%
  \BibitemOpen
  \bibfield  {author} {\bibinfo {author} {\bibfnamefont {Shohini}\ \bibnamefont
  {Ghose}}, \bibinfo {author} {\bibfnamefont {Rene}\ \bibnamefont {Stock}},
  \bibinfo {author} {\bibfnamefont {Poul}\ \bibnamefont {Jessen}}, \bibinfo
  {author} {\bibfnamefont {Roshan}\ \bibnamefont {Lal}}, \ and\ \bibinfo
  {author} {\bibfnamefont {Andrew}\ \bibnamefont {Silberfarb}},\ }\bibfield
  {title} {\enquote {\bibinfo {title} {{Chaos, entanglement, and decoherence in
  the quantum kicked top}},}\ }\href {\doibase 10.1103/PhysRevA.78.042318}
  {\bibfield  {journal} {\bibinfo  {journal} {Physical Review A}\ }\textbf
  {\bibinfo {volume} {78}},\ \bibinfo {pages} {042318} (\bibinfo {year}
  {2008})}\BibitemShut {NoStop}%
\bibitem [{\citenamefont {Kumari}\ and\ \citenamefont
  {Ghose}(2018)}]{Kumari2018}%
  \BibitemOpen
  \bibfield  {author} {\bibinfo {author} {\bibfnamefont {Meenu}\ \bibnamefont
  {Kumari}}\ and\ \bibinfo {author} {\bibfnamefont {Shohini}\ \bibnamefont
  {Ghose}},\ }\bibfield  {title} {\enquote {\bibinfo {title}
  {{Quantum-classical correspondence in the vicinity of periodic orbits}},}\
  }\href {\doibase 10.1103/PhysRevE.97.052209} {\bibfield  {journal} {\bibinfo
  {journal} {Physical Review E}\ }\textbf {\bibinfo {volume} {97}},\ \bibinfo
  {pages} {052209} (\bibinfo {year} {2018})}\BibitemShut {NoStop}%
\bibitem [{\citenamefont {Kumari}\ and\ \citenamefont
  {Ghose}(2019)}]{Kumari2019}%
  \BibitemOpen
  \bibfield  {author} {\bibinfo {author} {\bibfnamefont {Meenu}\ \bibnamefont
  {Kumari}}\ and\ \bibinfo {author} {\bibfnamefont {Shohini}\ \bibnamefont
  {Ghose}},\ }\bibfield  {title} {\enquote {\bibinfo {title} {{Untangling
  entanglement and chaos}},}\ }\href {\doibase 10.1103/PhysRevA.99.042311}
  {\bibfield  {journal} {\bibinfo  {journal} {Physical Review A}\ }\textbf
  {\bibinfo {volume} {99}},\ \bibinfo {pages} {042311} (\bibinfo {year}
  {2019})}\BibitemShut {NoStop}%
\bibitem [{\citenamefont {Chaudhury}\ \emph {et~al.}(2009)\citenamefont
  {Chaudhury}, \citenamefont {Smith}, \citenamefont {Anderson}, \citenamefont
  {Ghose},\ and\ \citenamefont {Jessen}}]{Chaudhury2009}%
  \BibitemOpen
  \bibfield  {author} {\bibinfo {author} {\bibfnamefont {S.}~\bibnamefont
  {Chaudhury}}, \bibinfo {author} {\bibfnamefont {A.}~\bibnamefont {Smith}},
  \bibinfo {author} {\bibfnamefont {B.~E.}\ \bibnamefont {Anderson}}, \bibinfo
  {author} {\bibfnamefont {S.}~\bibnamefont {Ghose}}, \ and\ \bibinfo {author}
  {\bibfnamefont {P.~S.}\ \bibnamefont {Jessen}},\ }\bibfield  {title}
  {\enquote {\bibinfo {title} {{Quantum signatures of chaos in a kicked
  top}},}\ }\href {\doibase 10.1038/nature08396} {\bibfield  {journal}
  {\bibinfo  {journal} {Nature}\ }\textbf {\bibinfo {volume} {461}},\ \bibinfo
  {pages} {768--771} (\bibinfo {year} {2009})}\BibitemShut {NoStop}%
\bibitem [{\citenamefont {Neill}\ \emph {et~al.}(2016)\citenamefont {Neill},
  \citenamefont {Roushan}, \citenamefont {Fang}, \citenamefont {Chen},
  \citenamefont {Kolodrubetz}, \citenamefont {Chen}, \citenamefont {Megrant},
  \citenamefont {Barends}, \citenamefont {Campbell}, \citenamefont {Chiaro}
  \emph {et~al.}}]{Neill2016}%
  \BibitemOpen
  \bibfield  {author} {\bibinfo {author} {\bibfnamefont {Charles}\ \bibnamefont
  {Neill}}, \bibinfo {author} {\bibfnamefont {P}~\bibnamefont {Roushan}},
  \bibinfo {author} {\bibfnamefont {M}~\bibnamefont {Fang}}, \bibinfo {author}
  {\bibfnamefont {Y}~\bibnamefont {Chen}}, \bibinfo {author} {\bibfnamefont
  {M}~\bibnamefont {Kolodrubetz}}, \bibinfo {author} {\bibfnamefont
  {Z}~\bibnamefont {Chen}}, \bibinfo {author} {\bibfnamefont {A}~\bibnamefont
  {Megrant}}, \bibinfo {author} {\bibfnamefont {R}~\bibnamefont {Barends}},
  \bibinfo {author} {\bibfnamefont {B}~\bibnamefont {Campbell}}, \bibinfo
  {author} {\bibfnamefont {B}~\bibnamefont {Chiaro}},  \emph {et~al.},\
  }\bibfield  {title} {\enquote {\bibinfo {title} {Ergodic dynamics and
  thermalization in an isolated quantum system},}\ }\href@noop {} {\bibfield
  {journal} {\bibinfo  {journal} {Nature Physics}\ }\textbf {\bibinfo {volume}
  {12}},\ \bibinfo {pages} {1037--1041} (\bibinfo {year} {2016})}\BibitemShut
  {NoStop}%
\bibitem [{\citenamefont {Trail}\ \emph {et~al.}(2008)\citenamefont {Trail},
  \citenamefont {Madhok},\ and\ \citenamefont {Deutsch}}]{Trail2008}%
  \BibitemOpen
  \bibfield  {author} {\bibinfo {author} {\bibfnamefont {Collin~M.}\
  \bibnamefont {Trail}}, \bibinfo {author} {\bibfnamefont {Vaibhav}\
  \bibnamefont {Madhok}}, \ and\ \bibinfo {author} {\bibfnamefont {Ivan~H.}\
  \bibnamefont {Deutsch}},\ }\bibfield  {title} {\enquote {\bibinfo {title}
  {{Entanglement and the generation of random states in the quantum chaotic
  dynamics of kicked coupled tops}},}\ }\href {\doibase
  10.1103/PhysRevE.78.046211} {\bibfield  {journal} {\bibinfo  {journal}
  {Physical Review E}\ }\textbf {\bibinfo {volume} {78}},\ \bibinfo {pages}
  {046211} (\bibinfo {year} {2008})}\BibitemShut {NoStop}%
\bibitem [{\citenamefont {Herrmann}\ \emph {et~al.}(2020)\citenamefont
  {Herrmann}, \citenamefont {Kieler}, \citenamefont {Fritzsch},\ and\
  \citenamefont {B\"acker}}]{Herman2020}%
  \BibitemOpen
  \bibfield  {author} {\bibinfo {author} {\bibfnamefont {Tabea}\ \bibnamefont
  {Herrmann}}, \bibinfo {author} {\bibfnamefont {Maximilian F.~I.}\
  \bibnamefont {Kieler}}, \bibinfo {author} {\bibfnamefont {Felix}\
  \bibnamefont {Fritzsch}}, \ and\ \bibinfo {author} {\bibfnamefont {Arnd}\
  \bibnamefont {B\"acker}},\ }\bibfield  {title} {\enquote {\bibinfo {title}
  {Entanglement in coupled kicked tops with chaotic dynamics},}\ }\href
  {\doibase 10.1103/PhysRevE.101.022221} {\bibfield  {journal} {\bibinfo
  {journal} {Phys. Rev. E}\ }\textbf {\bibinfo {volume} {101}},\ \bibinfo
  {pages} {022221} (\bibinfo {year} {2020})}\BibitemShut {NoStop}%
\bibitem [{\citenamefont {Lombardi}\ and\ \citenamefont
  {Matzkin}(2011)}]{Lombardi2011}%
  \BibitemOpen
  \bibfield  {author} {\bibinfo {author} {\bibfnamefont {M.}~\bibnamefont
  {Lombardi}}\ and\ \bibinfo {author} {\bibfnamefont {A.}~\bibnamefont
  {Matzkin}},\ }\bibfield  {title} {\enquote {\bibinfo {title} {Entanglement
  and chaos in the kicked top},}\ }\href {\doibase 10.1103/PhysRevE.83.016207}
  {\bibfield  {journal} {\bibinfo  {journal} {Phys. Rev. E}\ }\textbf {\bibinfo
  {volume} {83}},\ \bibinfo {pages} {016207} (\bibinfo {year}
  {2011})}\BibitemShut {NoStop}%
\bibitem [{\citenamefont {Mu\~noz Arias}\ \emph
  {et~al.}(2020{\natexlab{a}})\citenamefont {Mu\~noz Arias}, \citenamefont
  {Poggi}, \citenamefont {Jessen},\ and\ \citenamefont {Deutsch}}]{munoz2019}%
  \BibitemOpen
  \bibfield  {author} {\bibinfo {author} {\bibfnamefont {Manuel~H.}\
  \bibnamefont {Mu\~noz Arias}}, \bibinfo {author} {\bibfnamefont {Pablo~M.}\
  \bibnamefont {Poggi}}, \bibinfo {author} {\bibfnamefont {Poul~S.}\
  \bibnamefont {Jessen}}, \ and\ \bibinfo {author} {\bibfnamefont {Ivan~H.}\
  \bibnamefont {Deutsch}},\ }\bibfield  {title} {\enquote {\bibinfo {title}
  {Simulating nonlinear dynamics of collective spins via quantum measurement
  and feedback},}\ }\href {\doibase 10.1103/PhysRevLett.124.110503} {\bibfield
  {journal} {\bibinfo  {journal} {Phys. Rev. Lett.}\ }\textbf {\bibinfo
  {volume} {124}},\ \bibinfo {pages} {110503} (\bibinfo {year}
  {2020}{\natexlab{a}})}\BibitemShut {NoStop}%
\bibitem [{\citenamefont {J{\"{o}}rg}\ \emph {et~al.}(2010)\citenamefont
  {J{\"{o}}rg}, \citenamefont {Krzakala}, \citenamefont {Kurchan},
  \citenamefont {Maggs},\ and\ \citenamefont {Pujos}}]{Jorg2010}%
  \BibitemOpen
  \bibfield  {author} {\bibinfo {author} {\bibfnamefont {T.}~\bibnamefont
  {J{\"{o}}rg}}, \bibinfo {author} {\bibfnamefont {F.}~\bibnamefont
  {Krzakala}}, \bibinfo {author} {\bibfnamefont {J.}~\bibnamefont {Kurchan}},
  \bibinfo {author} {\bibfnamefont {A.~C.}\ \bibnamefont {Maggs}}, \ and\
  \bibinfo {author} {\bibfnamefont {J.}~\bibnamefont {Pujos}},\ }\bibfield
  {title} {\enquote {\bibinfo {title} {{Energy gaps in quantum first-order
  mean-field–like transitions: The problems that quantum annealing cannot
  solve}},}\ }\href {\doibase 10.1209/0295-5075/89/40004} {\bibfield  {journal}
  {\bibinfo  {journal} {EPL (Europhysics Letters)}\ }\textbf {\bibinfo {volume}
  {89}},\ \bibinfo {pages} {40004} (\bibinfo {year} {2010})}\BibitemShut
  {NoStop}%
\bibitem [{\citenamefont {Bapst}\ and\ \citenamefont
  {Semerjian}(2012)}]{Bapst2012}%
  \BibitemOpen
  \bibfield  {author} {\bibinfo {author} {\bibfnamefont {Victor}\ \bibnamefont
  {Bapst}}\ and\ \bibinfo {author} {\bibfnamefont {Guilhem}\ \bibnamefont
  {Semerjian}},\ }\bibfield  {title} {\enquote {\bibinfo {title} {On quantum
  mean-field models and their quantum annealing},}\ }\href@noop {} {\bibfield
  {journal} {\bibinfo  {journal} {Journal of Statistical Mechanics: Theory and
  Experiment}\ }\textbf {\bibinfo {volume} {2012}},\ \bibinfo {pages} {P06007}
  (\bibinfo {year} {2012})}\BibitemShut {NoStop}%
\bibitem [{\citenamefont {Kong}\ and\ \citenamefont
  {Crosson}(2017)}]{Kong2017}%
  \BibitemOpen
  \bibfield  {author} {\bibinfo {author} {\bibfnamefont {Linghang}\
  \bibnamefont {Kong}}\ and\ \bibinfo {author} {\bibfnamefont {Elizabeth}\
  \bibnamefont {Crosson}},\ }\bibfield  {title} {\enquote {\bibinfo {title}
  {The performance of the quantum adiabatic algorithm on spike hamiltonians},}\
  }\href {\doibase 10.1142/S0219749917500113} {\bibfield  {journal} {\bibinfo
  {journal} {International Journal of Quantum Information}\ }\textbf {\bibinfo
  {volume} {15}},\ \bibinfo {pages} {1750011} (\bibinfo {year} {2017})},\
  \Eprint {http://arxiv.org/abs/https://doi.org/10.1142/S0219749917500113}
  {https://doi.org/10.1142/S0219749917500113} \BibitemShut {NoStop}%
\bibitem [{\citenamefont {Matsuura}\ \emph {et~al.}(2017)\citenamefont
  {Matsuura}, \citenamefont {Nishimori}, \citenamefont {Vinci}, \citenamefont
  {Albash},\ and\ \citenamefont {Lidar}}]{Matsuura2017}%
  \BibitemOpen
  \bibfield  {author} {\bibinfo {author} {\bibfnamefont {Shunji}\ \bibnamefont
  {Matsuura}}, \bibinfo {author} {\bibfnamefont {Hidetoshi}\ \bibnamefont
  {Nishimori}}, \bibinfo {author} {\bibfnamefont {Walter}\ \bibnamefont
  {Vinci}}, \bibinfo {author} {\bibfnamefont {Tameem}\ \bibnamefont {Albash}},
  \ and\ \bibinfo {author} {\bibfnamefont {Daniel~A.}\ \bibnamefont {Lidar}},\
  }\bibfield  {title} {\enquote {\bibinfo {title} {{Quantum-annealing
  correction at finite temperature: Ferromagnetic p -spin models}},}\ }\href
  {\doibase 10.1103/PhysRevA.95.022308} {\bibfield  {journal} {\bibinfo
  {journal} {Physical Review A}\ }\textbf {\bibinfo {volume} {95}},\ \bibinfo
  {pages} {022308} (\bibinfo {year} {2017})}\BibitemShut {NoStop}%
\bibitem [{\citenamefont {Mu\~noz Arias}\ \emph
  {et~al.}(2020{\natexlab{b}})\citenamefont {Mu\~noz Arias}, \citenamefont
  {Deutsch}, \citenamefont {Jessen},\ and\ \citenamefont
  {Poggi}}]{Munoz-Arias2020}%
  \BibitemOpen
  \bibfield  {author} {\bibinfo {author} {\bibfnamefont {Manuel~H.}\
  \bibnamefont {Mu\~noz Arias}}, \bibinfo {author} {\bibfnamefont {Ivan~H.}\
  \bibnamefont {Deutsch}}, \bibinfo {author} {\bibfnamefont {Poul~S.}\
  \bibnamefont {Jessen}}, \ and\ \bibinfo {author} {\bibfnamefont {Pablo~M.}\
  \bibnamefont {Poggi}},\ }\bibfield  {title} {\enquote {\bibinfo {title}
  {Simulation of the complex dynamics of mean-field $p$-spin models using
  measurement-based quantum feedback control},}\ }\href {\doibase
  10.1103/PhysRevA.102.022610} {\bibfield  {journal} {\bibinfo  {journal}
  {Phys. Rev. A}\ }\textbf {\bibinfo {volume} {102}},\ \bibinfo {pages}
  {022610} (\bibinfo {year} {2020}{\natexlab{b}})}\BibitemShut {NoStop}%
\bibitem [{\citenamefont {Meyer}(1970)}]{Meyer1970}%
  \BibitemOpen
  \bibfield  {author} {\bibinfo {author} {\bibfnamefont {K.~R.}\ \bibnamefont
  {Meyer}},\ }\bibfield  {title} {\enquote {\bibinfo {title} {Generic
  bifurcation of periodic points},}\ }\href
  {http://www.jstor.org/stable/1995662} {\bibfield  {journal} {\bibinfo
  {journal} {Transactions of the American Mathematical Society}\ }\textbf
  {\bibinfo {volume} {149}},\ \bibinfo {pages} {95--107} (\bibinfo {year}
  {1970})}\BibitemShut {NoStop}%
\bibitem [{\citenamefont {MacKay}(1983)}]{Mackay1983}%
  \BibitemOpen
  \bibfield  {author} {\bibinfo {author} {\bibfnamefont {Robert~S}\
  \bibnamefont {MacKay}},\ }\enquote {\bibinfo {title} {Period doubling as a
  universal route to stochasticity},}\ in\ \href@noop {} {\emph {\bibinfo
  {booktitle} {Long Time Prediction in Dynamics}}}\ (\bibinfo  {publisher}
  {John Wiley~\& Sons},\ \bibinfo {year} {1983})\ p.\ \bibinfo {pages}
  {127}\BibitemShut {NoStop}%
\bibitem [{\citenamefont {Henon}(1969)}]{Henon1969}%
  \BibitemOpen
  \bibfield  {author} {\bibinfo {author} {\bibfnamefont {Michel}\ \bibnamefont
  {Henon}},\ }\bibfield  {title} {\enquote {\bibinfo {title} {Numerical study
  of quadratic area-preserving mappings},}\ }\href@noop {} {\bibfield
  {journal} {\bibinfo  {journal} {Quarterly of applied mathematics}\ ,\
  \bibinfo {pages} {291--312}} (\bibinfo {year} {1969})}\BibitemShut {NoStop}%
\bibitem [{\citenamefont {Sim\'o}(1982)}]{Simo1981}%
  \BibitemOpen
  \bibfield  {author} {\bibinfo {author} {\bibfnamefont {Carles}\ \bibnamefont
  {Sim\'o}},\ }\bibfield  {title} {\enquote {\bibinfo {title} {Stability of
  degenerate fixed points of analytic area preserving mappings},}\ }in\ \href
  {http://www.numdam.org/item/AST_1983__98-99__184_0} {\emph {\bibinfo
  {booktitle} {Bifurcation, th\'eorie ergodique et applications - 22-26 juin
  1981}}},\ \bibinfo {series and number} {\bibinfo {series} {Ast\'erisque}\
  No.\ \bibinfo {number} {98-99}}\ (\bibinfo  {publisher} {Soci\'et\'e
  math\'ematique de France},\ \bibinfo {year} {1982})\ pp.\ \bibinfo {pages}
  {184--194}\BibitemShut {NoStop}%
\bibitem [{Note1()}]{Note1}%
  \BibitemOpen
  \bibinfo {note} {In our delta-kicked Hamiltonian in Eq. (\ref
  {eqn:kicked_p_spin_hamil}) we have dropped a minus sign compared to the
  effective Hamiltonian obtained from the Trotterization of the $p$-spin
  evolution. This is done in order to be faithful with the conventions in
  Haake's original work~\cite {Haake1987}, as in the present work we aim to
  stress the differences between the kicked top and its generalizations. Notice
  however that for the present study, of dynamical character, the choice of
  sign does not alter the observed phenomenology. It does change the character
  of the ground state phase diagram, which will be important in the context of
  analog quantum simulation of $p$-spin models, study that will be address in a
  future work.}\BibitemShut {Stop}%
\bibitem [{Note2()}]{Note2}%
  \BibitemOpen
  \bibinfo {note} {The classical phase space is restricted to the surface of
  the unit sphere, and hence one can also write the map in Eq. (\ref
  {eqn:classical_kicked_p_spin}) in terms of the angular variables of spherical
  coordinates $(\theta , \phi )$, where these are the polar and azimuthal
  angle, respectively.}\BibitemShut {Stop}%
\bibitem [{\citenamefont {Reichl}(2004)}]{Reichl}%
  \BibitemOpen
  \bibfield  {author} {\bibinfo {author} {\bibfnamefont {Linda}\ \bibnamefont
  {Reichl}},\ }\href {\doibase 10.1007/978-1-4757-4350-0} {\emph {\bibinfo
  {title} {The transition to chaos: conservative classical systems and quantum
  manifestations}}}\ (\bibinfo  {publisher} {Springer-Verlag New York},\
  \bibinfo {year} {2004})\BibitemShut {NoStop}%
\bibitem [{\citenamefont {Schuster}(1995)}]{Schuster1995}%
  \BibitemOpen
  \bibfield  {author} {\bibinfo {author} {\bibfnamefont {Heinz~Georg}\
  \bibnamefont {Schuster}},\ }\href@noop {} {\emph {\bibinfo {title}
  {{Deterministic chaos : an introduction}}}}\ (\bibinfo  {publisher} {VCH},\
  \bibinfo {year} {1995})\ p.\ \bibinfo {pages} {291}\BibitemShut {NoStop}%
\bibitem [{\citenamefont {MacKay}(1993)}]{Mackay1993}%
  \BibitemOpen
  \bibfield  {author} {\bibinfo {author} {\bibfnamefont {Robert~Sinclair}\
  \bibnamefont {MacKay}},\ }\href@noop {} {\emph {\bibinfo {title}
  {Renormalisation in area-preserving maps}}},\ Vol.~\bibinfo {volume} {6}\
  (\bibinfo  {publisher} {World Scientific},\ \bibinfo {year}
  {1993})\BibitemShut {NoStop}%
\bibitem [{\citenamefont {M{\"o}ser}(1962)}]{Moser1962}%
  \BibitemOpen
  \bibfield  {author} {\bibinfo {author} {\bibfnamefont {J}~\bibnamefont
  {M{\"o}ser}},\ }\bibfield  {title} {\enquote {\bibinfo {title} {On invariant
  curves of area-preserving mappings of an annulus},}\ }\href@noop {}
  {\bibfield  {journal} {\bibinfo  {journal} {Nachr. Akad. Wiss. G{\"o}ttingen,
  II}\ ,\ \bibinfo {pages} {1--20}} (\bibinfo {year} {1962})}\BibitemShut
  {NoStop}%
\bibitem [{\citenamefont {Aharonov}\ and\ \citenamefont
  {Elias}(1990)}]{Aharonov1990}%
  \BibitemOpen
  \bibfield  {author} {\bibinfo {author} {\bibfnamefont {Dov}\ \bibnamefont
  {Aharonov}}\ and\ \bibinfo {author} {\bibfnamefont {Uri}\ \bibnamefont
  {Elias}},\ }\bibfield  {title} {\enquote {\bibinfo {title} {Parabolic fixed
  points, invariant curves and action-angle variables},}\ }\href {\doibase
  10.1017/S0143385700005526} {\bibfield  {journal} {\bibinfo  {journal}
  {Ergodic Theory and Dynamical Systems}\ }\textbf {\bibinfo {volume} {10}},\
  \bibinfo {pages} {231–245} (\bibinfo {year} {1990})}\BibitemShut {NoStop}%
\bibitem [{Note3()}]{Note3}%
  \BibitemOpen
  \bibinfo {note} {Meaning that additional fixed points could exist arbitrarily
  close to the original one and the original fixed point might not be robust to
  small perturbations (see for instance~\cite {Mackay1983})}\BibitemShut
  {NoStop}%
\bibitem [{Note4()}]{Note4}%
  \BibitemOpen
  \bibinfo {note} {Considering $p=2$ with $\alpha =\pi /2$ and the fixed points
  on the poles, one has that ${\protect \rm Tr}(\protect \mathbf {M}) = \pm k$
  for $(0,\pm 1,0)$. Therefore, at $k=2$ the poles are parabolic fixed points.
  The north pole undergoes a tangent bifurcation, the south pole undergoes a
  period doubling bifurcation. Additionally, at $\alpha =\pi /2$, $F^2$ is
  invariant under $R_x(\pi )$, which forces the north pole to undergo a period
  doubling bifurcation as well. This is one of the main results of Haake~\cite
  {Haake1987}, and a good example of the importance of parabolic fixed
  points.\label {fn_example}}\BibitemShut {NoStop}%
\bibitem [{Note5()}]{Note5}%
  \BibitemOpen
  \bibinfo {note} {This is nothing but a restatement of invariance under
  $R_y(\pi )$.}\BibitemShut {Stop}%
\bibitem [{\citenamefont {Lee~Rodgers}\ and\ \citenamefont
  {Nicewander}(1988)}]{Lee1988}%
  \BibitemOpen
  \bibfield  {author} {\bibinfo {author} {\bibfnamefont {Joseph}\ \bibnamefont
  {Lee~Rodgers}}\ and\ \bibinfo {author} {\bibfnamefont {W~Alan}\ \bibnamefont
  {Nicewander}},\ }\bibfield  {title} {\enquote {\bibinfo {title} {Thirteen
  ways to look at the correlation coefficient},}\ }\href@noop {} {\bibfield
  {journal} {\bibinfo  {journal} {The American Statistician}\ }\textbf
  {\bibinfo {volume} {42}},\ \bibinfo {pages} {59--66} (\bibinfo {year}
  {1988})}\BibitemShut {NoStop}%
\bibitem [{\citenamefont {Lichtenberg}\ and\ \citenamefont
  {Lieberman}(1992)}]{Lichtenberg1992}%
  \BibitemOpen
  \bibfield  {author} {\bibinfo {author} {\bibfnamefont {Allan~J.}\
  \bibnamefont {Lichtenberg}}\ and\ \bibinfo {author} {\bibfnamefont {M.~A.}\
  \bibnamefont {Lieberman}},\ }\href
  {https://books.google.com/books/about/Regular{\_}and{\_}chaotic{\_}dynamics.html?id=2ssPAQAAMAAJ}
  {\emph {\bibinfo {title} {{Regular and chaotic dynamics}}}}\ (\bibinfo
  {publisher} {Springer-Verlag},\ \bibinfo {year} {1992})\ p.\ \bibinfo {pages}
  {692}\BibitemShut {NoStop}%
\bibitem [{\citenamefont {Wimberger}(2014)}]{Wimberger}%
  \BibitemOpen
  \bibfield  {author} {\bibinfo {author} {\bibfnamefont {Sandro}\ \bibnamefont
  {Wimberger}},\ }\href
  {https://books.google.com/books/about/Nonlinear{\_}Dynamics{\_}and{\_}Quantum{\_}Chaos.html?id=nNIkBAAAQBAJ}
  {\emph {\bibinfo {title} {{Nonlinear dynamics and quantum chaos : an
  introduction}}}}\ (\bibinfo  {publisher} {Springer},\ \bibinfo {year}
  {2014})\ p.\ \bibinfo {pages} {206}\BibitemShut {NoStop}%
\bibitem [{\citenamefont {Zaslavsky}\ \emph {et~al.}(1991)\citenamefont
  {Zaslavsky}, \citenamefont {Sagdeev}, \citenamefont {Usikov},\ and\
  \citenamefont {Chernikov}}]{Zaslavsky1991}%
  \BibitemOpen
  \bibfield  {author} {\bibinfo {author} {\bibfnamefont {G.~M.}\ \bibnamefont
  {Zaslavsky}}, \bibinfo {author} {\bibfnamefont {R.~Z.}\ \bibnamefont
  {Sagdeev}}, \bibinfo {author} {\bibfnamefont {D.~A.}\ \bibnamefont {Usikov}},
  \ and\ \bibinfo {author} {\bibfnamefont {A.~A.}\ \bibnamefont {Chernikov}},\
  }\href {\doibase 10.1017/CBO9780511599996} {\emph {\bibinfo {title} {Weak
  Chaos and Quasi-Regular Patterns}}},\ Cambridge Nonlinear Science Series\
  (\bibinfo  {publisher} {Cambridge University Press},\ \bibinfo {year}
  {1991})\BibitemShut {NoStop}%
\bibitem [{\citenamefont {Birkhoff}(1935)}]{Birkhoff1935}%
  \BibitemOpen
  \bibfield  {author} {\bibinfo {author} {\bibfnamefont {G.~D.}\ \bibnamefont
  {Birkhoff}},\ }\bibfield  {title} {\enquote {\bibinfo {title} {{Nouvelles
  recherches sur les syst{\`{e}}mes dynamiques}},}\ }\href@noop {} {\bibfield
  {journal} {\bibinfo  {journal} {Mem. Pont. Acad. Novi Lyncaei}\ }\textbf
  {\bibinfo {volume} {1}},\ \bibinfo {pages} {85--216} (\bibinfo {year}
  {1935})}\BibitemShut {NoStop}%
\bibitem [{\citenamefont {Arnol'd}(1964)}]{Arnold1964}%
  \BibitemOpen
  \bibfield  {author} {\bibinfo {author} {\bibfnamefont {V.~I.}\ \bibnamefont
  {Arnol'd}},\ }\bibfield  {title} {\enquote {\bibinfo {title} {{Instability of
  dynamical systems with many degrees of freedom}},}\ }\href@noop {} {\bibfield
   {journal} {\bibinfo  {journal} {Dokl. Akad. Nauk SSSR}\ }\textbf {\bibinfo
  {volume} {156}},\ \bibinfo {pages} {9--12} (\bibinfo {year}
  {1964})}\BibitemShut {NoStop}%
\bibitem [{\citenamefont {Benettin}\ and\ \citenamefont
  {Strelcyn}(1978)}]{Benettin1978}%
  \BibitemOpen
  \bibfield  {author} {\bibinfo {author} {\bibfnamefont {G.}~\bibnamefont
  {Benettin}}\ and\ \bibinfo {author} {\bibfnamefont {J.~M.}\ \bibnamefont
  {Strelcyn}},\ }\bibfield  {title} {\enquote {\bibinfo {title} {{Numerical
  experiments on the free motion of a point mass moving in a plane convex
  region: Stochastic transition and entropy}},}\ }\href {\doibase
  10.1103/PhysRevA.17.773} {\bibfield  {journal} {\bibinfo  {journal} {Physical
  Review A}\ }\textbf {\bibinfo {volume} {17}},\ \bibinfo {pages} {773--785}
  (\bibinfo {year} {1978})}\BibitemShut {NoStop}%
\bibitem [{Note6()}]{Note6}%
  \BibitemOpen
  \bibinfo {note} {Transition to chaos via a cascade of period doubling
  bifurcations was observed initially in dissipative systems~\cite
  {Feigenbaum1979,Sander2012}, for instance the Logistic map}\BibitemShut
  {NoStop}%
\bibitem [{\citenamefont {Bountis}(1981)}]{Bountis1981}%
  \BibitemOpen
  \bibfield  {author} {\bibinfo {author} {\bibfnamefont {Tassos~C.}\
  \bibnamefont {Bountis}},\ }\bibfield  {title} {\enquote {\bibinfo {title}
  {{Period doubling bifurcations and universality in conservative systems}},}\
  }\href {\doibase 10.1016/0167-2789(81)90041-5} {\bibfield  {journal}
  {\bibinfo  {journal} {Physica D: Nonlinear Phenomena}\ }\textbf {\bibinfo
  {volume} {3}},\ \bibinfo {pages} {577--589} (\bibinfo {year}
  {1981})}\BibitemShut {NoStop}%
\bibitem [{\citenamefont {Greene}\ \emph {et~al.}(1981)\citenamefont {Greene},
  \citenamefont {MacKay}, \citenamefont {Vivaldi},\ and\ \citenamefont
  {Feigenbaum}}]{Greene1981}%
  \BibitemOpen
  \bibfield  {author} {\bibinfo {author} {\bibfnamefont {J.M.}\ \bibnamefont
  {Greene}}, \bibinfo {author} {\bibfnamefont {R.S.}\ \bibnamefont {MacKay}},
  \bibinfo {author} {\bibfnamefont {F.}~\bibnamefont {Vivaldi}}, \ and\
  \bibinfo {author} {\bibfnamefont {M.J.}\ \bibnamefont {Feigenbaum}},\
  }\bibfield  {title} {\enquote {\bibinfo {title} {{Universal behaviour in
  families of area-preserving maps}},}\ }\href {\doibase
  10.1016/0167-2789(81)90034-8} {\bibfield  {journal} {\bibinfo  {journal}
  {Physica D: Nonlinear Phenomena}\ }\textbf {\bibinfo {volume} {3}},\ \bibinfo
  {pages} {468--486} (\bibinfo {year} {1981})}\BibitemShut {NoStop}%
\bibitem [{\citenamefont {Kolmogorov}(1958)}]{Kolmogorov1958}%
  \BibitemOpen
  \bibfield  {author} {\bibinfo {author} {\bibfnamefont {A.~N.}\ \bibnamefont
  {Kolmogorov}},\ }\bibfield  {title} {\enquote {\bibinfo {title} {{A new
  metric invariant of transient dynamical systems and automorphisms in Lebesgue
  spaces}},}\ }\href@noop {} {\bibfield  {journal} {\bibinfo  {journal} {Dokl.
  Akad. Nauk SSSR}\ }\textbf {\bibinfo {volume} {119}},\ \bibinfo {pages}
  {861--864} (\bibinfo {year} {1958})}\BibitemShut {NoStop}%
\bibitem [{\citenamefont {Latora}\ and\ \citenamefont
  {Baranger}(1999)}]{Latora1999}%
  \BibitemOpen
  \bibfield  {author} {\bibinfo {author} {\bibfnamefont {Vito}\ \bibnamefont
  {Latora}}\ and\ \bibinfo {author} {\bibfnamefont {Michel}\ \bibnamefont
  {Baranger}},\ }\bibfield  {title} {\enquote {\bibinfo {title}
  {{Kolmogorov-Sinai Entropy Rate versus Physical Entropy}},}\ }\href {\doibase
  10.1103/PhysRevLett.82.520} {\bibfield  {journal} {\bibinfo  {journal}
  {Physical Review Letters}\ }\textbf {\bibinfo {volume} {82}},\ \bibinfo
  {pages} {520--523} (\bibinfo {year} {1999})}\BibitemShut {NoStop}%
\bibitem [{\citenamefont {Boffetta}\ \emph {et~al.}(2002)\citenamefont
  {Boffetta}, \citenamefont {Cencini}, \citenamefont {Falcioni},\ and\
  \citenamefont {Vulpiani}}]{Boffetta2002}%
  \BibitemOpen
  \bibfield  {author} {\bibinfo {author} {\bibfnamefont {G.}~\bibnamefont
  {Boffetta}}, \bibinfo {author} {\bibfnamefont {M.}~\bibnamefont {Cencini}},
  \bibinfo {author} {\bibfnamefont {M.}~\bibnamefont {Falcioni}}, \ and\
  \bibinfo {author} {\bibfnamefont {A.}~\bibnamefont {Vulpiani}},\ }\bibfield
  {title} {\enquote {\bibinfo {title} {{Predictability: a way to characterize
  complexity}},}\ }\href {\doibase 10.1016/S0370-1573(01)00025-4} {\bibfield
  {journal} {\bibinfo  {journal} {Physics Reports}\ }\textbf {\bibinfo {volume}
  {356}},\ \bibinfo {pages} {367--474} (\bibinfo {year} {2002})}\BibitemShut
  {NoStop}%
\bibitem [{\citenamefont {{V. I. Oseledets}}(1968)}]{Ose1968}%
  \BibitemOpen
  \bibfield  {author} {\bibinfo {author} {\bibnamefont {{V. I. Oseledets}}},\
  }\bibfield  {title} {\enquote {\bibinfo {title} {{A multiplicative ergodic
  theorem. Characteristic Ljapunov, exponents of dynamical systems}},}\
  }\href@noop {} {\bibfield  {journal} {\bibinfo  {journal} {Trans. Moscow
  Math. Soc.}\ }\textbf {\bibinfo {volume} {19}},\ \bibinfo {pages} {197--231}
  (\bibinfo {year} {1968})}\BibitemShut {NoStop}%
\bibitem [{\citenamefont {Eckmann}\ and\ \citenamefont
  {Ruelle}(1985)}]{Eckmann1985}%
  \BibitemOpen
  \bibfield  {author} {\bibinfo {author} {\bibfnamefont {J.~P.}\ \bibnamefont
  {Eckmann}}\ and\ \bibinfo {author} {\bibfnamefont {D.}~\bibnamefont
  {Ruelle}},\ }\bibfield  {title} {\enquote {\bibinfo {title} {{Ergodic theory
  of chaos and strange attractors}},}\ }\href {\doibase
  10.1103/RevModPhys.57.617} {\bibfield  {journal} {\bibinfo  {journal}
  {Reviews of Modern Physics}\ }\textbf {\bibinfo {volume} {57}},\ \bibinfo
  {pages} {617--656} (\bibinfo {year} {1985})}\BibitemShut {NoStop}%
\bibitem [{\citenamefont {Constantoudis}\ and\ \citenamefont
  {Theodorakopoulos}(1997)}]{Constantoudis1997}%
  \BibitemOpen
  \bibfield  {author} {\bibinfo {author} {\bibfnamefont {V.}~\bibnamefont
  {Constantoudis}}\ and\ \bibinfo {author} {\bibfnamefont {N.}~\bibnamefont
  {Theodorakopoulos}},\ }\bibfield  {title} {\enquote {\bibinfo {title}
  {{Lyapunov exponent, stretching numbers, and islands of stability of the
  kicked top}},}\ }\href {\doibase 10.1103/PhysRevE.56.5189} {\bibfield
  {journal} {\bibinfo  {journal} {Physical Review E}\ }\textbf {\bibinfo
  {volume} {56}},\ \bibinfo {pages} {5189--5194} (\bibinfo {year}
  {1997})}\BibitemShut {NoStop}%
\bibitem [{\citenamefont {Benettin}\ \emph {et~al.}(1976)\citenamefont
  {Benettin}, \citenamefont {Galgani},\ and\ \citenamefont
  {Strelcyn}}]{Benettin1976}%
  \BibitemOpen
  \bibfield  {author} {\bibinfo {author} {\bibfnamefont {Giancarlo}\
  \bibnamefont {Benettin}}, \bibinfo {author} {\bibfnamefont {Luigi}\
  \bibnamefont {Galgani}}, \ and\ \bibinfo {author} {\bibfnamefont
  {Jean-Marie}\ \bibnamefont {Strelcyn}},\ }\bibfield  {title} {\enquote
  {\bibinfo {title} {{Kolmogorov entropy and numerical experiments}},}\ }\href
  {\doibase 10.1103/PhysRevA.14.2338} {\bibfield  {journal} {\bibinfo
  {journal} {Physical Review A}\ }\textbf {\bibinfo {volume} {14}},\ \bibinfo
  {pages} {2338--2345} (\bibinfo {year} {1976})}\BibitemShut {NoStop}%
\bibitem [{\citenamefont {Benettin}\ \emph {et~al.}(1980)\citenamefont
  {Benettin}, \citenamefont {Galgani}, \citenamefont {Giorgilli},\ and\
  \citenamefont {Strelcyn}}]{Benettin1980}%
  \BibitemOpen
  \bibfield  {author} {\bibinfo {author} {\bibfnamefont {Giancarlo}\
  \bibnamefont {Benettin}}, \bibinfo {author} {\bibfnamefont {Luigi}\
  \bibnamefont {Galgani}}, \bibinfo {author} {\bibfnamefont {Antonio}\
  \bibnamefont {Giorgilli}}, \ and\ \bibinfo {author} {\bibfnamefont
  {Jean-Marie}\ \bibnamefont {Strelcyn}},\ }\bibfield  {title} {\enquote
  {\bibinfo {title} {{Lyapunov Characteristic Exponents for smooth dynamical
  systems and for hamiltonian systems; a method for computing all of them. Part
  1: Theory}},}\ }\href {\doibase 10.1007/BF02128236} {\bibfield  {journal}
  {\bibinfo  {journal} {Meccanica}\ }\textbf {\bibinfo {volume} {15}},\
  \bibinfo {pages} {9--20} (\bibinfo {year} {1980})}\BibitemShut {NoStop}%
\bibitem [{\citenamefont {Geist}\ \emph {et~al.}(1990)\citenamefont {Geist},
  \citenamefont {Parlitz},\ and\ \citenamefont {Lauterborn}}]{Geist1990}%
  \BibitemOpen
  \bibfield  {author} {\bibinfo {author} {\bibfnamefont {K.}~\bibnamefont
  {Geist}}, \bibinfo {author} {\bibfnamefont {U.}~\bibnamefont {Parlitz}}, \
  and\ \bibinfo {author} {\bibfnamefont {W.}~\bibnamefont {Lauterborn}},\
  }\bibfield  {title} {\enquote {\bibinfo {title} {{Comparison of Different
  Methods for Computing Lyapunov Exponents}},}\ }\href {\doibase
  10.1143/PTP.83.875} {\bibfield  {journal} {\bibinfo  {journal} {Progress of
  Theoretical Physics}\ }\textbf {\bibinfo {volume} {83}},\ \bibinfo {pages}
  {875--893} (\bibinfo {year} {1990})}\BibitemShut {NoStop}%
\bibitem [{\citenamefont {Fortes}\ \emph {et~al.}(2019)\citenamefont {Fortes},
  \citenamefont {Garc{\'{i}}a-Mata}, \citenamefont {Jalabert},\ and\
  \citenamefont {Wisniacki}}]{Fortes2019}%
  \BibitemOpen
  \bibfield  {author} {\bibinfo {author} {\bibfnamefont {Emiliano~M.}\
  \bibnamefont {Fortes}}, \bibinfo {author} {\bibfnamefont {Ignacio}\
  \bibnamefont {Garc{\'{i}}a-Mata}}, \bibinfo {author} {\bibfnamefont
  {Rodolfo~A.}\ \bibnamefont {Jalabert}}, \ and\ \bibinfo {author}
  {\bibfnamefont {Diego~A.}\ \bibnamefont {Wisniacki}},\ }\bibfield  {title}
  {\enquote {\bibinfo {title} {{Gauging classical and quantum integrability
  through out-of-time-ordered correlators}},}\ }\href {\doibase
  10.1103/PhysRevE.100.042201} {\bibfield  {journal} {\bibinfo  {journal}
  {Physical Review E}\ }\textbf {\bibinfo {volume} {100}},\ \bibinfo {pages}
  {042201} (\bibinfo {year} {2019})}\BibitemShut {NoStop}%
\bibitem [{\citenamefont {Anishchenko}\ and\ \citenamefont
  {Astakhov}(2013)}]{Anishchenko2013}%
  \BibitemOpen
  \bibfield  {author} {\bibinfo {author} {\bibfnamefont {V~S}\ \bibnamefont
  {Anishchenko}}\ and\ \bibinfo {author} {\bibfnamefont {S~V}\ \bibnamefont
  {Astakhov}},\ }\bibfield  {title} {\enquote {\bibinfo {title}
  {{Poincar{\'{e}} recurrence theory and its applications to nonlinear
  physics}},}\ }\href {\doibase 10.3367/UFNe.0183.201310a.1009} {\bibfield
  {journal} {\bibinfo  {journal} {Physics-Uspekhi}\ }\textbf {\bibinfo {volume}
  {56}},\ \bibinfo {pages} {955--972} (\bibinfo {year} {2013})}\BibitemShut
  {NoStop}%
\bibitem [{\citenamefont {Haake}(2001)}]{Haake2001a}%
  \BibitemOpen
  \bibfield  {author} {\bibinfo {author} {\bibfnamefont {Fritz.}\ \bibnamefont
  {Haake}},\ }\href
  {https://books.google.com/books/about/Quantum{\_}Signatures{\_}of{\_}Chaos.html?id=Orv0BXoorFEC}
  {\emph {\bibinfo {title} {{Quantum signatures of chaos}}}}\ (\bibinfo
  {publisher} {Springer},\ \bibinfo {year} {2001})\ p.\ \bibinfo {pages}
  {479}\BibitemShut {NoStop}%
\bibitem [{\citenamefont {Emerson}\ \emph {et~al.}(2002)\citenamefont
  {Emerson}, \citenamefont {Weinstein}, \citenamefont {Lloyd},\ and\
  \citenamefont {Cory}}]{Emerson2002}%
  \BibitemOpen
  \bibfield  {author} {\bibinfo {author} {\bibfnamefont {Joseph}\ \bibnamefont
  {Emerson}}, \bibinfo {author} {\bibfnamefont {Yaakov~S.}\ \bibnamefont
  {Weinstein}}, \bibinfo {author} {\bibfnamefont {Seth}\ \bibnamefont {Lloyd}},
  \ and\ \bibinfo {author} {\bibfnamefont {D.~G.}\ \bibnamefont {Cory}},\
  }\bibfield  {title} {\enquote {\bibinfo {title} {{Fidelity Decay as an
  Efficient Indicator of Quantum Chaos}},}\ }\href {\doibase
  10.1103/PhysRevLett.89.284102} {\bibfield  {journal} {\bibinfo  {journal}
  {Physical Review Letters}\ }\textbf {\bibinfo {volume} {89}},\ \bibinfo
  {pages} {284102} (\bibinfo {year} {2002})}\BibitemShut {NoStop}%
\bibitem [{\citenamefont {Srednicki}(1999)}]{Srednicki1999}%
  \BibitemOpen
  \bibfield  {author} {\bibinfo {author} {\bibfnamefont {Mark}\ \bibnamefont
  {Srednicki}},\ }\bibfield  {title} {\enquote {\bibinfo {title} {{The approach
  to thermal equilibrium in quantized chaotic systems}},}\ }\href {\doibase
  10.1088/0305-4470/32/7/007} {\bibfield  {journal} {\bibinfo  {journal}
  {Journal of Physics A: Mathematical and General}\ }\textbf {\bibinfo {volume}
  {32}},\ \bibinfo {pages} {1163--1175} (\bibinfo {year} {1999})}\BibitemShut
  {NoStop}%
\bibitem [{\citenamefont {Torres-Herrera}\ and\ \citenamefont
  {Santos}(2017)}]{Torres-Herrera2017}%
  \BibitemOpen
  \bibfield  {author} {\bibinfo {author} {\bibfnamefont {E~J}\ \bibnamefont
  {Torres-Herrera}}\ and\ \bibinfo {author} {\bibfnamefont {Lea~F}\
  \bibnamefont {Santos}},\ }\bibfield  {title} {\enquote {\bibinfo {title}
  {{Dynamical manifestations of quantum chaos: correlation hole and bulge.}}}\
  }\href {\doibase 10.1098/rsta.2016.0434} {\bibfield  {journal} {\bibinfo
  {journal} {Philosophical transactions. Series A, Mathematical, physical, and
  engineering sciences}\ }\textbf {\bibinfo {volume} {375}} (\bibinfo {year}
  {2017}),\ 10.1098/rsta.2016.0434}\BibitemShut {NoStop}%
\bibitem [{\citenamefont {Zhuang}\ and\ \citenamefont {Wu}(2013)}]{Zhuang2013}%
  \BibitemOpen
  \bibfield  {author} {\bibinfo {author} {\bibfnamefont {Quntao}\ \bibnamefont
  {Zhuang}}\ and\ \bibinfo {author} {\bibfnamefont {Biao}\ \bibnamefont {Wu}},\
  }\bibfield  {title} {\enquote {\bibinfo {title} {{Equilibration of quantum
  chaotic systems}},}\ }\href {\doibase 10.1103/PhysRevE.88.062147} {\bibfield
  {journal} {\bibinfo  {journal} {Physical Review E}\ }\textbf {\bibinfo
  {volume} {88}},\ \bibinfo {pages} {062147} (\bibinfo {year}
  {2013})}\BibitemShut {NoStop}%
\bibitem [{\citenamefont {Zurek}\ and\ \citenamefont {Paz}(1995)}]{Zurek1995}%
  \BibitemOpen
  \bibfield  {author} {\bibinfo {author} {\bibfnamefont {Wojciech~Hubert}\
  \bibnamefont {Zurek}}\ and\ \bibinfo {author} {\bibfnamefont {Juan~Pablo}\
  \bibnamefont {Paz}},\ }\bibfield  {title} {\enquote {\bibinfo {title}
  {{Quantum chaos: a decoherent definition}},}\ }\href {\doibase
  10.1016/0167-2789(94)00271-Q} {\bibfield  {journal} {\bibinfo  {journal}
  {Physica D: Nonlinear Phenomena}\ }\textbf {\bibinfo {volume} {83}},\
  \bibinfo {pages} {300--308} (\bibinfo {year} {1995})}\BibitemShut {NoStop}%
\bibitem [{\citenamefont {Wang}\ \emph {et~al.}(2004)\citenamefont {Wang},
  \citenamefont {Ghose}, \citenamefont {Sanders},\ and\ \citenamefont
  {Hu}}]{Wang2004}%
  \BibitemOpen
  \bibfield  {author} {\bibinfo {author} {\bibfnamefont {Xiaoguang}\
  \bibnamefont {Wang}}, \bibinfo {author} {\bibfnamefont {Shohini}\
  \bibnamefont {Ghose}}, \bibinfo {author} {\bibfnamefont {Barry~C.}\
  \bibnamefont {Sanders}}, \ and\ \bibinfo {author} {\bibfnamefont {Bambi}\
  \bibnamefont {Hu}},\ }\bibfield  {title} {\enquote {\bibinfo {title}
  {Entanglement as a signature of quantum chaos},}\ }\href {\doibase
  10.1103/PhysRevE.70.016217} {\bibfield  {journal} {\bibinfo  {journal} {Phys.
  Rev. E}\ }\textbf {\bibinfo {volume} {70}},\ \bibinfo {pages} {016217}
  (\bibinfo {year} {2004})}\BibitemShut {NoStop}%
\bibitem [{\citenamefont {Lakshminarayan}(2001)}]{Arul2001}%
  \BibitemOpen
  \bibfield  {author} {\bibinfo {author} {\bibfnamefont {Arul}\ \bibnamefont
  {Lakshminarayan}},\ }\bibfield  {title} {\enquote {\bibinfo {title}
  {Entangling power of quantized chaotic systems},}\ }\href {\doibase
  10.1103/PhysRevE.64.036207} {\bibfield  {journal} {\bibinfo  {journal} {Phys.
  Rev. E}\ }\textbf {\bibinfo {volume} {64}},\ \bibinfo {pages} {036207}
  (\bibinfo {year} {2001})}\BibitemShut {NoStop}%
\bibitem [{\citenamefont {Kubotani}\ \emph {et~al.}(2006)\citenamefont
  {Kubotani}, \citenamefont {Toda},\ and\ \citenamefont
  {Adachi}}]{Kubotani2006}%
  \BibitemOpen
  \bibfield  {author} {\bibinfo {author} {\bibfnamefont {Hiroto}\ \bibnamefont
  {Kubotani}}, \bibinfo {author} {\bibfnamefont {Mikito}\ \bibnamefont {Toda}},
  \ and\ \bibinfo {author} {\bibfnamefont {Satoshi}\ \bibnamefont {Adachi}},\
  }\bibfield  {title} {\enquote {\bibinfo {title} {Universality in dynamical
  formation of entanglement for quantum chaos},}\ }\href {\doibase
  10.1103/PhysRevA.74.032314} {\bibfield  {journal} {\bibinfo  {journal} {Phys.
  Rev. A}\ }\textbf {\bibinfo {volume} {74}},\ \bibinfo {pages} {032314}
  (\bibinfo {year} {2006})}\BibitemShut {NoStop}%
\bibitem [{\citenamefont {Schack}\ and\ \citenamefont
  {Caves}(1996)}]{Schack1996}%
  \BibitemOpen
  \bibfield  {author} {\bibinfo {author} {\bibfnamefont {R{\"{u}}diger}\
  \bibnamefont {Schack}}\ and\ \bibinfo {author} {\bibfnamefont {Carlton~M.}\
  \bibnamefont {Caves}},\ }\bibfield  {title} {\enquote {\bibinfo {title}
  {{Information-theoretic characterization of quantum chaos}},}\ }\href
  {\doibase 10.1103/PhysRevE.53.3257} {\bibfield  {journal} {\bibinfo
  {journal} {Physical Review E}\ }\textbf {\bibinfo {volume} {53}},\ \bibinfo
  {pages} {3257--3270} (\bibinfo {year} {1996})}\BibitemShut {NoStop}%
\bibitem [{\citenamefont {Seshadri}\ \emph {et~al.}(2018)\citenamefont
  {Seshadri}, \citenamefont {Madhok},\ and\ \citenamefont
  {Lakshminarayan}}]{Seshadri2018}%
  \BibitemOpen
  \bibfield  {author} {\bibinfo {author} {\bibfnamefont {Akshay}\ \bibnamefont
  {Seshadri}}, \bibinfo {author} {\bibfnamefont {Vaibhav}\ \bibnamefont
  {Madhok}}, \ and\ \bibinfo {author} {\bibfnamefont {Arul}\ \bibnamefont
  {Lakshminarayan}},\ }\bibfield  {title} {\enquote {\bibinfo {title}
  {{Tripartite mutual information, entanglement, and scrambling in permutation
  symmetric systems with an application to quantum chaos}},}\ }\href {\doibase
  10.1103/PhysRevE.98.052205} {\bibfield  {journal} {\bibinfo  {journal}
  {Physical Review E}\ }\textbf {\bibinfo {volume} {98}},\ \bibinfo {pages}
  {052205} (\bibinfo {year} {2018})}\BibitemShut {NoStop}%
\bibitem [{\citenamefont {Madhok}\ \emph {et~al.}(2016)\citenamefont {Madhok},
  \citenamefont {Riofr{\'{i}}o},\ and\ \citenamefont {Deutsch}}]{Madhok2016}%
  \BibitemOpen
  \bibfield  {author} {\bibinfo {author} {\bibfnamefont {Vaibhav}\ \bibnamefont
  {Madhok}}, \bibinfo {author} {\bibfnamefont {Carlos~A}\ \bibnamefont
  {Riofr{\'{i}}o}}, \ and\ \bibinfo {author} {\bibfnamefont {Ivan~H}\
  \bibnamefont {Deutsch}},\ }\bibfield  {title} {\enquote {\bibinfo {title}
  {{Review: Characterizing and quantifying quantum chaos with quantum
  tomography}},}\ }\href {\doibase 10.1007/s12043-016-1259-x} {\bibfield
  {journal} {\bibinfo  {journal} {Pramana}\ }\textbf {\bibinfo {volume} {87}},\
  \bibinfo {pages} {65} (\bibinfo {year} {2016})}\BibitemShut {NoStop}%
\bibitem [{\citenamefont {Madhok}\ \emph {et~al.}(2014)\citenamefont {Madhok},
  \citenamefont {Riofr{\'{i}}o}, \citenamefont {Ghose},\ and\ \citenamefont
  {Deutsch}}]{Madhok2014}%
  \BibitemOpen
  \bibfield  {author} {\bibinfo {author} {\bibfnamefont {Vaibhav}\ \bibnamefont
  {Madhok}}, \bibinfo {author} {\bibfnamefont {Carlos~A.}\ \bibnamefont
  {Riofr{\'{i}}o}}, \bibinfo {author} {\bibfnamefont {Shohini}\ \bibnamefont
  {Ghose}}, \ and\ \bibinfo {author} {\bibfnamefont {Ivan~H.}\ \bibnamefont
  {Deutsch}},\ }\bibfield  {title} {\enquote {\bibinfo {title} {{Information
  Gain in Tomography–A Quantum Signature of Chaos}},}\ }\href {\doibase
  10.1103/PhysRevLett.112.014102} {\bibfield  {journal} {\bibinfo  {journal}
  {Physical Review Letters}\ }\textbf {\bibinfo {volume} {112}},\ \bibinfo
  {pages} {014102} (\bibinfo {year} {2014})}\BibitemShut {NoStop}%
\bibitem [{\citenamefont {Larkin}\ and\ \citenamefont
  {Ovchinnikov}(1969)}]{Larkin1969}%
  \BibitemOpen
  \bibfield  {author} {\bibinfo {author} {\bibfnamefont {A.~I.}\ \bibnamefont
  {Larkin}}\ and\ \bibinfo {author} {\bibfnamefont {Yu.~N.}\ \bibnamefont
  {Ovchinnikov}},\ }\bibfield  {title} {\enquote {\bibinfo {title}
  {{Quasiclassical Method in the Theory of Superconductivity}},}\ }\href
  {https://ui.adsabs.harvard.edu/abs/1969JETP...28.1200L/abstract} {\bibfield
  {journal} {\bibinfo  {journal} {Journal of Experimental and Theoretical
  Physics}\ }\textbf {\bibinfo {volume} {28}},\ \bibinfo {pages} {1200}
  (\bibinfo {year} {1969})}\BibitemShut {NoStop}%
\bibitem [{\citenamefont {Maldacena}\ \emph {et~al.}(2016)\citenamefont
  {Maldacena}, \citenamefont {Shenker},\ and\ \citenamefont
  {Stanford}}]{Maldacena2016}%
  \BibitemOpen
  \bibfield  {author} {\bibinfo {author} {\bibfnamefont {Juan}\ \bibnamefont
  {Maldacena}}, \bibinfo {author} {\bibfnamefont {Stephen~H.}\ \bibnamefont
  {Shenker}}, \ and\ \bibinfo {author} {\bibfnamefont {Douglas}\ \bibnamefont
  {Stanford}},\ }\bibfield  {title} {\enquote {\bibinfo {title} {{A bound on
  chaos}},}\ }\href {\doibase 10.1007/JHEP08(2016)106} {\bibfield  {journal}
  {\bibinfo  {journal} {Journal of High Energy Physics}\ }\textbf {\bibinfo
  {volume} {2016}},\ \bibinfo {pages} {106} (\bibinfo {year}
  {2016})}\BibitemShut {NoStop}%
\bibitem [{\citenamefont {Sachdev}\ and\ \citenamefont
  {Ye}(1993)}]{Sachdev1993}%
  \BibitemOpen
  \bibfield  {author} {\bibinfo {author} {\bibfnamefont {Subir}\ \bibnamefont
  {Sachdev}}\ and\ \bibinfo {author} {\bibfnamefont {Jinwu}\ \bibnamefont
  {Ye}},\ }\bibfield  {title} {\enquote {\bibinfo {title} {{Gapless spin-fluid
  ground state in a random quantum Heisenberg magnet}},}\ }\href {\doibase
  10.1103/PhysRevLett.70.3339} {\bibfield  {journal} {\bibinfo  {journal}
  {Physical Review Letters}\ }\textbf {\bibinfo {volume} {70}},\ \bibinfo
  {pages} {3339--3342} (\bibinfo {year} {1993})}\BibitemShut {NoStop}%
\bibitem [{\citenamefont {Swingle}(2018)}]{Swingle2018}%
  \BibitemOpen
  \bibfield  {author} {\bibinfo {author} {\bibfnamefont {Brian}\ \bibnamefont
  {Swingle}},\ }\bibfield  {title} {\enquote {\bibinfo {title} {Unscrambling
  the physics of out-of-time-order correlators},}\ }\href@noop {} {\bibfield
  {journal} {\bibinfo  {journal} {Nature Physics}\ }\textbf {\bibinfo {volume}
  {14}},\ \bibinfo {pages} {988--990} (\bibinfo {year} {2018})}\BibitemShut
  {NoStop}%
\bibitem [{\citenamefont {Riddell}\ and\ \citenamefont
  {S{\o}rensen}(2019)}]{Riddell2019}%
  \BibitemOpen
  \bibfield  {author} {\bibinfo {author} {\bibfnamefont {Jonathon}\
  \bibnamefont {Riddell}}\ and\ \bibinfo {author} {\bibfnamefont {Erik~S.}\
  \bibnamefont {S{\o}rensen}},\ }\bibfield  {title} {\enquote {\bibinfo {title}
  {{Out-of-time ordered correlators and entanglement growth in the random-field
  XX spin chain}},}\ }\href {\doibase 10.1103/PhysRevB.99.054205} {\bibfield
  {journal} {\bibinfo  {journal} {Physical Review B}\ }\textbf {\bibinfo
  {volume} {99}},\ \bibinfo {pages} {054205} (\bibinfo {year}
  {2019})}\BibitemShut {NoStop}%
\bibitem [{\citenamefont {Landsman}\ \emph {et~al.}(2019)\citenamefont
  {Landsman}, \citenamefont {Figgatt}, \citenamefont {Schuster}, \citenamefont
  {Linke}, \citenamefont {Yoshida}, \citenamefont {Yao},\ and\ \citenamefont
  {Monroe}}]{Landsman2019}%
  \BibitemOpen
  \bibfield  {author} {\bibinfo {author} {\bibfnamefont {K.~A.}\ \bibnamefont
  {Landsman}}, \bibinfo {author} {\bibfnamefont {C.}~\bibnamefont {Figgatt}},
  \bibinfo {author} {\bibfnamefont {T.}~\bibnamefont {Schuster}}, \bibinfo
  {author} {\bibfnamefont {N.~M.}\ \bibnamefont {Linke}}, \bibinfo {author}
  {\bibfnamefont {B.}~\bibnamefont {Yoshida}}, \bibinfo {author} {\bibfnamefont
  {N.~Y.}\ \bibnamefont {Yao}}, \ and\ \bibinfo {author} {\bibfnamefont
  {C.}~\bibnamefont {Monroe}},\ }\bibfield  {title} {\enquote {\bibinfo {title}
  {{Verified quantum information scrambling}},}\ }\href {\doibase
  10.1038/s41586-019-0952-6} {\bibfield  {journal} {\bibinfo  {journal}
  {Nature}\ }\textbf {\bibinfo {volume} {567}},\ \bibinfo {pages} {61--65}
  (\bibinfo {year} {2019})}\BibitemShut {NoStop}%
\bibitem [{\citenamefont {Pappalardi}\ \emph {et~al.}(2018)\citenamefont
  {Pappalardi}, \citenamefont {Russomanno}, \citenamefont
  {{\v{Z}}unkovi{\v{c}}}, \citenamefont {Iemini}, \citenamefont {Silva},\ and\
  \citenamefont {Fazio}}]{Pappalardi2018}%
  \BibitemOpen
  \bibfield  {author} {\bibinfo {author} {\bibfnamefont {Silvia}\ \bibnamefont
  {Pappalardi}}, \bibinfo {author} {\bibfnamefont {Angelo}\ \bibnamefont
  {Russomanno}}, \bibinfo {author} {\bibfnamefont {Bojan}\ \bibnamefont
  {{\v{Z}}unkovi{\v{c}}}}, \bibinfo {author} {\bibfnamefont {Fernando}\
  \bibnamefont {Iemini}}, \bibinfo {author} {\bibfnamefont {Alessandro}\
  \bibnamefont {Silva}}, \ and\ \bibinfo {author} {\bibfnamefont {Rosario}\
  \bibnamefont {Fazio}},\ }\bibfield  {title} {\enquote {\bibinfo {title}
  {{Scrambling and entanglement spreading in long-range spin chains}},}\ }\href
  {\doibase 10.1103/PhysRevB.98.134303} {\bibfield  {journal} {\bibinfo
  {journal} {Physical Review B}\ }\textbf {\bibinfo {volume} {98}},\ \bibinfo
  {pages} {134303} (\bibinfo {year} {2018})}\BibitemShut {NoStop}%
\bibitem [{\citenamefont {Stockmann}(2006)}]{Stockmann2006a}%
  \BibitemOpen
  \bibfield  {author} {\bibinfo {author} {\bibfnamefont {Hans-Jurgen}\
  \bibnamefont {Stockmann}},\ }\href@noop {} {\emph {\bibinfo {title} {{Quantum
  chaos}}}}\ (\bibinfo  {publisher} {Cambridge University Press},\ \bibinfo
  {year} {2006})\ p.\ \bibinfo {pages} {368}\BibitemShut {NoStop}%
\bibitem [{\citenamefont {Berry}\ and\ \citenamefont
  {Tabor}(1977)}]{Berry1977}%
  \BibitemOpen
  \bibfield  {author} {\bibinfo {author} {\bibfnamefont {M.~V.}\ \bibnamefont
  {Berry}}\ and\ \bibinfo {author} {\bibfnamefont {M.}~\bibnamefont {Tabor}},\
  }\bibfield  {title} {\enquote {\bibinfo {title} {{Level Clustering in the
  Regular Spectrum}},}\ }\href {\doibase 10.1098/rspa.1977.0140} {\bibfield
  {journal} {\bibinfo  {journal} {Proceedings of the Royal Society of London.
  Series A: Mathematical and Physical Sciences}\ }\textbf {\bibinfo {volume}
  {356}},\ \bibinfo {pages} {375--394} (\bibinfo {year} {1977})}\BibitemShut
  {NoStop}%
\bibitem [{\citenamefont {Bohigas}\ \emph {et~al.}(1984)\citenamefont
  {Bohigas}, \citenamefont {Giannoni},\ and\ \citenamefont
  {Schmit}}]{Bohigas1984}%
  \BibitemOpen
  \bibfield  {author} {\bibinfo {author} {\bibfnamefont {O.}~\bibnamefont
  {Bohigas}}, \bibinfo {author} {\bibfnamefont {M.~J.}\ \bibnamefont
  {Giannoni}}, \ and\ \bibinfo {author} {\bibfnamefont {C.}~\bibnamefont
  {Schmit}},\ }\bibfield  {title} {\enquote {\bibinfo {title}
  {{Characterization of Chaotic Quantum Spectra and Universality of Level
  Fluctuation Laws}},}\ }\href {\doibase 10.1103/PhysRevLett.52.1} {\bibfield
  {journal} {\bibinfo  {journal} {Physical Review Letters}\ }\textbf {\bibinfo
  {volume} {52}},\ \bibinfo {pages} {1--4} (\bibinfo {year}
  {1984})}\BibitemShut {NoStop}%
\bibitem [{\citenamefont {Oganesyan}\ and\ \citenamefont
  {Huse}(2007)}]{Oganesyan2007}%
  \BibitemOpen
  \bibfield  {author} {\bibinfo {author} {\bibfnamefont {Vadim}\ \bibnamefont
  {Oganesyan}}\ and\ \bibinfo {author} {\bibfnamefont {David~A.}\ \bibnamefont
  {Huse}},\ }\bibfield  {title} {\enquote {\bibinfo {title} {{Localization of
  interacting fermions at high temperature}},}\ }\href {\doibase
  10.1103/PhysRevB.75.155111} {\bibfield  {journal} {\bibinfo  {journal}
  {Physical Review B}\ }\textbf {\bibinfo {volume} {75}},\ \bibinfo {pages}
  {155111} (\bibinfo {year} {2007})}\BibitemShut {NoStop}%
\bibitem [{\citenamefont {Atas}\ \emph {et~al.}(2013)\citenamefont {Atas},
  \citenamefont {Bogomolny}, \citenamefont {Giraud},\ and\ \citenamefont
  {Roux}}]{Atas2013}%
  \BibitemOpen
  \bibfield  {author} {\bibinfo {author} {\bibfnamefont {Y.~Y.}\ \bibnamefont
  {Atas}}, \bibinfo {author} {\bibfnamefont {E.}~\bibnamefont {Bogomolny}},
  \bibinfo {author} {\bibfnamefont {O.}~\bibnamefont {Giraud}}, \ and\ \bibinfo
  {author} {\bibfnamefont {G.}~\bibnamefont {Roux}},\ }\bibfield  {title}
  {\enquote {\bibinfo {title} {{Distribution of the Ratio of Consecutive Level
  Spacings in Random Matrix Ensembles}},}\ }\href {\doibase
  10.1103/PhysRevLett.110.084101} {\bibfield  {journal} {\bibinfo  {journal}
  {Physical Review Letters}\ }\textbf {\bibinfo {volume} {110}},\ \bibinfo
  {pages} {084101} (\bibinfo {year} {2013})}\BibitemShut {NoStop}%
\bibitem [{Note7()}]{Note7}%
  \BibitemOpen
  \bibinfo {note} {This was expected, since in the semiclassical limit $N_s\gg
  1$ Floquet states will have a strong correspondence with classical phase
  space trajectories, as can be seen by looking at their phase space
  representation (for instance using the Husimi $Q$-function) and therefore
  they will inherit properties of the classical system, that we observe in the
  kinematic signatures.}\BibitemShut {Stop}%
\bibitem [{\citenamefont {Jalabert}\ \emph {et~al.}(2018)\citenamefont
  {Jalabert}, \citenamefont {Garc{\'{i}}a-Mata},\ and\ \citenamefont
  {Wisniacki}}]{Jalabert2018}%
  \BibitemOpen
  \bibfield  {author} {\bibinfo {author} {\bibfnamefont {Rodolfo~A.}\
  \bibnamefont {Jalabert}}, \bibinfo {author} {\bibfnamefont {Ignacio}\
  \bibnamefont {Garc{\'{i}}a-Mata}}, \ and\ \bibinfo {author} {\bibfnamefont
  {Diego~A.}\ \bibnamefont {Wisniacki}},\ }\bibfield  {title} {\enquote
  {\bibinfo {title} {{Semiclassical theory of out-of-time-order correlators for
  low-dimensional classically chaotic systems}},}\ }\href {\doibase
  10.1103/PhysRevE.98.062218} {\bibfield  {journal} {\bibinfo  {journal}
  {Physical Review E}\ }\textbf {\bibinfo {volume} {98}},\ \bibinfo {pages}
  {062218} (\bibinfo {year} {2018})}\BibitemShut {NoStop}%
\bibitem [{\citenamefont {Hashimoto}\ \emph {et~al.}(2017)\citenamefont
  {Hashimoto}, \citenamefont {Murata},\ and\ \citenamefont
  {Yoshii}}]{Hashimoto2017}%
  \BibitemOpen
  \bibfield  {author} {\bibinfo {author} {\bibfnamefont {Koji}\ \bibnamefont
  {Hashimoto}}, \bibinfo {author} {\bibfnamefont {Keiju}\ \bibnamefont
  {Murata}}, \ and\ \bibinfo {author} {\bibfnamefont {Ryosuke}\ \bibnamefont
  {Yoshii}},\ }\bibfield  {title} {\enquote {\bibinfo {title}
  {{Out-of-time-order correlators in quantum mechanics}},}\ }\href {\doibase
  10.1007/JHEP10(2017)138} {\bibfield  {journal} {\bibinfo  {journal} {Journal
  of High Energy Physics}\ }\textbf {\bibinfo {volume} {2017}},\ \bibinfo
  {pages} {138} (\bibinfo {year} {2017})}\BibitemShut {NoStop}%
\bibitem [{\citenamefont {D{\'{o}}ra}\ and\ \citenamefont
  {Moessner}(2017)}]{Dora2017}%
  \BibitemOpen
  \bibfield  {author} {\bibinfo {author} {\bibfnamefont {Bal{\'{a}}zs}\
  \bibnamefont {D{\'{o}}ra}}\ and\ \bibinfo {author} {\bibfnamefont {Roderich}\
  \bibnamefont {Moessner}},\ }\bibfield  {title} {\enquote {\bibinfo {title}
  {{Out-of-Time-Ordered Density Correlators in Luttinger Liquids}},}\ }\href
  {\doibase 10.1103/PhysRevLett.119.026802} {\bibfield  {journal} {\bibinfo
  {journal} {Physical Review Letters}\ }\textbf {\bibinfo {volume} {119}},\
  \bibinfo {pages} {026802} (\bibinfo {year} {2017})}\BibitemShut {NoStop}%
\bibitem [{\citenamefont {Kukuljan}\ \emph {et~al.}(2017)\citenamefont
  {Kukuljan}, \citenamefont {Grozdanov},\ and\ \citenamefont
  {Prosen}}]{Kukuljan2017}%
  \BibitemOpen
  \bibfield  {author} {\bibinfo {author} {\bibfnamefont {Ivan}\ \bibnamefont
  {Kukuljan}}, \bibinfo {author} {\bibfnamefont {Sa{\v{s}}o}\ \bibnamefont
  {Grozdanov}}, \ and\ \bibinfo {author} {\bibfnamefont {Toma{\v{z}}}\
  \bibnamefont {Prosen}},\ }\bibfield  {title} {\enquote {\bibinfo {title}
  {{Weak quantum chaos}},}\ }\href {\doibase 10.1103/PhysRevB.96.060301}
  {\bibfield  {journal} {\bibinfo  {journal} {Physical Review B}\ }\textbf
  {\bibinfo {volume} {96}},\ \bibinfo {pages} {060301} (\bibinfo {year}
  {2017})}\BibitemShut {NoStop}%
\bibitem [{\citenamefont {Rozenbaum}\ \emph {et~al.}(2017)\citenamefont
  {Rozenbaum}, \citenamefont {Ganeshan},\ and\ \citenamefont
  {Galitski}}]{Rozenbaum2017}%
  \BibitemOpen
  \bibfield  {author} {\bibinfo {author} {\bibfnamefont {Efim~B.}\ \bibnamefont
  {Rozenbaum}}, \bibinfo {author} {\bibfnamefont {Sriram}\ \bibnamefont
  {Ganeshan}}, \ and\ \bibinfo {author} {\bibfnamefont {Victor}\ \bibnamefont
  {Galitski}},\ }\bibfield  {title} {\enquote {\bibinfo {title} {{Lyapunov
  Exponent and Out-of-Time-Ordered Correlator's Growth Rate in a Chaotic
  System}},}\ }\href {\doibase 10.1103/PhysRevLett.118.086801} {\bibfield
  {journal} {\bibinfo  {journal} {Physical Review Letters}\ }\textbf {\bibinfo
  {volume} {118}},\ \bibinfo {pages} {086801} (\bibinfo {year}
  {2017})}\BibitemShut {NoStop}%
\bibitem [{\citenamefont {Garc{\'{i}}a-Mata}\ \emph {et~al.}(2018)\citenamefont
  {Garc{\'{i}}a-Mata}, \citenamefont {Saraceno}, \citenamefont {Jalabert},
  \citenamefont {Roncaglia},\ and\ \citenamefont
  {Wisniacki}}]{Garcia-Mata2018}%
  \BibitemOpen
  \bibfield  {author} {\bibinfo {author} {\bibfnamefont {Ignacio}\ \bibnamefont
  {Garc{\'{i}}a-Mata}}, \bibinfo {author} {\bibfnamefont {Marcos}\ \bibnamefont
  {Saraceno}}, \bibinfo {author} {\bibfnamefont {Rodolfo~A.}\ \bibnamefont
  {Jalabert}}, \bibinfo {author} {\bibfnamefont {Augusto~J.}\ \bibnamefont
  {Roncaglia}}, \ and\ \bibinfo {author} {\bibfnamefont {Diego~A.}\
  \bibnamefont {Wisniacki}},\ }\bibfield  {title} {\enquote {\bibinfo {title}
  {{Chaos Signatures in the Short and Long Time Behavior of the Out-of-Time
  Ordered Correlator}},}\ }\href {\doibase 10.1103/PhysRevLett.121.210601}
  {\bibfield  {journal} {\bibinfo  {journal} {Physical Review Letters}\
  }\textbf {\bibinfo {volume} {121}},\ \bibinfo {pages} {210601} (\bibinfo
  {year} {2018})}\BibitemShut {NoStop}%
\bibitem [{\citenamefont {Ch{\'{a}}vez-Carlos}\ \emph
  {et~al.}(2019)\citenamefont {Ch{\'{a}}vez-Carlos}, \citenamefont
  {L{\'{o}}pez-del Carpio}, \citenamefont {Bastarrachea-Magnani}, \citenamefont
  {Str{\'{a}}nsk{\'{y}}}, \citenamefont {Lerma-Hern{\'{a}}ndez}, \citenamefont
  {Santos},\ and\ \citenamefont {Hirsch}}]{Chavez-Carlos2019}%
  \BibitemOpen
  \bibfield  {author} {\bibinfo {author} {\bibfnamefont {Jorge}\ \bibnamefont
  {Ch{\'{a}}vez-Carlos}}, \bibinfo {author} {\bibfnamefont {B.}~\bibnamefont
  {L{\'{o}}pez-del Carpio}}, \bibinfo {author} {\bibfnamefont {Miguel~A.}\
  \bibnamefont {Bastarrachea-Magnani}}, \bibinfo {author} {\bibfnamefont
  {Pavel}\ \bibnamefont {Str{\'{a}}nsk{\'{y}}}}, \bibinfo {author}
  {\bibfnamefont {Sergio}\ \bibnamefont {Lerma-Hern{\'{a}}ndez}}, \bibinfo
  {author} {\bibfnamefont {Lea~F.}\ \bibnamefont {Santos}}, \ and\ \bibinfo
  {author} {\bibfnamefont {Jorge~G.}\ \bibnamefont {Hirsch}},\ }\bibfield
  {title} {\enquote {\bibinfo {title} {{Quantum and Classical Lyapunov
  Exponents in Atom-Field Interaction Systems}},}\ }\href {\doibase
  10.1103/PhysRevLett.122.024101} {\bibfield  {journal} {\bibinfo  {journal}
  {Physical Review Letters}\ }\textbf {\bibinfo {volume} {122}},\ \bibinfo
  {pages} {024101} (\bibinfo {year} {2019})}\BibitemShut {NoStop}%
\bibitem [{\citenamefont {Lerose}\ and\ \citenamefont
  {Pappalardi}(2020)}]{Lerose2020}%
  \BibitemOpen
  \bibfield  {author} {\bibinfo {author} {\bibfnamefont {Alessio}\ \bibnamefont
  {Lerose}}\ and\ \bibinfo {author} {\bibfnamefont {Silvia}\ \bibnamefont
  {Pappalardi}},\ }\bibfield  {title} {\enquote {\bibinfo {title} {Bridging
  entanglement dynamics and chaos in semiclassical systems},}\ }\href {\doibase
  10.1103/PhysRevA.102.032404} {\bibfield  {journal} {\bibinfo  {journal}
  {Phys. Rev. A}\ }\textbf {\bibinfo {volume} {102}},\ \bibinfo {pages}
  {032404} (\bibinfo {year} {2020})}\BibitemShut {NoStop}%
\bibitem [{\citenamefont {Yan}\ and\ \citenamefont
  {Chemissany}(2020)}]{Yan2020}%
  \BibitemOpen
  \bibfield  {author} {\bibinfo {author} {\bibfnamefont {Bin}\ \bibnamefont
  {Yan}}\ and\ \bibinfo {author} {\bibfnamefont {Wissam}\ \bibnamefont
  {Chemissany}},\ }\bibfield  {title} {\enquote {\bibinfo {title} {Quantum
  chaos on complexity geometry},}\ }\href@noop {} {\bibfield  {journal}
  {\bibinfo  {journal} {arXiv preprint arXiv:2004.03501}\ } (\bibinfo {year}
  {2020})}\BibitemShut {NoStop}%
\bibitem [{\citenamefont {Xu}\ \emph {et~al.}(2020)\citenamefont {Xu},
  \citenamefont {Scaffidi},\ and\ \citenamefont {Cao}}]{Xu2020}%
  \BibitemOpen
  \bibfield  {author} {\bibinfo {author} {\bibfnamefont {Tianrui}\ \bibnamefont
  {Xu}}, \bibinfo {author} {\bibfnamefont {Thomas}\ \bibnamefont {Scaffidi}}, \
  and\ \bibinfo {author} {\bibfnamefont {Xiangyu}\ \bibnamefont {Cao}},\
  }\bibfield  {title} {\enquote {\bibinfo {title} {Does scrambling equal
  chaos?}}\ }\href {\doibase 10.1103/PhysRevLett.124.140602} {\bibfield
  {journal} {\bibinfo  {journal} {Phys. Rev. Lett.}\ }\textbf {\bibinfo
  {volume} {124}},\ \bibinfo {pages} {140602} (\bibinfo {year}
  {2020})}\BibitemShut {NoStop}%
\bibitem [{\citenamefont {Kidd}\ \emph {et~al.}(2021)\citenamefont {Kidd},
  \citenamefont {Safavi-Naini},\ and\ \citenamefont {Corney}}]{Kidd2021}%
  \BibitemOpen
  \bibfield  {author} {\bibinfo {author} {\bibfnamefont {R.~A.}\ \bibnamefont
  {Kidd}}, \bibinfo {author} {\bibfnamefont {A.}~\bibnamefont {Safavi-Naini}},
  \ and\ \bibinfo {author} {\bibfnamefont {J.~F.}\ \bibnamefont {Corney}},\
  }\bibfield  {title} {\enquote {\bibinfo {title} {Saddle-point scrambling
  without thermalization},}\ }\href {\doibase 10.1103/PhysRevA.103.033304}
  {\bibfield  {journal} {\bibinfo  {journal} {Phys. Rev. A}\ }\textbf {\bibinfo
  {volume} {103}},\ \bibinfo {pages} {033304} (\bibinfo {year}
  {2021})}\BibitemShut {NoStop}%
\bibitem [{\citenamefont {Swingle}\ \emph {et~al.}(2016)\citenamefont
  {Swingle}, \citenamefont {Bentsen}, \citenamefont {Schleier-Smith},\ and\
  \citenamefont {Hayden}}]{Swingle2016}%
  \BibitemOpen
  \bibfield  {author} {\bibinfo {author} {\bibfnamefont {Brian}\ \bibnamefont
  {Swingle}}, \bibinfo {author} {\bibfnamefont {Gregory}\ \bibnamefont
  {Bentsen}}, \bibinfo {author} {\bibfnamefont {Monika}\ \bibnamefont
  {Schleier-Smith}}, \ and\ \bibinfo {author} {\bibfnamefont {Patrick}\
  \bibnamefont {Hayden}},\ }\bibfield  {title} {\enquote {\bibinfo {title}
  {{Measuring the scrambling of quantum information}},}\ }\href {\doibase
  10.1103/PhysRevA.94.040302} {\bibfield  {journal} {\bibinfo  {journal}
  {Physical Review A}\ }\textbf {\bibinfo {volume} {94}},\ \bibinfo {pages}
  {040302} (\bibinfo {year} {2016})}\BibitemShut {NoStop}%
\bibitem [{\citenamefont {Crooks}(2018)}]{Crooks2018}%
  \BibitemOpen
  \bibfield  {author} {\bibinfo {author} {\bibfnamefont {Gavin~E}\ \bibnamefont
  {Crooks}},\ }\bibfield  {title} {\enquote {\bibinfo {title} {Performance of
  the quantum approximate optimization algorithm on the maximum cut problem},}\
  }\href@noop {} {\bibfield  {journal} {\bibinfo  {journal} {arXiv preprint
  arXiv:1811.08419}\ } (\bibinfo {year} {2018})}\BibitemShut {NoStop}%
\bibitem [{\citenamefont {Zhou}\ \emph {et~al.}(2020)\citenamefont {Zhou},
  \citenamefont {Wang}, \citenamefont {Choi}, \citenamefont {Pichler},\ and\
  \citenamefont {Lukin}}]{Zhou2020}%
  \BibitemOpen
  \bibfield  {author} {\bibinfo {author} {\bibfnamefont {Leo}\ \bibnamefont
  {Zhou}}, \bibinfo {author} {\bibfnamefont {Sheng-Tao}\ \bibnamefont {Wang}},
  \bibinfo {author} {\bibfnamefont {Soonwon}\ \bibnamefont {Choi}}, \bibinfo
  {author} {\bibfnamefont {Hannes}\ \bibnamefont {Pichler}}, \ and\ \bibinfo
  {author} {\bibfnamefont {Mikhail~D.}\ \bibnamefont {Lukin}},\ }\bibfield
  {title} {\enquote {\bibinfo {title} {Quantum approximate optimization
  algorithm: Performance, mechanism, and implementation on near-term
  devices},}\ }\href {\doibase 10.1103/PhysRevX.10.021067} {\bibfield
  {journal} {\bibinfo  {journal} {Phys. Rev. X}\ }\textbf {\bibinfo {volume}
  {10}},\ \bibinfo {pages} {021067} (\bibinfo {year} {2020})}\BibitemShut
  {NoStop}%
\bibitem [{\citenamefont {Lysne}\ \emph {et~al.}(2020)\citenamefont {Lysne},
  \citenamefont {Kuper}, \citenamefont {Poggi}, \citenamefont {Deutsch},\ and\
  \citenamefont {Jessen}}]{Lysne2020}%
  \BibitemOpen
  \bibfield  {author} {\bibinfo {author} {\bibfnamefont {Nathan~K}\
  \bibnamefont {Lysne}}, \bibinfo {author} {\bibfnamefont {Kevin~W}\
  \bibnamefont {Kuper}}, \bibinfo {author} {\bibfnamefont {Pablo~M}\
  \bibnamefont {Poggi}}, \bibinfo {author} {\bibfnamefont {Ivan~H}\
  \bibnamefont {Deutsch}}, \ and\ \bibinfo {author} {\bibfnamefont {Poul~S}\
  \bibnamefont {Jessen}},\ }\bibfield  {title} {\enquote {\bibinfo {title}
  {Small, highly accurate quantum processor for intermediate-depth quantum
  simulations},}\ }\href@noop {} {\bibfield  {journal} {\bibinfo  {journal}
  {Physical Review Letters}\ }\textbf {\bibinfo {volume} {124}},\ \bibinfo
  {pages} {230501} (\bibinfo {year} {2020})}\BibitemShut {NoStop}%
\bibitem [{\citenamefont {Chirikov}(1979)}]{Chirikov1979}%
  \BibitemOpen
  \bibfield  {author} {\bibinfo {author} {\bibfnamefont {Boris~V}\ \bibnamefont
  {Chirikov}},\ }\bibfield  {title} {\enquote {\bibinfo {title} {{A universal
  instability of many-dimensional oscillator systems}},}\ }\href {\doibase
  10.1016/0370-1573(79)90023-1} {\bibfield  {journal} {\bibinfo  {journal}
  {Physics Reports}\ }\textbf {\bibinfo {volume} {52}},\ \bibinfo {pages}
  {263--379} (\bibinfo {year} {1979})}\BibitemShut {NoStop}%
\bibitem [{\citenamefont {Keinert}\ \emph {et~al.}(2015)\citenamefont
  {Keinert}, \citenamefont {Innmann}, \citenamefont {S\"{a}nger},\ and\
  \citenamefont {Stamminger}}]{Keinert2015}%
  \BibitemOpen
  \bibfield  {author} {\bibinfo {author} {\bibfnamefont {Benjamin}\
  \bibnamefont {Keinert}}, \bibinfo {author} {\bibfnamefont {Matthias}\
  \bibnamefont {Innmann}}, \bibinfo {author} {\bibfnamefont {Michael}\
  \bibnamefont {S\"{a}nger}}, \ and\ \bibinfo {author} {\bibfnamefont {Marc}\
  \bibnamefont {Stamminger}},\ }\bibfield  {title} {\enquote {\bibinfo {title}
  {Spherical fibonacci mapping},}\ }\href {\doibase 10.1145/2816795.2818131}
  {\bibfield  {journal} {\bibinfo  {journal} {ACM Trans. Graph.}\ }\textbf
  {\bibinfo {volume} {34}} (\bibinfo {year} {2015}),\
  10.1145/2816795.2818131}\BibitemShut {NoStop}%
\bibitem [{\citenamefont {Feigenbaum}(1979)}]{Feigenbaum1979}%
  \BibitemOpen
  \bibfield  {author} {\bibinfo {author} {\bibfnamefont {Mitchell~J}\
  \bibnamefont {Feigenbaum}},\ }\bibfield  {title} {\enquote {\bibinfo {title}
  {The universal metric properties of nonlinear transformations},}\ }\href@noop
  {} {\bibfield  {journal} {\bibinfo  {journal} {Journal of Statistical
  Physics}\ }\textbf {\bibinfo {volume} {21}},\ \bibinfo {pages} {669--706}
  (\bibinfo {year} {1979})}\BibitemShut {NoStop}%
\bibitem [{\citenamefont {Sander}\ and\ \citenamefont
  {Yorke}(2012)}]{Sander2012}%
  \BibitemOpen
  \bibfield  {author} {\bibinfo {author} {\bibfnamefont {Evelyn}\ \bibnamefont
  {Sander}}\ and\ \bibinfo {author} {\bibfnamefont {James~A}\ \bibnamefont
  {Yorke}},\ }\bibfield  {title} {\enquote {\bibinfo {title} {Connecting
  period-doubling cascades to chaos},}\ }\href@noop {} {\bibfield  {journal}
  {\bibinfo  {journal} {International Journal of Bifurcation and Chaos}\
  }\textbf {\bibinfo {volume} {22}},\ \bibinfo {pages} {1250022} (\bibinfo
  {year} {2012})}\BibitemShut {NoStop}%
\end{thebibliography}%
\end{document}